\documentclass[
reprint,
superscriptaddress,
 nofootinbib,
 amsmath,amssymb,
 aps,
notitlepage
]{revtex4-1}

\usepackage{graphicx}% Include figure files
\usepackage{dcolumn}% Align table columns on decimal point
\usepackage{bm}% bold math
\usepackage[colorlinks=true]{hyperref}% add hypertext capabilities
\usepackage{multirow}
\usepackage{subfigure}
\begin{document}

\preprint{APS/123-QED}

\title{Parkes Pulsar Timing Array constraints on ultralight scalar-field dark matter}% Force line breaks with \\

\author{Nataliya K. Porayko}
\email{nporayko@mpifr-bonn.mpg.de}
\affiliation{Max-Planck-Institut f{\"u}r Radioastronomie, Auf dem H{\"u}gel 69, D-53121 Bonn, Germany}%

\author{Xingjiang Zhu}%
 \email{xingjiang.zhu@monash.edu}
\affiliation{School of Physics and Astronomy, Monash University, Clayton, VIC 3800, Australia}
\affiliation{School of Physics, University of Western Australia, Crawley, WA 6009, Australia}
\affiliation{OzGrav: Australian Research Council Centre of Excellence for Gravitational Wave Discovery}%

\author{Yuri Levin}
\affiliation{Center for Theoretical Physics, Department of Physics, Columbia University, New York, NY 10027}
\affiliation{Center for Computational Astrophysics, Flatiron Institute, New York, NY 10010}
\affiliation{School of Physics and Astronomy, Monash University, Clayton, VIC 3800, Australia}

\author{Lam Hui}
\affiliation{Center for Theoretical Physics, Department of Physics, Columbia University, New York, NY 10027}

\author{George Hobbs}
\affiliation{CSIRO Astronomy and Space Science, P.O. Box 76, Epping, NSW 1710, Australia}

\author{Aleksandra Grudskaya}
\affiliation{Sternberg Astronomical Institute, Lomonosov Moscow State University, Universitetskii pr. 13
Moscow 119234, Russia}

\author{Konstantin Postnov}
\affiliation{Sternberg Astronomical Institute, Lomonosov Moscow State University, Universitetskii pr. 13
Moscow 119234, Russia}
\affiliation{Kazan Federal University, Kremlevskaya 18, 420008 Kazan, Russia}

\author{Matthew Bailes}
\affiliation{Centre for Astrophysics and Supercomputing, Swinburne University of Technology, P.O. Box 218, Hawthorn, VIC 3122, Australia}
\affiliation{OzGrav: Australian Research Council Centre of Excellence for Gravitational Wave Discovery}%

\author{N. D. Ramesh Bhat}
\affiliation{International Centre for Radio Astronomy Research, Curtin University, Bentley, WA 6102, Australia}

%\author{Sarah Burke-Spolaor}
%\affiliation{Department of Physics and Astronomy, West Virginia University, P.O. Box 6315, Morgantown, WV 26506, USA}

\author{William Coles}
\affiliation{Department of Electrical and Computer Engineering, University of California at San Diego, La Jolla, CA 92093, USA}

\author{Shi Dai}
\affiliation{CSIRO Astronomy and Space Science, P.O. Box 76, Epping, NSW 1710, Australia}

\author{James Dempsey}
\affiliation{CSIRO Information Management and Technology, PO Box 225, Dickson, ACT 2602, Australia}

\author{Michael J. Keith}
\affiliation{Jodrell Bank Centre for Astrophysics, University of Manchester, Manchester M13 9PL, UK}

\author{Matthew Kerr}
\affiliation{Space Science Division, Naval Research Laboratory, Washington, DC 20375, USA}

\author{Michael Kramer}
\affiliation{Max-Planck-Institut f{\"u}r Radioastronomie, Auf dem H{\"u}gel 69, D-53121 Bonn, Germany}
\affiliation{Jodrell Bank Centre for Astrophysics, University of Manchester, Manchester M13 9PL, UK}

\author{Paul D. Lasky}
\affiliation{School of Physics and Astronomy, Monash University, Clayton, VIC 3800, Australia}
\affiliation{OzGrav: Australian Research Council Centre of Excellence for Gravitational Wave Discovery}%

\author{Richard N. Manchester}
\affiliation{CSIRO Astronomy and Space Science, P.O. Box 76, Epping, NSW 1710, Australia}

\author{Stefan Os{\l}owski}
\affiliation{Centre for Astrophysics and Supercomputing, Swinburne University of Technology, P.O. Box 218, Hawthorn, VIC 3122, Australia}

\author{Aditya Parthasarathy}
\affiliation{Centre for Astrophysics and Supercomputing, Swinburne University of Technology, P.O. Box 218, Hawthorn, VIC 3122, Australia}

\author{Vikram Ravi}
\affiliation{Cahill Center for Astronomy and Astrophysics, MC 249-17, California Institute of Technology, Pasadena, CA 91125, USA}

\author{Daniel J. Reardon}
\affiliation{Centre for Astrophysics and Supercomputing, Swinburne University of Technology, P.O. Box 218, Hawthorn, VIC 3122, Australia}
\affiliation{OzGrav: Australian Research Council Centre of Excellence for Gravitational Wave Discovery}%

\author{Pablo A. Rosado}
\affiliation{Centre for Astrophysics and Supercomputing, Swinburne University of Technology, P.O. Box 218, Hawthorn, VIC 3122, Australia}

\author{Christopher J. Russell}
\affiliation{CSIRO Scientific Computing, Australian Technology Park, Locked Bag 9013, Alexandria, NSW 1435, Australia}

\author{Ryan M. Shannon}
\affiliation{Centre for Astrophysics and Supercomputing, Swinburne University of Technology, P.O. Box 218, Hawthorn, VIC 3122, Australia}
\affiliation{OzGrav: Australian Research Council Centre of Excellence for Gravitational Wave Discovery}%

\author{Ren\'ee Spiewak}
\affiliation{Centre for Astrophysics and Supercomputing, Swinburne University of Technology, P.O. Box 218, Hawthorn, VIC 3122, Australia}

\author{Willem van Straten}
\affiliation{Institute for Radio Astronomy \& Space Research, Auckland University of Technology, Private Bag 92006, Auckland 1142, New Zealand}

\author{Lawrence Toomey}
\affiliation{CSIRO Astronomy and Space Science, P.O. Box 76, Epping, NSW 1710, Australia}

\author{Jingbo Wang}
\affiliation{Xinjiang Astronomical Observatory, Chinese Academy of Sciences, 150 Science 1-Street, Urumqi, Xinjiang 830011, China}

\author{Linqing Wen}
\affiliation{School of Physics, University of Western Australia, Crawley, WA 6009, Australia}
\affiliation{OzGrav: Australian Research Council Centre of Excellence for Gravitational Wave Discovery}%

\author{Xiaopeng You}
\affiliation{School of Physical Science and Technology, Southwest University, Chongqing, 400715, China}

\collaboration{The PPTA Collaboration}%\noaffiliation

\date{\today}% It is always \today, today,

\begin{abstract}
It is widely accepted that dark matter contributes about a quarter of the critical mass-energy density in our Universe. The nature of dark matter is currently unknown, with the mass of possible constituents spanning nearly one hundred orders of magnitude. The ultralight scalar field dark matter, consisting of extremely light bosons with $m \sim 10^{-22}$ eV and often called ``fuzzy'' dark matter, provides intriguing solutions to some challenges at sub-Galactic scales for the standard cold dark matter model. As shown by Khmelnitsky and Rubakov, such a scalar field in the Galaxy would produce an oscillating gravitational potential with nanohertz frequencies, resulting in periodic variations in the times of arrival of radio pulses from pulsars. The Parkes Pulsar Timing Array (PPTA) has been monitoring 20 millisecond pulsars at two to three weeks intervals for more than a decade. In addition to the detection of nanohertz gravitational waves, PPTA offers the opportunity for direct searches for fuzzy dark matter in an astrophysically feasible range of masses. We analyze the latest PPTA data set which includes timing observations for 26 pulsars made between 2004 and 2016. We perform a search in this data set for evidence of ultralight dark matter in the Galaxy using Bayesian and Frequentist methods. No statistically significant detection has been made.
We, therefore, place upper limits on the local dark matter density. Our limits, improving on previous searches by a factor of 2 to 5, constrain the dark matter density of ultralight bosons with $m \leq 10^{-23}$ eV to be below $6\,\text{GeV}\,\text{cm}^{-3}$ with 95\% confidence in the Earth neighborhood.
Finally, we discuss the prospect of probing the astrophysically favored mass range $m \gtrsim 10^{-22}$ eV with next-generation pulsar timing facilities.
\end{abstract}

\maketitle

%\tableofcontents

\section{Introduction}
\label{sec:Intro}
Dark matter, a concept established in the early 1930s for the purpose of explaining the observed enigmatic dynamics of disk galaxies and motion of galaxies in clusters \cite{1933AcHPh...6..110Z,1937ApJ....86..217Z,1936ApJ....83...23S}, is nowadays considered to be an essential ingredient of the Universe. It is instrumental in explaining a wide range of  astrophysical phenomena, such as strong gravitational lensing of elliptical galaxies \cite{2003ApJ...583..606K}, the dynamics of interacting clusters \cite{2004ApJ...604..596C} and the large-scale structure of the Universe \cite{2004ApJ...606..702T}. The latest analysis of temperature and polarization anisotropies of the cosmic microwave background \cite{2016A&A...594A..13P} suggested that the Universe contains 26\% dark matter, which is five times more than ordinary baryonic matter such as stars and galaxies.

The most popular dark matter candidates are weakly interacting massive
particles (WIMPs) and QCD (quantum chromodynamics) axions. We refer to both as standard cold
dark matter, or simply CDM. The CDM paradigm has met with impressive success in
matching observational data on large cosmological scales (see
\cite{2005PhR...405..279B, 2012AnP...524..535P}, for reviews). 
Recently, there has been an increased number of ideas about dark
matter that go beyond the standard paradigm, building
on old ideas in some cases
(see e.g. \cite{Battaglieri:2017aum} for an overview).

One such idea--an ultralight axion or axion-like particle--can be thought of as a generalization of the QCD axion. An axion is an angular
field; i.e., the field range is finite and periodic with a periodicity
$2\pi F_{\rm axion}$ with $F_{\rm axion}$ often referred to as the axion decay constant.
A simple axion Lagrangian has a standard kinetic term, and 
a self-interaction potential $V$ generated by non-perturbative effects (that can be approximated by instanton potential),
\begin{equation}
\label{Vphi}
V(\phi) = m^2 F_{\rm axion}^2 [1 - {\,\rm cos} (\phi/F_{\rm axion})] \, ,
\end{equation}
where $m$ is the mass of the axion $\phi$.
The non-perturbative effects are typically highly suppressed
(e.g. exponentially suppressed by an instanton action), leading to
a fairly low energy scale $\sqrt{m F_{\rm axion}}$. 
In the early Universe, the scalar field is frozen at its primordial
value, generically expected to be order of $F_{\rm axion}$. When the
Hubble expansion rate drops below the mass scale $m$, the scalar field
oscillates with an amplitude that redshifts with the expansion of the
Universe. Averaging over oscillation cycles, $\phi$ behaves like CDM
with a relic density of (see e.g. \cite{2010PhRvD..81l3530A,2017PhRvD..95d3541H})\footnote{The relic density computation follows the classic arguments of
\cite{Preskill:1982cy,Abbott:1982af,Dine:1982ah}, which were developed
for the QCD axion.}
\begin{equation}
\Omega_{\rm axion} \sim 0.1 \left( {\frac{m}{10^{-22} {\,\rm eV}}
  }\right)^{1/2}
\left( {\frac{F_{\rm axion}}{ 10^{17} {\,\rm GeV}}} \right)^2 \, .
\label{omegaAxion}
\end{equation}
String theory contains many axion candidates with $F_{\rm axion}$ somewhere in the range
$10^{16} - 10^{18}$ GeV \cite{Svrcek:2006yi}. 
Equation~(\ref{omegaAxion}) tells us that a very low $m$ is preferred
if the axion were to account for dark matter.
It should be emphasized though that there is a fairly large possible
range for $m$; in fact, the relic abundance is more sensitive
to $F_{\rm axion}$ than to $m$.
A lighter mass, e.g., $m \sim 10^{-23}$ eV, can be easily accommodated by a slightly higher $F_{\rm axion}$, though it is disfavored by astrophysical observations such as the existence and structure of dwarf galaxies\footnote{Note that the requisite $\sqrt{m F_{\rm axion}}$ is much less than the QCD scale; hence this is not the QCD axion.}.

Such an ultralight axion has a macroscopic de Broglie wavelength
$\lambda_{\text{dB}}$, given by
\begin{equation}
\frac{\lambda_{\text{dB}}}{2\pi} =\frac{\hbar}{mv} \approx 60 \, \text{pc}\left(\frac{10^{-22}\text{eV}}{m}\right)\left(\frac{10^{-3}c}{v}\right)\, ,
\end{equation}
where $v$ is the velocity,
implying wave-like phenomena on astronomically accessible scales,
unlike standard CDM. 
In linear perturbation theory, the wave-like property leads to a
suppression of power on small scales (small compared to the
Jeans scale, which is a geometric mean of the Compton and Hubble
scale). It is this property that motivated Hu, Barkana and Gruzinov 
\cite{2000PhRvL..85.1158H}
to propose an ultralight boson as an alternative to standard CDM, and to
coin the term ``fuzzy dark matter" (FDM). The term FDM refers generally to a scalar dark matter particle with
a very small mass, such that its de Broglie wavelength is macroscopic.
An ultralight axion is a particularly compelling realization.
Our constraints derived in this paper apply to the ultralight axion,
as well as the broader class of FDM.

The thinking was that the
suppression of power on small scales would help resolve certain small-scale problems of CDM, which generally have to do with CDM predicting
too much small-scale structure compared to that observed. 
There is a vast literature on this subject, but it remains a matter of
debate as to whether the perceived small-scale structure problems of
CDM are in fact amenable to astrophysical solutions (such as
feedback processes modifying the mass distribution within Galactic
halos); see \cite{Bullock:2017xww} for a review.

There exist several different bounds on the FDM model.
One class of bounds comes from measurements of the
linear power spectrum at high redshifts, such as from the microwave
background (e.g. \cite{2018MNRAS.tmp..273H}), and from the Lyman-alpha
forest \cite{2017PhRvL.119c1302I,2017PhRvD..96l3514K}.
In particular, the Lyman-alpha forest data appear to disfavor a FDM
mass lighter than about $10^{-21}$ eV. 
Another example of a bound of this kind come from
21-cm observations -- the recent detection of a global 21-cm
absorption signal at redshift around $18$ \cite{Bowman:2018yin} puts
a lower limit on the FDM mass similar to the Lyman-alpha forest bound
\cite{Schneider:2018xba,Lidz:2018fqo, 2018arXiv180901679S}.
Yet another class of bounds comes
from dynamical data on the density profiles of galaxies e.g.
\cite{Calabrese:2016hmp,Deng:2018jjz,2018arXiv180500122B}. 
Many of these bounds are subject to their own astrophysical uncertainties.
For instance, the Lyman-alpha forest bound is predicated upon the
correct modeling of fluctuations from such as the ionizing background, the temperature and feedback processes. The 21-cm bound relies on assumptions about star formation (that it tracks the halo formation
and that the fraction of baryons that form stars is less than about $5
\%$), and of course assumes the validity of the detection.
Constraints from rotation curve measurements generally make assumptions about how feedback processes, such as from stellar explosions, affect (or do not affect) density profiles.

Recently, a number of authors, based on numerical simulations and
analytical arguments, pointed out additional testable
astrophysical implications of FDM, especially in the nonlinear regime
\cite{2014NatPh..10..496S,Mocz:2015sda,Veltmaat:2016rxo,
2017PhRvD..95d3541H,Nori:2018hud,Veltmaat:2018dfz}.
A particularly interesting probe of ultralight dark matter using
pulsar timing arrays (PTAs) was pointed
out by Khmelnitsky and Rubakov \cite{2014JCAP...02..019K}.
Through purely gravitational coupling, scalar field dark matter induces periodic oscillations in gravitational potentials with frequency twice the field mass $f\sim 2m \sim 5 \times 10^{-8}\,\text{Hz}\,(m/10^{-22}\text{eV})$.
The oscillating gravitational potentials along the line of sight of
pulsars cause sinusoidal variations in the times of arrival (ToAs) of
radio pulses. 
%From the physical point of view, this is the classical Sachs-Wolfe effect \cite{1967ApJ...147...73S}.
The frequency of such variations lies right in the sensitivity band of
PTAs. This way of detecting or constraining FDM is completely
independent of other methods (and their assumptions), and provides a
useful check.
As shown in \cite{2014JCAP...02..019K, Natasha14, DeMartino:2017qsa, Blas2017} and later in this paper, the current PTA data can only be sensitive to very low-mass FDM ($m<10^{-23}$ eV). We will discuss what
would be required to probe the higher and cosmologically more favorable masses.

The concept of a PTA is to regularly monitor ToAs of pulses from an array of the most rotationally stable millisecond pulsars \cite{1978SvA....22...36S,Detweiler79,Hellings_Downs,1990ApJ...361..300F}. Measured ToAs are fitted with a deterministic timing model that accounts for the pulsar spin behavior and for the geometrical effects due to the motion of the pulsar and the Earth. The difference between the observed ToAs and those predicted by the best-fit timing model are called ``timing residuals". By analyzing the pulsar timing residuals, we can obtain the information about other physical processes that affect the propagation of radio pulses through the Galaxy, for instance, the presence of ultralight scalar field dark matter in the Galaxy.

The Parkes Pulsar Timing Array (PPTA) \cite{2013PASA...30...17M} uses the 64-m Parkes radio telescope in Australia. Building on earlier pulsar timing observations at Parkes, it started in 2005 to time 20 millisecond pulsars at a regular interval of two to three weeks. PPTA and its counterparts in North America (NANOGrav) \cite{2013CQGra..30v4008M} and Europe (EPTA) \cite{2013CQGra..30v4009K} have joined together to form the International Pulsar Timing Array (IPTA) \cite{2010CQGra..27h4013H,IPTA16}, aiming for a more sensitive data set. The IPTA currently observes around 70 pulsars using the world's most powerful radio telescopes.

The first PPTA data release was published in 2013 \cite{2013PASA...30...17M}. It included six years of observations for 20 pulsars. This data set was used to search for a stochastic gravitational wave (GW) background \cite{2013Sci...342..334S}, continuous GWs \cite{2014MNRAS.444.3709Z} and GW bursts with memory \cite{2015MNRAS.446.1657W}. The second data release is still being actively developed, but for this paper, we have made use of a data set that contains observations made between 2004 and 2016 with five new pulsars added since 2010.  An early subset of this data was used to place the most constraining limit to date on the amplitude of a stochastic GW background in the nHz regime \cite{2015Sci...349.1522S}.

In this work we search for evidence of ultralight scalar field dark matter in the Galaxy using the PPTA data. A similar study was carried out, through Bayesian analysis, by Porayko and Postnov \cite{Natasha14}, using the NANOGrav 5-yr 17-pulsar data set published in \cite{NANOGrav5yr}. Our work improves on that of \cite{Natasha14} in several ways. First, we make use of an independent data set with much longer data span and smaller errors in the timing residuals. Second, we use an up-to-date Bayesian inference packages for PTA data analysis--PAL2 \cite{PAL2} and NX01 \cite{steve_taylor_2017_250258}--and include proper treatment of the noise processes. Re-analyzing the NANOGrav data with the improved analysis, we find that the sensitivity was overestimated by a factor of 10 in \cite{Natasha14}. Third, we also adopt a standard Frequentist searching method and obtain consistent results with Bayesian analysis.

Our paper is organized as follows. In Sec. \ref{sec:sigmodel}, we describe pulsar timing residuals expected in the presence of ultralight scalar field dark matter in the Galaxy. In Sec. \ref{sec:data-noise}, we introduce our data set, the likelihood function and our Bayesian and Frequentist methods to model the noise properties of PPTA data. We also present results of our noise analysis. In Sec. \ref{sec:methods-results}, we describe our search techniques and apply them to the PPTA data set. As we find no significant signals, we set upper limits on the local density of FDM in the Galaxy. In Sec. \ref{sec:future}, we discuss how the sensitivity will be improved in the future. Finally, we provide concluding remarks in Sec. \ref{sec:conclu}.

\section{The pulsar timing residuals from fuzzy dark matter}
\label{sec:sigmodel}
In this section we briefly describe the magnitude and time dependence of timing residuals induced by the scalar field dark matter in the Galaxy.
A detailed derivation can be found in \cite{2014JCAP...02..019K}.

Because of the huge occupation number, the collection of ultralight dark matter particles behaves like a classical scalar field $\phi$.
To a very good approximation, here we ignore quartic self-interaction and coupling of ultralight dark matter particles to other fields\footnote{In the axion context, the oscillation amplitude of $\phi$ gradually diminishes due to the expansion of the universe, making the quadratic $m^2 \phi^2 / 2$ an excellent approximation to the potential in Eq.~(\ref{Vphi}).} \citep{2018PhRvD..97g5017K, 2018PhRvD..97d3001B}.
The scalar action in this case can be written as
\begin{equation}
S_\phi =\int d^4 x \sqrt{-g} 
\left[\frac{1}{2}g^{\mu \nu}D_{\nu}\phi D_{\mu} \phi-\frac{1}{2}m^2
  \phi^2 \right], 
\end{equation}
to which the standard Einstein-Hilbert action for the metric should be
added.
The $\phi$ equation of motion is the Klein-Grodon-Fock
Equation: $(\Box_{g}+m^2)\phi(x)=0$. 
We are interested in a computation of $\phi$ and the metric
$g_{\mu\nu}$ inside the Galaxy. The metric is approximately Minkowski
plus corrections at the level of $10^{-6}$. 
To good approximation, $\phi$ everywhere in the Galaxy oscillates
at an angular frequency $mc^2/\hbar$ (corrections due to the momentum
of the particles and the gravitational potential are small).
The energy-momentum tensor 
to the leading order diagonal and its spatial components (pressure)
oscillate at twice the field particle mass. 
This produces time-dependent gravitational potentials $g_{00}=1+2\Phi(t)$ and $g_{ij}=-1-2\Psi(t) \delta_{ij}$ in the metric tensor (in the Newtonian covariant form) with leading oscillating contributions at a frequency
\begin{equation}
f=\frac{2m c^{2}}{h} \approx 4.8\times 10^{-8} \left(\frac{m}{10^{-22}\,\text{eV}}\right)\, \text{Hz}\, .
\end{equation}
The amplitude of oscillating parts of the potentials $\Psi$ and $\Phi$ are a factor of $(v/c)^2$ smaller than the time-independent parts $\Phi_0=-\Psi_0 \sim G\rho_{\text{SF}}\lambda_{dB}^2$, where $\rho_{\text{SF}}$ is the local scalar field dark matter density.
For cosmologically favoured boson masses $\sim 10^{-22}$ eV, the frequency is fortuitously located in the sensitivity range of PTAs.

As in the case of GWs \cite{Detweiler79}, a pulsar emitting pulses propagating in a time-dependent metric experiences a frequency shift $\delta\nu$, which is related to timing residuals \cite{2014JCAP...02..019K}
\begin{eqnarray}
&& s(t)=\int_{0}^{t}\frac{\delta \nu}{\nu}dt= 
\frac{\Psi_c(x_{e})}{2\pi f}\sin[2\pi f
   t+2\alpha(x_{e})] \nonumber \\
&& \quad \quad \quad \quad \quad
-\frac{\Psi_c(x_{p})}{2\pi f}\sin\left[2\pi f \left(t-\frac{d_{p}}{c}\right)+2\alpha(x_{p})\right] \nonumber \\
&& \quad \quad \quad \quad \quad + \left(\frac{\Psi+\Phi}{2\pi f}\right) \mathcal{O}\left(\frac{v}{c}\right),
\label{eq:signalt}
\end{eqnarray}
where $d_p$ is the distance to the pulsar and $\Psi_c$ is the amplitude of cosine component of the oscillating part of the energy-momentum tensor. The subsequent terms in Eq.~(\ref{eq:signalt}) are suppressed with respect to $\Psi_c$ by a factor $v/c \simeq 10^{-3}$, and to the leading order the signal $s(t)$ does not depend on the oscillating part of the potential $\Phi$. %\footnote{Note, that time-independent parts of the gravitational potentials are equal ($\Phi_{0}=\Psi_{0}$) and by a factor of $(v/c)^{-2}=10^6$ larger than $\Psi_c$.}

As one can see in Eq.~(\ref{eq:signalt}), the dark matter signal also has ``Earth" and ``pulsar" terms. Oscillation frequencies at the Earth and at the pulsar are identical, which makes it analogous to the case of nonevolving continuous GWs \cite{ZhuPTA16}. The scalar-field oscillation phases on the Earth $\alpha(x_{e})$ and pulsar $\alpha(x_{p})$ generally take different values; but they become correlated when the Earth and a pulsar are located within the coherence de Broglie wavelength $\lambda_{\text{dB}}$.

The amplitude $\Psi_c$, which can be effectively probed in pulsar timing experiments, depends on the local density of dark matter $\rho_{\text{SF}}$,
\begin{equation}
\Psi_c=\frac{G \rho_{\text{SF}}}{\pi f^2}\approx 6.1 \times 10^{-18} \left(\frac{m}{10^{-22}\,\text{eV}}\right)^{-2} \left(\frac{\rho_{\text{SF}}}{\rho_{0}}\right)\, ,
\label{eq:Psicf}
\end{equation}
where $\rho_{0}=0.4 \, \text{GeV cm}^{-3}$ is the measured local dark matter density \cite{2012ApJ...756...89B,DMdensityReview14,DMdensity18}.
The root-mean-square (rms) amplitude of induced pulsar-timing residuals is
\begin{equation}
\delta t \approx 0.02 \,\text{ns}\left(\frac{m}{10^{-22}\,\text{eV}}\right)^{-3}\left(\frac{\rho_{\text{SF}}}{0.4\, \text{GeV cm}^{-3}}\right)\, .
\label{eq:deltatf}
\end{equation}
The expected signal amplitude scales strongly with the boson mass. At $10^{-22}$ eV and above, the signal is negligibly small. For mass below $10^{-23}$ eV, the induced rms residuals ($\gtrsim 20$ ns) are comparable to current timing precision for the best pulsars, as we discuss in Sec. \ref{sec:data-obs}.

In this work, we assume the Earth term and pulsar terms have the same amplitude $\Psi_c$. This is a reasonable approximation since most PPTA pulsars are relatively close ($\lesssim 1$ kpc) to the Earth (see Table \ref{tab:PPTA26p}). We discuss effects of the dark matter density variability in Sec. \ref{sec:future}. Under this assumption, Eq.~(\ref{eq:signalt}) can be written into a more compact form,
\begin{equation}
s(t)=\frac{\Psi_{c}}{\pi f}\sin(\alpha_{e}-\theta_{p})\cos(2\pi f t+\alpha_{e}+\theta_{p})\, ,
\label{eq:signalt1}
\end{equation}
where we have defined $\alpha_{e}=\alpha(x_{e})$ and $\theta_{p}=\alpha_{p}-\pi f d_{p}/c$ with $\alpha_{p}=\alpha(x_{p})$. Defining $\theta_{p}$ this way allows us searching for a single phase parameter per pulsar. One should note, however, that the parameter pair $(\alpha_{e},\, \theta_{p})$ is indistinguishable from $(\alpha_{e}\pm \pi,\, \theta_{p}\pm \pi)$.

\section{PPTA data and noise properties}
\label{sec:data-noise}
\subsection{Observations and timing analysis}
\label{sec:data-obs}

Here we provide a brief overview of the data set used in this work. The data set is available from the CSIRO pulsar data archive\footnote{\url{https://doi.org/10.4225/08/5afff8174e9b3}}.
The observing systems and data processing techniques are similar to the first data release (DR1) as described in Ref. 	\cite{2013PASA...30...17M}. Table \ref{tab:PPTA26p} summarizes key characteristics of the PPTA data set, including the median ToA uncertainties, weighted rms values of timing residuals, data spans and the number of observations.
	
Our data set consists of observations for 26 pulsars collected between 2004, February 5 and January 31, 2016 using the Parkes telescope. It includes DR1 data that were acquired between March 2005 and March 2011 for 20 pulsars, along with some earlier data for some pulsars that were observed in the 20-cm observing band prior to the official start of the PPTA project. Currently, the PPTA observes 25 pulsars, with PSR J1732$-$5049 having been removed from the pulsar sample in 2011 because we were unable to obtain high quality data sets for this pulsar.
The observing cadence is normally once every two to three weeks. In each session, every pulsar was observed in three radio bands (10, 20 and 50 cm) with a typical integration time of one hour. Twenty of these pulsars were monitored for more than ten years up to twelve years; only five pulsars have data spans less than five years. For this data set, the median ToA uncertainties vary from 149 ns (PSR J0437$-$4715) to 4.67 $\mu$s (PSR J2124$-$3358); the weighted rms residuals in this data set vary from 152 ns (PSR J0437$-$4715) to 16.53 $\mu$s (PSR J1824$-$2452A). PSRs J1939+2134 and J1824-2452A were excluded from the search analysis, as they show strong evidence for a large unmodeled red-noise component\footnote{This is evident as their rms residuals are much larger than the median ToA uncertainties given in Table \ref{tab:PPTA26p}. This may be accounted for using system- and band-specific noise terms \citep{2016MNRAS.458.2161L}.}. For our purpose, we find these two pulsars make little contribution to the sensitivity.

\begin{table}
 \begin{tabular}{lcccccc}
  \hline
  \multirow{2}{*}{Pulsar Name} & $\sigma$ & rms & $T_{\rm{obs}}$ & \multirow{2}{*}{Range} & \multirow{2}{*}{$N_{\rm{obs}}$} & $d_p$\\
  & ($\mu$s) & ($\mu$s) & (yr) & & & (kpc)\\
  \hline
J0437$-$4715 & 0.15 & 0.15 & 11.98 & 2004.02$-$2016.01 & 3820 & 0.16 \\
J0613$-$0200 & 1.20 & 1.38 & 11.98 & 2004.02$-$2016.01 & 969 & 0.78 \\
J0711$-$6830 & 3.29 & 1.58 & 11.98 & 2004.02$-$2016.01 & 1017 & 0.11\\
J1017$-$7156 & 0.97 & 0.76 & 5.54  & 2010.07$-$2016.01 & 524 & 0.26\\
J1022+1001 & 2.23 & 2.11 & 11.98  &  2004.02$-$2016.01 & 1008 & 1.13\\
J1024$-$0719 & 3.39 & 3.61 & 11.87 & 2004.02$-$2015.12 & 679 & 1.22\\
J1045$-$4509 & 3.82 & 3.35 & 11.98 & 2004.02$-$2016.01 & 854 & 0.34\\
J1125$-$6014 & 1.59 & 1.29 & 10.12 & 2005.12$-$2016.01 & 203 & 0.99\\
J1446$-$4701 & 1.81 & 1.47 & 5.19 &  2010.11$-$2016.01 & 161 & 1.57\\
J1545$-$4550 & 1.08 & 1.01 & 4.74  & 2011.05$-$2016.01 & 215 & 2.25\\
J1600$-$3053 & 0.91 & 0.71 & 11.98 & 2004.02$-$2016.01 & 969 & 1.80\\
J1603$-$7202 & 2.13 & 1.43 & 11.98 & 2004.02$-$2016.01 & 747 & 0.53\\
J1643$-$1224 & 1.75 & 2.96 & 11.98 & 2004.02$-$2016.01 & 713 & 0.74\\
J1713+0747 & 0.38 & 0.24 & 11.98  &  2004.02$-$2016.01 & 880 & 1.18\\
J1730$-$2304 & 2.01 & 1.48 & 11.98 & 2004.02$-$2016.02 & 655 & 0.62\\
J1732$-$5049 & 2.55 & 2.75 & 7.23  & 2004.03$-$2011.12 & 144 & 1.87
\\J1744$-$1134 & 0.68 & 0.61 & 11.98 & 2004.02$-$2016.01 & 855 & 0.40\\
J1824$-$2452A & 2.67 & 16.5 & 10.36 & 2005.05$-$2015.10 & 339 & 5.50\\
J1832$-$0836 & 0.53 & 0.25 & 2.86  & 2012.11$-$2015.10  & 68 & 0.81\\
J1857+0943  & 2.00 & 1.93 & 11.98  & 2004.02$-$2016.01 & 580 & 1.20\\
J1909$-$3744 & 0.25 & 0.16 & 11.98 & 2004.02$-$2016.01 & 1670 & 1.14\\
J1939+2134 &  0.36 & 1.43 & 11.87  & 2004.03$-$2016.01 & 591 & 3.50\\
J2124$-$3358 & 4.67 & 2.52 & 11.98 & 2004.02$-$2016.01  & 889 & 0.41\\
J2129$-$5721 & 1.82 & 1.19 & 11.65 & 2004.06$-$2016.01 & 540 & 3.20\\
J2145$-$0750 & 1.71 & 1.16 & 11.86 & 2004.03$-$2016.01 & 881 & 0.53\\
J2241$-$5236 & 0.44 & 0.28  & 5.98 & 2010.02$-$2016.01 & 615 & 0.96\\
  \hline
 \end{tabular}
  \caption{Key characteristics of the PPTA data set: $\sigma$ - median ToA uncertainty, rms - weighted root-mean-square of timing residuals, $T_{\rm{obs}}$ - data span and its start and end months, $N_{\rm{obs}}$ - number of observations, $d_p$ - pulsar distance taken from the ATNF Pulsar Catalogue \cite{ATNF05Pulsar}. }
 \label{tab:PPTA26p}
\end{table}

During pulsar timing observations, ToAs are first referred to a local hydrogen maser frequency standard at the observatory. These ToAs are further transformed to Coordinated Universal Time (UTC) and then to a Terrestrial Time (TT) as published by the Bureau International des Poids et Mesures. For the current data set, we used TT(BIPM2015) and adopted the JPL DE418 \cite{DE418} solar system ephemeris (SSE) model to project ToAs to the solar-system barycenter. Potential errors in SSE models are accounted for in our Bayesian analysis (Sec. \ref{sec:Buplim}).

Before performing the search for dark matter signals, we fit pulsar ToAs with a timing model using the standard \texttt{TEMPO2} software package \cite{TEMPO2,TEMPO2model}. Typical parameters included in this fit are the pulsar sky location (RAJ and DecJ), spin frequency and spin-down rate, dispersion measure, proper motion, parallax and (when applicable) binary orbital parameters. Additionally, constant offsets or jumps were fitted among ToAs collected with different receiver/backend systems. Below we describe our methods to estimate the noise properties of the PPTA data.

\subsection{The likelihood function}
\label{sec:anal_meth}
% In order to investigate the sensitivity of PPTA DR2 dataset to the signal $s(t)$, generated by the fuzzy dark matter Halo, in the presence of noise we have applied two different statistical methods. One is known as frequentist approach, within which the "true" parameters of our model are postulated to be unknown, but fixed. In contrast to frequentist, within Bayesian approach we are interested in reconstruction of the posterior probability of the unknown parameters of the model, which are in this case considered as random. Both approaches are used broadly in astrophysics and particularly, in pulsar timing analysis, and have been extensively discussed in the literature (e.g. \cite{2009MNRAS.395.1005V}, \cite{2011MNRAS.414.3117V}, \cite{2013PhRvD..87j4021L}).

The likelihood function for pulsar timing residuals, marginalized over the $m$ timing model parameters, can be written as \cite{2009MNRAS.395.1005V, vanHaasteren13},
\begin{equation}
\begin{split}
\mathcal{L}(\bm{\vartheta}, \bm{\psi}| \bm{\delta t} )=\frac{\sqrt{\det (M^T C^{-1} M)^{-1}}}{\sqrt{(2\pi)^{n-m} \det C}} 
\\
\times \exp\left[-\frac{1}{2} ( \bm{\delta t}-\bm{s}')^T C' (\bm{ \delta t}-\bm{s}')\right],
\label{likel}
\end{split}
\end{equation}
where $\bm{\delta t}$ is a vector of timing residuals with length $n$, $\bm{s}'$ is the deterministic signal vector, including the dark matter signal as described in Sec. \ref{sec:sigmodel} and deterministic systematics, $M$ is the $(n\times m)$ design matrix or regression matrix of the linear model \cite{Numerical_Recipes} that describes how ToAs depend on timing model parameters\footnote{It can be obtained with the \texttt{TEMPO2 designmatrix} plugin.}.
The noise covariance matrix $C=C_{\text{WN}}+C_{\text{SN}}+C_{\text{DM}}$ includes contributions from uncorrelated white noise ($C_{\text{WN}}$), time-correlated spin noise ($C_{\text{SN}}$) and dispersion measure variations ($C_{\text{DM}}$).
In Eq.~(\ref{likel}), we have defined $C'=C^{-1}-C^{-1}M(M^T C^{-1} M)^{-1}M^T C^{-1}$. The covariance matrix $C$ depends on the set of noise parameters $\bm{\vartheta}$, and $\bm{\psi}$ denotes deterministic signal parameters so that $\bm{s}'=\bm{s}'(\bm{\psi})$. As a result, this form of the likelihood, which was first implemented in \cite{2009MNRAS.395.1005V}, depends both on $\bm{\vartheta}$ and $\bm{\psi}$, and provides the possibility of proper treatment of the signal in the presence of correlated noise and systematics.
The likelihood in Eq.~(\ref{likel}) can be further reduced to a more compact form (see Ref. \cite{vanHaasteren13} for details),
\begin{equation}
\begin{split}
\label{likelg}
\mathcal{L}(\bm{\vartheta}, \bm{\psi}|\delta \bm{t})=\frac{1}{\sqrt{(2\pi)^{n-m} \det(G^TCG)}} 
\\
\times \exp\left[-\frac{1}{2} ( \bm{ \delta t}-\textbf{s}')^T G(G^T C G)^{-1} G^T (\bm{\delta t}-\textbf{s}')\right],
\end{split}
\end{equation}
where the $n\times (n-m)$ dimension matrix $G$ is obtained through the singular-value decomposition of the design matrix $M$. Specifically, $M=USV^{*}$ where $U$ and $V$ are unitary matrices with $n \times n$ and $m \times m$ dimension respectively, and $S$ is an $n \times m$ diagonal matrix containing singular values of $M$. The $G$ matrix is obtained such that $U=(U_{1}\, G)$ with $U_{1}$ and $G$ consisting of the first $m$ and the remaining $n-m$ columns of $U$ respectively.

In this work, we assume that only the dark matter signal, noise processes (that will be described in the next subsection) and deterministic systematics, associated with SSE errors, contribute to the data. We neglect errors in terrestrial time standards and other common noise processes (such as a stochastic GW background). Therefore, the likelihood function for the full PTA can be expressed as a product:
\begin{equation}
\mathcal{L}(\bm{\vartheta}, \bm{\psi}|\delta \bm{t})=\prod_{i=1}^{N_{p}} \mathcal{L}(\bm{\vartheta}_{i}, \bm{\psi}_{i}|\delta \bm{t}_{i})\, ,
\label{eq:likePTA}
\end{equation}
where $N_{p}$ is the number of pulsars in the timing array.

\subsection{Noise modeling}
\label{sec:noiseanalysis}
% \subsection{Single pulsar analysis}
For each pulsar in the PPTA data set, we estimate its noise properties using both Bayesian and Frequentist approaches.
We present a general description of possible noise sources here.

%The current status on pulsar timing noise modeling, supplemented by the detailed descriptions of the noise properties and methods of parameter estimation, can be found in e.g. \cite{2016MNRAS.457.4421C}, \cite{2016MNRAS.458.2161L}.
% Here we will just remind the reader about the main concepts, used to account for pulsar timing noise. 
Stochastic noise processes can be divided into the time-correlated and uncorrelated components. The uncorrelated (white) noise is represented by the uncertainties of the measured ToAs derived through cross-correlation of the pulsar template and the integrated profile.
However, it is common that ToA uncertainties underestimate the white noise present in pulsar timing data. This might be caused by, e.g. radio frequency interference, pulse profile changes or instrumental artifacts. Two parameters, namely, {\tt EFAC} (Error FACtor) and {\tt EQUAD} (Error added in QUADrature), are included to account for excess white noise. They are introduced for each observing system used in the data set. Following standard conventions, different parameterizations are used for {\tt EFAC} and {\tt EQUAD}. In {\tt TEMPO2} and for our Frequentist analysis, the re-scaled ToA uncertainties ($\sigma_{s}$) are related to their original values ($\sigma$) by
\begin{equation}
\sigma_{s}^{2}=\texttt{T2EFAC}^2(\sigma^2 + \texttt{T2EQUAD}^2).
\end{equation}
In Bayesian analysis, we use the following relation
\begin{equation}
\sigma_{s}^{2}=(\texttt{EFAC}\, \sigma)^2 + \texttt{EQUAD}^2.
\end{equation} 

Numerous studies \cite{1972ApJ...175..217B,Blandford84rednoise,2010MNRAS.402.1027H} have found evidence for additional low-frequency noise in pulsar timing data. This time-correlated stochastic process is dominated by two components: \textit{achromatic} (i.e, independent of radio frequency) spin noise and \textit{chromatic} (i.e, dependent on radio frequency) such as dispersion measure (DM) variations. The former is intrinsic to the pulsar and might be related to pulsar rotational instabilities. The latter is associated with the interstellar medium which introduces time delays in pulsar ToAs. As pulsar travels in the tangent plane, the line of sight intersects spatially variable interstellar medium characterized by different column electron densities. For current receivers, the bandpass is generally not broad enough to resolve these kind of variations in each individual observation. Therefore, a typical strategy is to observe pulsars at widely separated radio bands, allowing the correction of DM variations.

Below we discuss details of noise modeling in the Bayesian and Frequentist frameworks.

\subsubsection{Bayesian framework}
\label{sec:bayes1}

The Bayesian framework provides a consistent approach to the estimation of a set of parameters $\bm{\Theta}$ by updating the initial distribution of those parameters $P_{\text{pr}}(\bm{\Theta})$ as more information becomes available. Bayes' theorem states:
\begin{equation}
P_{\text{pst}}(\bm{\Theta}|\bm{D} )=\frac{\mathcal{L}(\bm{\Theta}|\bm{D})P_{\text{pr}}(\bm{\Theta})}{Z},
\end{equation}
where $P_{\text{pst}}(\bm{\Theta}|\bm{D} )$ stands for the posterior (or updated) distribution of the parameters $\bm{\Theta}$, given the data (or external information) $\bm{D}$, $\mathcal{L}(\bm{\Theta}|\bm{D})$ is the likelihood function, and $Z$ is known as Bayesian evidence and defined as:
\begin{equation}
\label{eq:evid}
Z=\int \mathcal{L}(\bm{\Theta}|\bm{D})P_{\text{pr}}(\bm{\Theta})d^n(\bm{\Theta})
\end{equation}
The Bayesian evidence is a normalizing factor for parameter estimation problem and is a key criterion for model selection and decision making. Here $Z$ does not depend on $\bm{\Theta}$ and it holds that $P_{\text{pst}}(\bm{\Theta}|\bm{D} )\propto \mathcal{L}(\bm{\Theta}|\bm{D})P_{pr}(\bm{\Theta})$. When applied for the case of PTAs, data $\bm{D}$ includes an array of pulsar timing ToAs $\bm{\delta t}$, $\bm{\Theta}$ includes $[\bm{\vartheta}, \bm{\psi}]$ and the likelihood $\mathcal{L}(\bm{\Theta}|\bm{D})$ is given by Eq.~(\ref{likel}). The set of parameters, used for the Bayesian analysis, and the corresponding priors are described in Table \ref{tab:prior}.

\begin{table*}
\caption{List of parameters and prior distributions used for the Bayesian analysis. U and log-U stand for uniform and log-uniform priors, respectively.}
\label{tab:prior}
 \begin{tabular}{lccc}
  \hline
  \hline
  Parameter & Description &  Prior & Comments \\
  \hline
  \multicolumn{4}{c}{Noise parameters ($\bm{\vartheta}$)} \\
  \texttt{EFAC} & White-noise modifier per backend & U[0, 10] & fixed for setting limits \\
  \texttt{EQUAD} & Additive white noise per backend & log-U[$-$10, $-$4] & fixed for setting limits \\
  $A_\text{SN}$ & Spin-noise amplitude &  log-U[$-$20, $-$11] (search) & one parameter per pulsar \\
   & &  U[$10^{-20}$, $10^{-11}$] (limit) & \\
  $\gamma_{SN}$ & Spin-noise spectral index & U[0, 7] & one parameter per pulsar \\
  $A_{DM}$ & DM-noise amplitude & log-U[$-$20, $-$6.5] (search) & one parameter per pulsar\\
   & & U[$10^{-20}$, $10^{-6.5}$] (limit) & \\
  $\gamma_{DM}$ & DM-noise spectral index & U[0, 7] & one parameter per pulsar \\
  \hline
  \multicolumn{4}{c}{Signal parameters ($\bm{\psi}$)} \\
  $\Psi_{c}$ & Oscillation amplitude &  log-U[$-$20, $-$12] (search) & one parameter per PTA \\
  &  &  U[$10^{-20}$, $10^{-12}$] (limit) &  \\
  $\alpha_{\text{e}}$ & Oscillation phase on Earth &  U[0, 2$\pi$] & one parameter per PTA \\
  $\theta_{\text{p}}$ & $\theta_{p}=\alpha_{p}-\pi f d_{p}/c$ & U[0, 2$\pi$] & one parameter per pulsar \\
  $f$ (Hz) & Oscillation frequency & log-U[$-$9, $-$7]  & delta function for setting limits \\
  \hline
    \multicolumn{4}{c}{\texttt{BayesEphem} parameters ($\bm{\psi}^{\text{sys}}$)} \\
  $z_{\text{drift}}$ & Drift-rate of Earth's orbit about ecliptic z-axis &  U[$-10^{-9}$, $10^{-9}$] rad yr$^{-1}$ & one parameter per PTA \\
  $\Delta M_{\text{jupiter}}$ & Perturbation of Jupiter's mass & $\mathcal{N}(0, 1.5 \times 10^{-11}) M_{\odot}$   & one parameter per PTA \\
  $\Delta M_{\text{saturn}}$ & Perturbation of Saturn's mass & $\mathcal{N}(0, 8.2 \times 10^{-12}) M_{\odot}$ & one parameter per PTA \\
  $\Delta M_{\text{uranus}}$ & Perturbation of Uranus' mass & $\mathcal{N}(0, 5.7 \times 10^{-11}) M_{\odot}$  & one parameter per PTA \\
   $\Delta M_{\text{neptune}}$ & Perturbation of Neptune's mass & $\mathcal{N}(0, 7.9 \times 10^{-11}) M_{\odot}$  & one parameter per PTA \\
   $PCA_{i}$ & Principal components of Jupiter's orbit & U[$-$0.05, 0.05] & six parameters per PTA \\
  \hline
  \hline
  \end{tabular}
\end{table*}

For computational purposes, the noise covariance matrix $C$ from Eq.~(\ref{likel}) can be split as a sum of a diagonal matrix $C_{\text{WN}}$ and a large dense matrix $K=C_{\text{SN}}+C_{\text{DM}}=F \Phi F^T$, where $\Phi=\Phi_{\text{SN}}+\Phi_{\text{DM}}$ is the diagonal matrix ($2k \times 2k$), $k<<n$, where $k$ is the number of terms in the approximation sum. By using the Woodbury matrix lemma\footnote{$(N+F \Phi F^T)^{-1} = N^{-1} - N^{-1} F (\Phi^{-1} +F^T N^{-1} F) ^{-1} F^T N^{-1}$} \cite{Hager}, the computationally heavy inversion of covariance matrix $C$, involving $\mathcal{O}(n^3)$ operations, is reduced to lower rank diagonal matrix inversion $\Phi^{-1}$. More details on this technique can be found in \cite{2015MNRAS.446.1170V}, \cite{2014ApJ...794..141A}. 

In this work we have used the so-called ``Fourier-sum" prescription (or ``time-frequency" method), introduced in \cite{2013PhRvD..87j4021L}. In this case, the Fourier transform matrix $\bm{F}$ for achromatic processes can be written as:
\begin{equation}
\begin{split}
\bm{F}=(\bm{F^s} \bm{F^c}),\\
F^s_{ji}=\sin(2\pi \nu_{i} t_{j}), F^c_{ji}=\cos(2\pi \nu_{i} t_{j})\, ,
\label{eq:Ftf}
\end{split}
\end{equation}
where $\nu_{i}=i/T$, where $T$ is the whole timespan of the PPTA data set, 11.98 years. The dimensionality of the Fourier matrix $\bm{F}$ is ($n\times 2k$), where $k$ is number of frequency components, which in our case is 30. The noise vector for a specific noise process can be expressed as $\tau_{j}=\sum \limits_{i}F_{ji}a_i=\sum \limits_{i} a^s_{i}\sin{2\pi \nu_{i} t_{j}}+a^c_{i}\cos{2\pi \nu_{i} t_{j}}$, where $\bm{a}=(\bm{a^s}, \bm{a^c})$ is the vector of Fourier coefficients. 

The covariance matrix of Fourier coefficients $\Phi$ can be derived from the covariance matrix of the theoretical power spectrum of a specific type of noise. Within Bayesian framework, we use the following parametrization for power-law noise:
\begin{equation}
\label{eq:PfredBayes}
P(f)=\frac{A^2}{12 \pi^2} \text{yr}^{3} \left(\frac{f}{\text{yr}^{-1}}\right)^{-\gamma}\, .
\end{equation}
Therefore, the elements of the matrix $\Phi$, which are identical for both spin and DM noises, are expressed as:
\begin{equation}
\Phi_{ij}=\frac{A^2}{12 \pi^2} \frac{\nu_{i}^{-\gamma}}{T}\text{yr}^{3}\delta_{ij}\, ,
\end{equation}
where $i, j$ iterates over different Fourier frequencies and $\delta_{ij}$ is a Kronecker delta. If multiband observations are available, the degeneracy between the spin noise and DM contributions can be broken, because of the dependency of the amplitude of the DM variations on the observational frequency $f_{o}$. This dependency enters in the Fourier transform matrix as:
\begin{equation}
\begin{split}
\bm{F_{\textrm{DM}}}=(\bm{F^s_{\textrm{DM}}} \bm{F^c}_{\textrm{DM}}),\\
F^s_{\text{DM},ji}=\frac{\sin(2\pi \nu_{i} t_{j})}{Kf_{o,j}^2}, F^c_{\text{DM},ji}=\frac{\cos(2\pi \nu_{i} t_{j})}{Kf_{o,j}^2}\, ,
\label{eq:FmatrixDM}
\end{split}
\end{equation}
where $K=2.41 \times 10^{-16}$Hz$^{-2}$cm$^{-3}$pc s$^{-1}$ and $f_{o, j}$ is the radio observing frequency at time $t_{j}$. Using this terminology, the time delay $\delta t$ between signal received at radio frequency $f_{0}$ and one received at $f\to\infty$ is given by $\delta t=K^{-1}f_{0}^{-2}\text{DM}=4.15 \times 10^6f_{0}^{-2}\text{DM}\text{ ms }$. Note that the linear and quadratic trends in DM variations get absorbed by timing model parameters DM1 and DM2, which are included in the Bayesian timing model. The inclusion of the DM derivatives in our analysis absolves us from the spectral leakage problem \cite{Lentati14}.

The formalism, described in this subsection, was implemented in a range of publicly available codes. For the single pulsar analysis we have used \href{https://github.com/jellis18/PAL2}{\texttt{PAL2}} Software -- a package for the Bayesian processing of the pulsar timing data. Efficient sampling from the posteriors is performed by the Bayesian inference tool \texttt{MULTINEST} \cite{MULTINEST09}, running in constant efficiency mode -- a computational technique that allows one to maintain the user-defined sampling efficiency for high-dimensional problems (see Ref. \cite{2013arXiv1306.2144F} for more details). For each PPTA pulsar we perform separately a full noise modeling analysis, simultaneously including all stochastic components discussed above. The noise parameters $\bm{\vartheta}$, estimated within single pulsar analysis, are given in Table \ref{tab:PPTA_bayes}. The marginalized posterior probabilities for the six most sensitive pulsars in PPTA (see Sec. \ref{sec:Buplim}) are presented in Appendix \ref{app:a}.

% As for the Bayesian analysis DM variations are modeled with chromatic two-parameter power-law
% described by the amplitude $A_{\text{DM}}$ and slope $\gamma_{DM}$, while the linear and quadratic trends in DM have been absorbed by timing model parameters DM1 and DM2, which have been included in our timing model. As it was shown in [], the inclusion of the DM derivatives in our analysis absolves us from the spectral leakage problem.
% Within Bayesian approach, the noise model, besides of the white noise, includes two power-law processes: achromatic spin noise, modeled with $A_{\text{SN}}$, $\gamma_{\text{SN}}$, and chromatic DM noise, parametrised with $A_{\text{DM}}$, $\gamma_{\text{DM}}$.

As was shown in \cite{KeithDM13,ColesESE}, and later confirmed in \cite{2016MNRAS.458.2161L}, data for PSR J1603$-$7202 and PSR J1713+0747 show significant evidence for nonstationary extreme scattering events (ESEs), which are usually associated with the passage of high density plasma ``blobs" along the line of sight of a pulsar. ESEs are modeled as deterministic signals $t_{\text{ESE,i}}$ \cite{2016MNRAS.458.2161L}:
\begin{equation}
t_{\text{ESE},i}=\frac{\mathcal{S}(t_{i}, \bm{A}_{\text{ESE}}, \mathcal{W})}{K f^2_{o,i}}\, ,
\end{equation}
by making use of shapelet basis function expansion:
\begin{equation}
\begin{split}
\mathcal{S}(t, \bm{A}_{\text{ESE}}, \mathcal{W})=\sum_{j=0}^{j_{\text{max}}}A_{\text{ESE,j}}B_{j}(t, \mathcal{W})\, ,\\
B_{j}(t, \mathcal{W})=[2^j j!\mathcal{W} \sqrt{\pi}]^{-1/2}H_{j}\left(\frac{t-t_{0}}{\mathcal{W}}\right)\exp\left[-\frac{(t-t_{0})^2}{2\mathcal{W}^2}\right]\, ,
\end{split}
\end{equation}
where $t_{0}$ is the epoch of ESE, $\mathcal{W}$ stands for the characteristic length scale of ESE, $H_{j}$ is the $j$-th Hermitian polynomial, $j_{\text{max}}$ is the number of terms used in the expansion, which is 3 in our case, $\bm{A}_{\text{ESE}}$ is a vector of shapelet amplitudes.
The inclusion of nonstationary ESEs in the noise model (see Table \ref{tab:PPTA_bayes}) leads to smaller DM spectral amplitudes $A_{\text{DM}}$ and slightly steeper slopes, characterised by $\gamma_{\text{DM}}$, which is consistent with results presented in \cite{2016MNRAS.458.2161L}.

\subsubsection{Frequentist methods}
\label{sec:freq1}

In the Frequentist framework, we use the method that was originally introduced in \cite{YouDM07} and further improved in \cite{KeithDM13} for correcting DM variations.
The basic idea works as follows. Timing residuals are separated into two components, one dependent on the radio wavelength, namely, dispersion measure variations -- DM(t), and the other independent of the radio wavelength. The latter could contain red noise, GWs or dark matter signals.
Since pulsar timing data are irregularly sampled, we use a linear interpolation scheme to estimate DM(t) at regular intervals. For the PPTA data, we estimate one DM(t) every 60-180 days using observations taken at three bands (10, 20, 50 cm). The time epochs and the estimated DM offsets are stored as {\tt DMOFF} parameters in the \texttt{TEMPO2 .par} files. 
We model the red spin noise on data that have been corrected for DM variations, in which case, the noise covariance matrix contains only the white noise and spin noise terms.

Following the {\tt TEMPO2} convention, for our Frequentist analysis the intrinsic spin noise is parameterized using the following power-law spectrum
\begin{equation}
\label{eq:PPTAred}
P(f)=\frac{P_0}{\left[1+\left(\frac{f}{f_c}\right)^2\right]^{\alpha/2}},
\end{equation}
where $P_0$ is an overall amplitude (normally expressed in yr$^{3}$), $f_c$ is the so-called corner frequency, $\alpha$ is the power-law exponent. The covariance matrix for such a red noise process is given by
\begin{eqnarray}
\label{eq:CVMred}
\mathcal{C}(\tau)&=&\int_{0}^{\infty}P(f)\cos \tau f {\rm{d}}f\\ \nonumber
&=&\frac{2^{(1-\alpha)/2}}{f_{c}^{-(1+\alpha)/2}}\frac{P_{0}\sqrt{\pi}\tau^{(\alpha-1)/2}J_{\frac{1-\alpha}{2}}\left(f_{c}\tau \right)}{\Gamma\left(\frac{\alpha}{2}\right)},
\end{eqnarray}
where $\tau=2\pi |t_{i}-t_{j}|$ with $t_{i}$ and $t_{j}$ being the ToA at the $i$-th and $j$-th observation respectively, $J$ is the modified Bessel function of second kind and $\Gamma$ is the Gamma function.

We follow the method described in \cite{Coles11red} to estimate red noise properties iteratively. We fit a power-law model of the form given by Eq. (\ref{eq:PPTAred}) to the power spectrum of timing residuals, leading to an initial estimate of the noise covariance matrix. We then use the Cholesky decomposition of this matrix to transform the data. The power spectrum of the transformed residuals should be white. We repeat the above procedure to obtain improved estimates of the spectrum. The iteration is considered converged if the whitened data show a sufficiently flat spectrum for which the spectral leakage is not dominant. The results are usually validated with simulations. We list our best estimates of red noise parameters in Table \ref{tab:PPTA_bayes}.

\begin{table*}
\caption{Noise properties of PPTA pulsars, determined through Bayesian and Frequentist analyses. The comparison of the results for intrinsic spin noise determined via two methods, can be performed when $f_{c}T<<1$, such as $P_{0}\to A_{SN}^2/(12\pi^2f_c^2)$. Dashed lines indicate either that noise parameters are not constrained, i.e., flat posterior probabilities (Bayesian) or that no spin noise is detected (Frequentist). In the two ``note" columns, C is for ``constrained" distributions, whereas SC stands for ``semiconstrained" distributions which exhibit long tails and significant deviation from Gaussianity (possibly due to correlation with other parameters); See Fig. \ref{fig:dm_noise} in Appendix \ref{app:a} for illustrations. The last two rows list results when parameters for nonstationary ESEs are included. Only pulsars with a $\dag$ symbol next to their names are used for setting Bayesian upper limits.}
 \label{tab:PPTA_bayes}
 \begin{tabular}{lccccccccc}
  \hline
  \hline
  \multirow{2}{*}{Pulsar Name} &\multicolumn{6}{|c}{Bayesian}  & \multicolumn{3}{|c|}{Frequentist}\\
   & $\log10(A_{SN})$ & $\gamma_{SN}$ & \textit{note} & $\log10(A_{DM})$ & $\gamma_{DM}$ & \textit{note} & $\alpha$ & $f_{c}($yr$^{-1})$ & $P_{0}($yr$^{3})$ \\
  \hline
  J0437$-$4715$^{\dag}$ & $-13.96^{+0.05}_{-0.05}$ & $2.0^{+0.2}_{-0.2}$ & C &$-10.90^{+0.04}_{-0.04}$ & $3.2^{+0.2}_{-0.2}$ & C & 3.5 & 0.08& $2.37\times 10^{-27}$\\
  J0613$-$0200& $-16.89^{+1.9}_{-1.9}$ & $3.4^{+2.0}_{-2.0}$ & SC &$-10.62^{+0.05}_{-0.05}$ & $2.1^{+0.3}_{-0.3}$ & C & 2.5 & 0.08 & $1.30\times 10^{-26}$ \\
  J0711$-$6830 & $-14.1^{+0.5}_{-0.4}$ & $4.2^{+1.2}_{-1.1}$ & C &$-12.1^{+0.8}_{-1.7}$ & $3.9^{+1.6}_{-1.7}$ & SC & 4.0 & 0.08 & $3.98\times 10^{-26}$ \\
  J1017$-$7156 & $-13.5^{+0.3}_{-0.6}$ & $3.6^{+1.9}_{-1.5}$ & C &$-10.12^{+0.06}_{-0.06}$ & $3.2^{+0.4}_{-0.4}$ & C & 6.0 & 1.0 & $9.54\times 10^{-28}$ \\
  J1022$+$1001 & $-16.9^{+2.4}_{-1.7}$ & $2.9^{+2.1}_{-2.0}$ & SC& $-11.3^{+0.3}_{-0.4}$ & $3.2^{+1.2}_{-0.8}$ & C & 2.0 & 0.08 & $3.04\times 10^{-26}$ \\
  J1024$-$0719 & $-14.6^{+0.4}_{-0.5}$ & $6.1^{+0.6}_{-0.9}$ & SC & $-11.6^{+0.4}_{-0.6}$ & $4.2^{+1.3}_{-1.2}$ & C& 3.0 & 0.08 & $4.30\times 10^{-25}$ \\
  J1045$-$4509 & $-12.85^{+0.2}_{-0.5}$ & $2.0^{+1.1}_{-0.6}$ & C & $-9.73^{+0.04}_{-0.04}$ & $2.8^{+0.3}_{-0.3}$ & C& 3.0 & 0.3 & $7.44\times 10^{-27}$ \\
  J1125$-$6014 & $-14.5^{+0.4}_{-0.4}$ & $6.0^{+0.7}_{-0.7}$ & C & $-11.6^{+0.5}_{-0.5}$ & $4.3^{+1.1}_{-1.2}$ & C& 3.0 & 0.2 & $5.79\times 10^{-27}$ \\
  J1446$-$4701 & $\cdots$ & $\cdots$ &  & $\cdots$ & $\cdots$ & & $\cdots$ & $\cdots$ & $\cdots$\\
  J1545$-$4550 & $\cdots$ & $\cdots$ & & $-10.8^{+0.3}_{-0.4}$ & $4.6^{+1.3}_{-1.3}$ & C & 3.0 & 0.1 & $1.66\times 10^{-26}$\\
  J1600$-$3053$^{\dag}$ & $-16.8^{+1.7}_{-1.9}$ & $3.3^{+2.1}_{-1.9}$ & SC & $-10.6^{+0.08}_{-0.09}$ & $2.7^{+0.3}_{-0.3}$ & C & 2.0 & 0.08 &$1.05\times 10^{-27}$ \\
  J1603$-$7202 & $-13.3^{+0.2}_{-0.5}$ & $2.4^{+1.2}_{-0.7}$ & C & $-10.20^{+0.05}_{-0.05}$ &  $2.5^{+0.3}_{-0.3}$ & C & 3.0 & 0.08 & $8.39\times 10^{-26}$\\
  J1643$-$1224 & $-12.40^{+0.05}_{-0.05}$ & $1.5^{+0.4}_{-0.3}$ & C &$-9.81^{+0.04}_{-0.04}$ & $1.6^{+0.3}_{-0.3}$ & C & 1.5 & 0.08 & $3.43\times 10^{-26}$\\
  J1713$+$0747 & $-13.5^{+0.1}_{-0.1}$ & $2.4^{+0.3}_{-0.3}$ & C  &$-10.79^{+0.07}_{-0.06}$ & $1.7^{+0.3}_{-0.3}$ & C & $\cdots$ & $\cdots$ & $\cdots$ \\
  J1730$-$2304 & $-17.2^{+1.7}_{-1.7}$ & $3.2^{+2.0}_{-2.0}$ & C & $-11.2^{+0.3}_{-0.4}$ & $3.6^{+0.9}_{-0.7}$ & C & 2.0 & 0.08 & $2.17\times 10^{-26}$\\
  J1732$-$5049 & $-16.1^{+2.3}_{-2.3}$ & $3.3^{+2.1}_{-1.9}$ & SC & $-10.6^{+0.6}_{-5.7}$ & $3.2^{+1.7}_{-1.3}$ & SC & $\cdots$ & $\cdots$ & $\cdots$ \\
  J1744$-$1134$^{\dag}$ & $-13.33^{+0.06}_{-0.06}$ & $1.2^{+0.3}_{-0.3}$ & SC & $-11.5^{+0.3}_{-0.5}$ & $3.3^{+1.2}_{-0.7}$ & SC & 6.0 & 1.0 & $2.55\times 10^{-28}$ \\
  J1824$-$2452A  & $-12.60^{+0.07}_{-0.12}$ & $3.7^{+1.4}_{-0.4}$ & SC & $-9.74^{+0.07}_{-0.06}$ & $2.5^{+0.4}_{-0.4}$ & C & 4.0 & 0.1 & $1.22\times 10^{-23}$\\
  J1832$-$0836 & $\cdots$ & $\cdots$ &  & $\cdots$ & $\cdots$ & & $\cdots$ & $\cdots$ & $\cdots$ \\
  J1857$+$0943 & $-15.1^{+1.1}_{-2.4}$ & $4.0^{+1.7}_{-2.0}$ & SC & $-10.6^{+0.1}_{-0.2}$ & $2.3^{+0.5}_{-0.5}$ & C & $\cdots$ & $\cdots$ & $\cdots$\\
  J1909$-$3744$^{\dag}$ & $-14.5^{+0.5}_{-0.7}$ & $2.4^{+1.1}_{-0.8}$ & C & $-11.09^{+0.04}_{-0.04}$ & $1.6^{+0.3}_{-0.2}$ & C & 2.5 & 0.07 & $7.54\times 10^{-28}$\\
  J1939$+$2134 & $-13.34^{+0.1}_{-0.2}$ & $3.2^{+0.6}_{-0.4}$ & C & $-10.25^{+0.04}_{-0.04}$ & $3.1^{+1.8}_{-1.5}$ & C & 4.0 & 0.08 & $2.50\times 10^{-25}$\\
  J2124$-$3358 & $\cdots$ & $\cdots$ & & $-11.9^{+0.9}_{-4.5}$ & $2.8^{+0.9}_{-0.9}$ & SC & 5.0 & 1.0 & $5.64\times 10^{-27}$\\
  J2129$-$5721 & $-16.9^{+1.8}_{-1.8}$ & $3.2^{+2.0}_{-2.0}$ & SC & $-10.9^{+0.1}_{-0.1}$ & $2.3^{+0.5}_{-0.5}$ & C & 2 & 0.08 & $1.37\times 10^{-26}$\\
  J2145$-$0750 & $-13.04^{+0.06}_{-0.06}$ & $1.4^{+0.2}_{-0.2}$ & C & $-11.1^{+0.2}_{-0.2}$ & $2.9^{+0.6}_{-0.6}$ & C & 1.0 & 0.08 & $5.13\times 10^{-27}$\\
  J2241$-$5236$^{\dag}$ & $-13.48^{+0.08}_{-0.1}$ & $1.4^{+0.6}_{-0.5}$ & C & $-12.8^{+1.0}_{-4.8}$ & $3.9^{+2.1}_{-2.4}$ & SC & $\cdots$ & $\cdots$ & $\cdots$ \\
  \hline
  & \multicolumn{6}{c}{Including extreme scattering events} & & & \\
   J1603$-$7202 & $-13.3^{+0.2}_{-0.2}$ & $2.3^{+0.5}_{-0.6}$ & C & $-10.55^{+0.08}_{-0.08}$ & $2.6^{+0.3}_{-0.3}$ & C &  &  & \\ 
   J1713$+$0747$^{\dag}$ & $-13.50^{+0.08}_{-0.08}$ & $2.3^{+0.3}_{-0.3}$ & C & $-11.2^{+0.1}_{-0.1}$ & $2.5^{+0.4}_{-0.4}$ & C &  &  & \\

  \hline
  \hline
 \end{tabular}
\end{table*}

\section{Search techniques and Results}
\label{sec:methods-results}
\subsection{Bayesian analysis}
\label{sec:Buplim}
Within a Bayesian framework, the signal detection problem is addressed through model selection. Given the observational data, we wish to choose between two mutually exclusive hypotheses: the null hypothesis $\mathcal{H}_{0}$ that the signal is absent and the alternative hypothesis $\mathcal{H}_{1}$ that the signal is present. We compute the evidences $\mathcal{Z}$, defined in Eq.~(\ref{eq:evid}), of the two hypotheses, $\mathcal{H}_0$ and $\mathcal{H}_1$. Assuming a priori equal probability for both hypotheses, the following evidence ratio (commonly called Bayes factor) quantifies the support of $\mathcal{H}_1$ against $\mathcal{H}_0$
\begin{equation}
\mathcal{B}=\frac{\mathcal{Z}_{1}}{\mathcal{Z}_{0}}=\frac{\int \mathcal{L}(\bm{\vartheta}, \bm{\psi}, \bm{\psi}^{\text{sys}}|\delta\bm{t}) P_{\text{pr}}(\bm{\vartheta}, \bm{\psi}, \bm{\psi}^{\text{sys}})\text{d}\bm{\vartheta} \text{d}\bm{\psi}^{sys}\text{d}\bm{\psi}}{\int \mathcal{L}(\bm{\vartheta},\bm{\psi}^{\text{sys}}|\delta\bm{t})P_{\text{pr}}(\bm{\vartheta}, \bm{\psi}^{\text{sys}})\text{d}\bm{\vartheta}\text{d}\bm{\psi}^{\text{sys}}},
\end{equation}
where $\bm{\psi}^{\text{sys}}$ are the parameters of the deterministic systematics, SSE errors in our case, which should be distinguished from dark matter signal parameters $\bm{\psi}$.
In order to obtain accurate evidence estimates, we carry out numerical integration via \texttt{MULTINEST} with enabled importance nested sampling in constant efficiency mode.
With the current PPTA data, we find a log Bayes factor $\ln\mathcal{B}$ of 2.1 in the frequency range [$10^{-9}, 8 \times 10^{-8}$] Hz, implying that our data are consistent with containing only noise. When we extend the search frequency to $10^{-7}$ Hz, the signal hypothesis is favored against the null hypothesis with $\ln\mathcal{B}=70$. We suspect this is caused by the unmodeled perturbations of the mass and orbital elements of Mercury, for which the synodic period is $\sim 116$ days, corresponding to a frequency of $10^{-7}$ Hz. We defer the investigation of this feature to a future work.

In order to set an upper limit on the signal amplitude within the Bayesian framework, we perform the parameter estimation routine. By sampling from the posterior probabilities of model parameters, we can numerically marginalize over nuisance parameters, and get the posterior distribution for the amplitude $\Psi_c$. We define the $95\%$ Bayesian upper limit $\bar{\Psi}_{c}$, such that $95\%$ of the samples from the posterior probability lie within the range $[0,\bar{\Psi}_{c}]$:
\begin{eqnarray}
0.95&=&\int_0^{\bar{\Psi}_{c}} \text{d} \Psi_c\int \text{d} \bm{\psi}'\text{d}\bm{\vartheta} \mathcal{L}(\Psi_{c}, \bm{\psi}', \bm{\vartheta}|\delta \bm{t})P_{\text{pr}}(\Psi_{c}) \nonumber \\ && P_{\text{pr}}(\bm{\psi}')P_{\text{pr}}(\bm{\vartheta}).
\end{eqnarray}
We split the frequency range between $10^{-9}$ and $10^{-7}$ Hz into a number of small bins and find $\bar{\Psi}_{c}$ for each bin separately.

To reduce the computational costs of numerical marginalization, a common practice is to fix the noise model parameters to their maximum likelihood values \cite{2016MNRAS.455.1665B, 2014ApJ...794..141A}, determined from single pulsar analysis. However such a procedure can possibly lead to upper limits biased by a factor of $\lesssim 2$ \cite{2014ApJ...794..141A}. In this work we allow both signal and correlated noise parameters to vary simultaneously. The white noise EFACs and EQUADs, which should have little or no correlation with dark matter parameters, are fixed to the maximum-likelihood values obtained from single pulsar analysis. 

Recently, it was shown that the search for a stochastic GW background can be seriously affected by the uncertainties in the SSE \cite{NANOGrav11yrLim,2016MNRAS.455.4339T}. We employ a physical model \texttt{BayesEphem} to account for the SSE uncertainties that are most relevant for pulsar timing. The \texttt{BayesEphem} model has 11 parameters, including 4 parameters which describe the perturbations in the masses of outer planets, 1 parameter which is associated with the uncertainty in the semi-major axis of Earth-Moon barycenter orbit, and 6 parameters that characterize  the perturbation of the Earth's orbit due to errors in the Jovian average orbital elements. The \texttt{BayesEphem} modeling technique is described in \cite{NANOGrav11yrLim} in detail, and implemented in publicly available software packages, such as \href{https://github.com/nanograv/enterprise}{\texttt{enterprise}} and \href{http://stevertaylor.github.io/NX01/}{\texttt{NX01}}. The latter was used to put robust constrains on the amplitude of the FDM in this work.

\begin{figure}
     \includegraphics[width=\columnwidth]{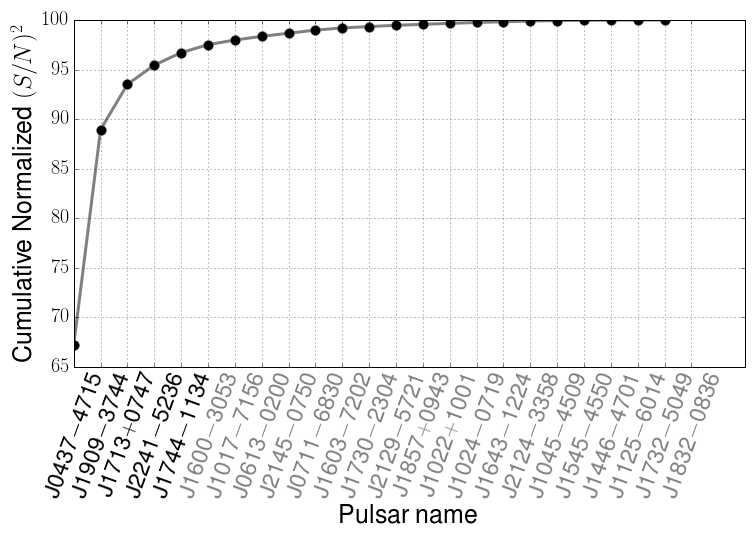}
    \caption{Cumulative normalized $(S/N)^2$. The pulsars are ranked according to their contribution to the PPTA sensitivity between $5\times10^{-9}-2\times10^{-8}$ Hz (see text for details).}
    \label{fig:grade_puls}
\end{figure}

The number of free parameters for the PPTA data set is $5\times N_{p}+ 3+11 = 144$ (see Table \ref{tab:prior}), where $N_{p}$ is the number of pulsars in PTA.
In order to further reduce the computational costs, we have formed the ``restricted data set" by choosing the five best pulsars. As shown in Fig. \ref{fig:grade_puls}, they contribute to more than 95\% sensitivity of the full PPTA. Here pulsars are ranked according to their contribution to the squared signal-to-noise ratio $(S/N)^2$; see Eq.~(\ref{eq:snr}) in the next section. We carry out the calculations by adding detectable signals to 1000 noise realizations, sampled from individual pulsar noise posterior distribution obtained in Sec. \ref{sec:bayes1}.

\subsubsection{Validation of the search results}

\begin{figure}
\includegraphics[width=\columnwidth]{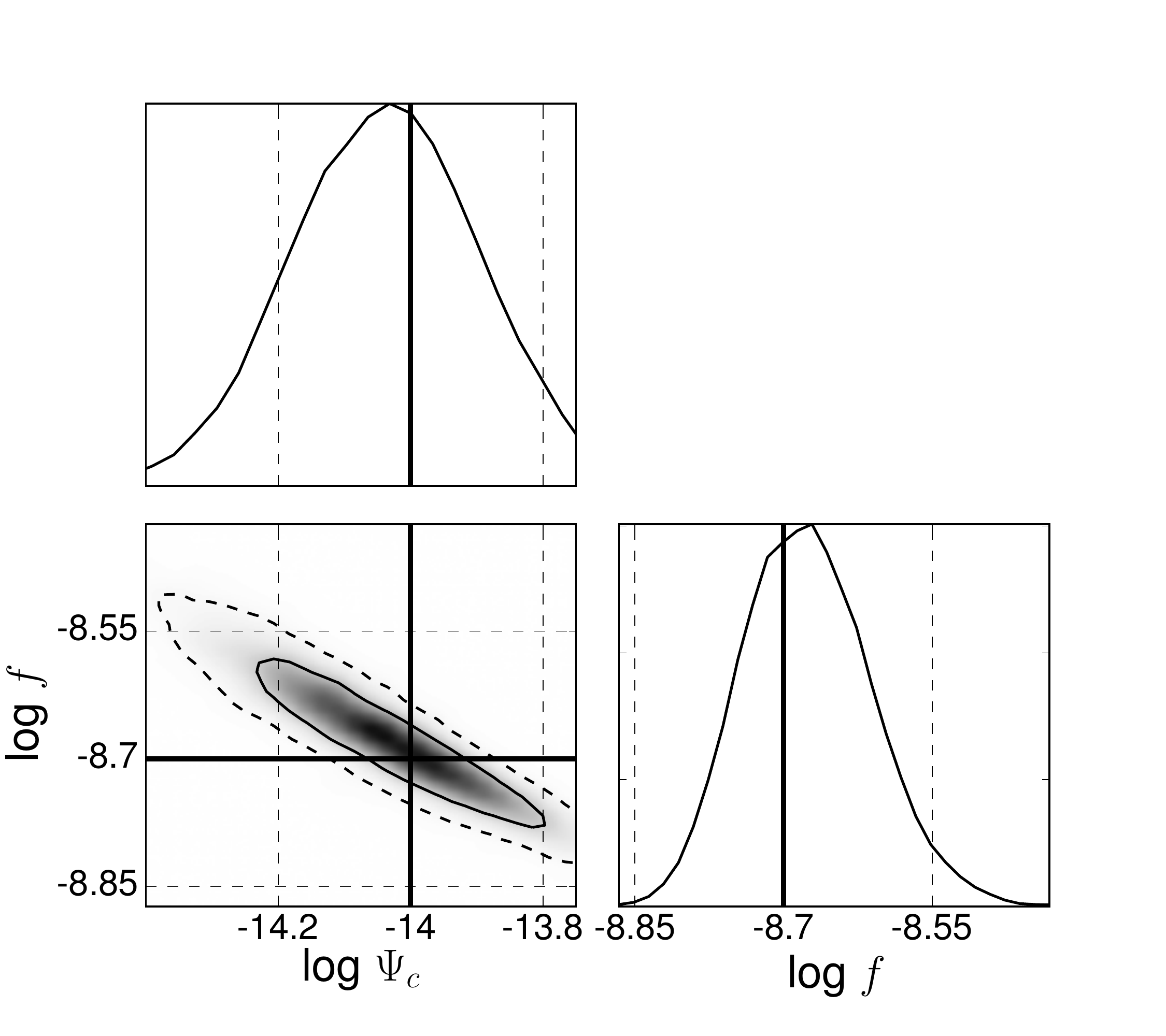}
    \caption{The marginalized posterior distributions for the amplitude $\Psi_c$ and frequency $f$ for a signal injection in the actual PPTA data. The thick black lines mark the injected values and the contours are 1- and 2-$\sigma$ credible regions.}
    \label{fig:rec_sig}
\end{figure}

In order to validate our upper limits and test the robustness of our algorithms, we have injected a signal with $f=2\times 10^{-9}$ Hz and amplitude $\Psi_c=10^{-14}$ into our restricted data set. At this frequency, the amplitude of the injected signal is comparable to the Bayesian upper limit. In order to recover this signal we run the full Bayesian analysis, simultaneously accounting for both dark matter signal and noise. The posterior probabilities are demonstrated in Fig. \ref{fig:rec_sig}, indicating successful recovery of the injected signal.

\subsection{Frequentist analysis}
In a Frequentist framework, signal detection is essentially a statistical hypothesis testing problem; we wish to choose between the null hypothesis $\mathcal{H}_{0}$ and the signal hypothesis $\mathcal{H}_{1}$ based on the observations. The task is to find an optimal statistic that maximizes the signal detection probability at a fixed false alarm probability. Following the Neyman-Pearson criterion, the log-likelihood ratio is an optimal statistic
\begin{equation}
\ln \Lambda \equiv \ln \frac{\mathcal{L}(\mathcal{H}_{1}|\delta \mathbf{t})}{\mathcal{L}(\mathcal{H}_{0}|\delta \mathbf{t})} = \sum_{i=1}^{N_{p}}\left[(\delta \mathbf{t}_{i}|\textbf{s}_{i})-\frac{1}{2}(\textbf{s}_{i}|\textbf{s}_{i})\right]\, ,
\label{eq:loglr}
\end{equation}
where we have used Eqs. (\ref{likelg})-(\ref{eq:likePTA}) to derive the second equality above, and the inner product between two time series $\textbf{x}$ and $\textbf{y}$ is defined as
\begin{equation}
(\textbf{x}|\textbf{y})=\textbf{x}^T G(G^T C G)^{-1} G^T \textbf{y}\, .
\label{eq:innerp}
\end{equation}

It is useful to define the signal-to-noise ratio in the following form
\begin{equation}
S/N=\sqrt[]{2\langle\ln \Lambda \rangle}=\left[\sum_{i=1}^{N_{p}} (\textbf{s}_{i}|\textbf{s}_{i}) \right]^{1/2}\, ,
\label{eq:snr}
\end{equation}
where $\langle...\rangle$ stands for the expectation value over a large number of noise realizations. In this work, we adopt $2\ln\Lambda$ as our detection statistic. For our Frequentist analysis, noise model parameters are fixed at their maximum likelihood values. The signal parameters in question are: the amplitude of dark matter induced gravitational-potential oscillations $\Psi_c$, oscillation frequency $f$, phase parameters $\alpha_{e}$ and $\theta_{p}$; see Eq.~(\ref{eq:signalt1}). It turns out that the statistic can be analytically maximized over $\Psi_c$ and thus the parameter space that needs to be numerically searched over is $N_{p}+2$ dimensional. For our data this corresponds to 28 dimensions, making a grid-based search unfeasible. We employ the Particle Swarm Optimization technique \cite{PSO95}, which has been demonstrated to be very effective for searches for continuous GWs with PTAs \cite{WangYan15pta,ZhuPTA16}. The detection statistic follows a $\chi^2$ distribution with one degree of freedom for noise-only data.

Since we find no evidence for statistically significant signals in the data, which is consistent with results from the Bayesian analysis as described in the previous subsection, we set upper limits on the dimensionless amplitude $\Psi_c$. We compute the 95\% confidence upper limits for a number of frequency bins between $10^{-9}$ and $10^{-7}$ Hz. At each frequency, we compute the $S/N$ for $10^{3}$ simulated signals with random phase parameters and a fixed $\Psi_c$. The 95\% confidence upper limit on $\bar{\Psi}_{c}$ corresponds to the amplitude at which 95\% of signals result in $S/N\geq 2.4$. Here the $S/N$ threshold is chosen such that the expectation value for our detection statistic in the presence of signals, given by $1+(S/N)^2$, is greater than the detection threshold that corresponds to 1\% false alarm probability. It implies that: if there was a signal with an amplitude higher than $\bar{\Psi}_{c}$ present in the data, it would have been detectable with more than 95\% probability.

\subsection{Upper limits}

\begin{figure}
\includegraphics[width=\columnwidth]{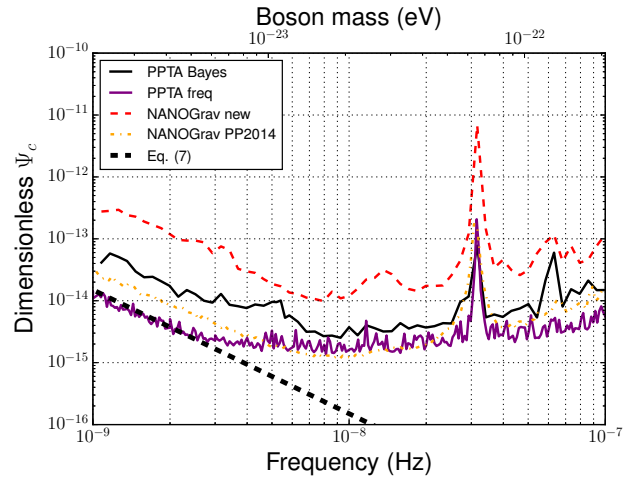}
    \caption{Upper limits on the signal amplitude $\Psi_c$, generated by the scalar field dark matter in the Galaxy, as a function of frequency (boson mass). The purple solid line shows results from Frequentist analysis of the full data set of 24 pulsars, while the black solid line demonstrates the upper limits derived within a Bayesian framework (only the five best pulsars were used). These are compared with previous studies using the NANOGrav 5-yr data set: dash-dotted orange -- upper limits set in \cite{Natasha14}, dashed red -- upper limits recalculated in this work. The thick black dashed line shows the model amplitude $\Psi_c$, assuming $\rho_{\text{SF}}=0.4\,\text{GeV}\,\text{cm}^{-3}$, given by Eq. (\ref{eq:Psicf}).}
    \label{fig:upper_lim}
\end{figure}

Figure \ref{fig:upper_lim} shows the 95\% upper limits on the amplitude $\Psi_c$, calculated within Bayesian (black solid line) and frequentist frameworks (purple solid line). As one can see, Bayesian upper limits are a factor of 2-3 worse than frequentist upper limits in the low-frequency regime, while in the mid-to-high frequency range both methods produce comparable results. The difference might be predominantly attributed to the covariance between signal and noise (especially the red spin noise). Frequentist upper limits were calculated by fixing noise parameters at their maximum likelihood values, whereas we search simultaneously over signal and noise parameters in the Bayesian analysis.

The Bayesian upper limits, obtained with 5-year NANOGrav data set \cite{NANOGrav5yr}, are also plotted as the thin dash-dotted (taken from \cite{Natasha14}) and dashed (recalculated in this paper) lines. We note that upper limits presented in Ref. \cite{Natasha14} were underestimated by a factor of 10 due to the less conservative\footnote{We note that uniform priors result in upper limits that are a factor of $\sim 5$ higher than log-uniform priors.} choice of prior (log-uniform) probability of the amplitude $\Psi_c$, as well as the noninclusion of DM variations and additional white noise terms (\texttt{EFAC} and \texttt{EQUAD}).
From Fig. \ref{fig:upper_lim}, one can see that our data set is a factor of 5 more sensitive to the dark matter signal than NANOGrav 5-year data at low frequencies, corresponding to boson masses $m\lesssim 10^{-23}$ eV. In the intermediate regime, the improvement is about a factor of 2. This is expected because of our much longer data span and higher observing cadence. It is interesting to note that the upper limit curves in Fig. \ref{fig:upper_lim} exhibit similar frequency dependencies to the sky-averaged upper limits for continuous GWs (see, e.g., \cite{2014MNRAS.444.3709Z}). In Appendix \ref{app:b}, we present Frequentist upper limits obtained by including in the analysis only Earth terms. We also show how Bayesian upper limits are modified if different fixed SSE models are used.

\section{Future prospects}
\label{sec:future}
\begin{figure}
\includegraphics[width=\columnwidth]{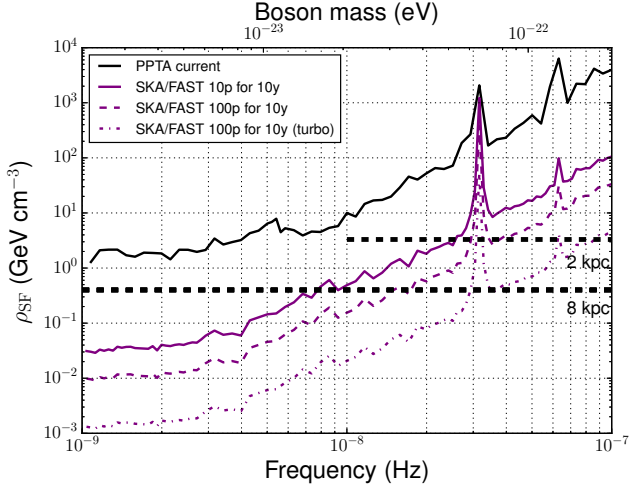}
    \caption{Upper limits on the dark matter density $\rho$ in the Galaxy. The current PPTA upper limits (black solid line) are shown along with projected limits in the FAST/SKA era (purple lines, all assuming 10-yr data span): a) 10 pulsars, 14-day cadence, 30-min integration, b) 100 pulsars, 14-day cadence, 30-min integration, and c) 100 pulsars, 1-day cadence, 2-hours integration (turbo). The black dashed lines show the dark matter density in the Halo at 8 kpc ($\rho_{\text{SF}}=0.4\,\text{GeV}\,\text{cm}^{-3}$) and 2 kpc ($\rho_{\text{SF}}=3.4\,\text{GeV}\,\text{cm}^{-3}$) from the Galactic Center, assuming NFW profile. The 8 kpc line demonstrates the predicted dark matter density, applicable to current PPTA pulsars and the Earth, while the 2 kpc line applies to pulsars located at 2 kpc distance from the Galactic Center. For boson masses $m \lesssim 4\times10^{-23}$ eV the size of the solitonic core becomes larger than 2 kpc \cite{2014NatPh..10..496S}, and the dark matter density will deviate from the NFW prediction towards higher values (see text for details).}
    \label{fig:future_lim}
\end{figure}

In this section we discuss the future improvement in sensitivity of PTAs to the dark matter signal. In particular, the Five-hundred-meter Aperture Spherical Telescope (FAST \cite{FAST11}) in China, MeerKAT \cite{MeerTime} -- a precursor for the planned Square Kilometre Array (SKA \cite{LazioSKAPTA}) -- and ultimately the SKA, are expected to significantly increase the sensitivities of PTAs. With broad frequency bands and massive collecting areas, the radiometer noise for some of the brightest pulsars can be reduced from current 100 ns level down to below 10 ns \cite{HobbsFAST}. However, it might be too optimistic to assume a white noise level of 10 ns because of the so-called jitter noise, which is thought to be associated with the intrinsic and stochastic variability in the shape of individual pulses \cite{StefanJitter}. Such a limitation implies that the timing precision stops improving for the brightest pulsars even when better instruments are used. The level of jitter noise can be approximately estimated with the following relation \cite{ShannonJitter12}
\begin{equation}
\label{eq:jitt_ex}
\sigma_{J}\approx 0.2 W \sqrt{\frac{P}{T_{\text{int}}}},
\end{equation}
where $T_{\text{int}}$ is the time of integration, $W$ and $P$ are the pulse width and pulse period, respectively. Note that the only way to reduce jitter noise is to increase $T_{\text{int}}$.
In comparison, the radiometer noise is given by \cite{HobbsFAST}
\begin{equation}
\sigma_{\text{r}}\approx \frac{W}{S/N}\approx \frac{WS_{\text{sys}}}{S_{\text{mean}}\sqrt{2\Delta f T_{\text{int}}}}\sqrt{\frac{W}{P-W}}\, ,
\label{eq:sigma_radio}
\end{equation}
where $S/N$ is the pulse profile signal-to-noise ratio, $S_{\text{sys}}$ is the system-equivalent flux density,
%$T_{\text{sys}}$ and $G$ is the system temperature and telescope gain respectively, 
$S_{\text{mean}}$ is the pulsar mean flux density and $\Delta f$ is the observing bandwidth. We adopt nominal SKA parameters\footnote{SKA1 system baseline V2 description \url{https://www.skatelescope.org/}}, $S_{\text{sys}}=1.8$ Jy,
%$T_{\text{sys}}=20$ K, $G=16.5\,\text{KJy}^{-1}$, 
$\Delta f=770$ MHz and set a fiducial $T_{\text{int}}=30$ minutes.

\begin{table}
  \caption{White noise for 10 PPTA pulsars in the FAST/SKA era.}
   \label{tab:PPTAfast10p}
 \begin{tabular}{lccc}
  \hline
  Pulsar Name & $\sigma_{\text{r}}$ (ns) & $\sigma_{J}$ (ns) & $\sigma$ (ns) \\
  \hline
%J0437$-$4715 & 0.1 & 50.4 & 50.5 \\
J0437$-$4715 & 0.06 & 50.4 & 50.4 \\
%J1017$-$7156 & 6.9 & 13.7 & 20.6  \\
J1017$-$7156 & 4.6 & 13.7 & 14.5  \\
%J1446$-$4701 & 39.0 & 22.1 & 61.1 \\
J1446$-$4701 & 26.0 & 22.1 & 34.1 \\
%J1545$-$4550 & 23.4 &36.1 & 59.5 \\
J1545$-$4550 & 15.6 &36.1 & 39.3 \\
%J1600$-$3053 & 4.4 & 26.6 & 31.0 \\
J1600$-$3053 & 2.9 & 26.6 & 26.8 \\
%J1713+0747 & 1.2 & 35.1 & 36.3 \\
J1713+0747 & 0.8 & 35.1 & 35.1 \\
%J1744$-$1134 & 5.9 & 41.2 & 47.1\\
J1744$-$1134 & 3.9 & 41.2 & 41.4\\
%J1832$-$0836 & 5.6 & 14.2 & 19.8 \\
J1832$-$0836 & 3.7 & 14.2 & 14.8 \\
%J1909$-$3744 & 1.8 & 11.2 & 13.0\\
J1909$-$3744 & 1.2 & 11.2 & 11.3\\
%J2241$-$5236 & 2.2 & 15.4 & 17.6 \\
J2241$-$5236 & 1.5 & 15.4 & 15.5 \\
  \hline
 \end{tabular}
\end{table}

% For 10 of the PSRs (J1939+2134, J1909-3744, J1017-7156, J1832-0836, J2241-5236, J1446-4701, J1600-3053, J1713+0747, J1545-4550, J1744-1134), regularly monitored with PPTA, the level of the jitter noise, estimated with Eq (\ref{eq:jitt_ex}), is less than 30 ns. Therefore, for SKA/FAST PTA sensitivity curve simulations, we have chosen those 10 PSRs, which were found to be less effected by jitter noise.

Table \ref{tab:PPTAfast10p} lists white noise budgets ($\sigma_{\text{r}}$, $\sigma_{J}$ and the total white noise $\sigma$) expected in the FAST/SKA era for ten PPTA pulsars that have the lowest value of $\sigma$. As one can see, for the SKA, jitter noise will dominate over the radiometer noise for the majority of bright pulsars.
In order to realistically estimate the PTA sensitivity in the FAST/SKA era, we use the total white noise given in Table \ref{tab:PPTAfast10p} plus the intrinsic spin noise (where appropriate) with parameters determined from the Bayesian analysis.

Figure \ref{fig:future_lim} shows forecasted upper limits on the density of FDM in the Galaxy for three cases, all assuming a data span of ten years.
Case a) is a conservative PTA that includes only ten pulsars as listed in Table \ref{tab:PPTAfast10p} and an observing cadence of once every 14 days. Upper limits in this case are obtained by running full Bayesian analysis of simulated data.
We analytically scale this limit curve to two more ambitious cases\footnote{Note that the scaling factor should be a good approximation at high frequencies where red noise plays a less important role.}.
We increase the number of pulsars to 100 in case b), leading to a factor of $\sqrt[]{10}$ improvement.
For case c), we further increase the cadence to once every day and adopt an integration time of two hours, providing another factor of $\sqrt[]{4\times 14}$ improvement. Case c) might be an interesting option in the SKA era since small radio telescopes (compared to SKA/FAST), such as Parkes, can be dedicated for high-cadence and long integration observations of the brighter pulsars.

As one can see from Fig. \ref{fig:future_lim}, we will be able to constrain the contribution of FDM to the local dark matter density below 10\% for $m\lesssim 10^{-23}$ eV in ten years under the conservative assumption for SKA sensitivity. However, it is more challenging for boson masses above $10^{-22}$ eV; we estimate that decade-long observations of hundreds of pulsars timed at nearly daily cadence with precision $\lesssim 20$ ns are necessary to place interesting limits.

There are a couple of ways to improve our analysis. First, the coherence between pulsar terms and Earth terms can be used to enhance the sensitivity. When a pulsar and the Earth are located within a de Broglie wavelength $\lambda_{dB}$, the oscillation phases, which have been assumed to be independent in the current analysis, are correlated. However, for $m\gtrsim 10^{-22}$ eV, this effect will have no impact on the current results, since $\lambda_{dB}=60\, \text{pc}\, (10^{-22}\textrm{eV}/m)$ and no pulsars have been found within 60 pc to the Earth. Another interesting point is that pulsars that are close to each other within $\lambda_{dB}$ also experience phase-coherent oscillations \cite{DeMartino:2017qsa}. We plan to explore these features in a future work.

Second, the oscillation amplitude $\Psi_{c}$ is proportional to the local dark matter density. Thus, in contrast to the amplitude of the Earth term, the amplitude of the pulsar term varies from pulsar to pulsar; see Eq.~(\ref{eq:Psicf}). In $\Lambda$-FDM cosmological simulations \cite{DeMartino:2017qsa,2014NatPh..10..496S}, it was shown that due to wave interference the dark matter forms gritty pattern with typical granule size of around $\lambda_{\text{dB}}$. When averaged over  $\gg \lambda_{\text{dB}}$ scales, the periphery ($>1$ kpc) density profile is similar to the classical Navarro-Frenk-White (NFW) profile, whereas a distinct density peak is seen in the central regions (usually called solitonic core, see \cite{2014NatPh..10..496S} for details).

\begin{figure}
    \includegraphics[width=\columnwidth]{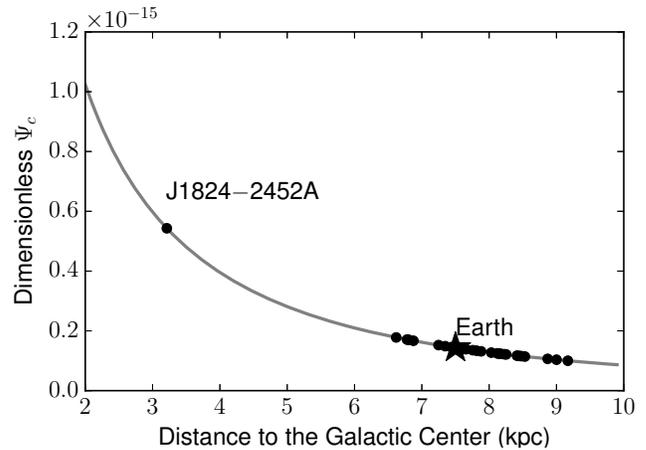}
    \caption{The amplitude of the expected dark matter signal for different pulsars, assuming NFW dark matter density profile. The mass of the scalar dark matter particles is assumed to be $2 \times 10^{-23}$ eV.}
    \label{fig:dm_NFW}
\end{figure}

Figure \ref{fig:dm_NFW} shows the expected signal amplitude for PPTA pulsars assuming the NFW dark matter density profile \cite{NFWdarkmatter} with parameters from \cite{2012PASJ...64...75S}. As one can see, pulsars closer to the Galactic Center provide better sensitivity to the dark matter signal. The amplitude of the dark matter signal becomes even larger than NFW prediction within the central solitonic core ($\lesssim 1$kpc) \cite{DeMartino:2017qsa}.
For the current PPTA sample, PSR J1824$-$2452A is expected to have the largest signal amplitude, a factor of $\sim 5$ larger than other pulsars\footnote{The density of the scalar field dark matter in globular clusters is not expected to deviate significantly from the general trend as $\lambda_{\text{dB}}$ is larger than typical sizes of globular clusters. Thus, the amplitude of the oscillation at J1824$-$2452A, located in a globular cluster, is expected to follow the NFW prediction.}. However, this pulsar is nearly the worst timer in PPTA (see Table \ref{tab:PPTA26p} and Fig. \ref{fig:grade_puls}). Existing and future pulsar surveys might help find high quality millisecond pulsars close to the Galactic Center and thus provide better sensitivity to the dark matter searches \cite{2004NewAR..48..993K}.

\section{Conclusions}
\label{sec:conclu}
Pulsar timing is a powerful tool to study a wide variety of astrophysical phenomena. By exploiting precision timing observations from an array of the most stable millisecond pulsars, PTAs allow us to measure minute correlations in the ToAs of different pulsars. Like continuous GWs from individual supermassive binary black holes, FDM in the Galaxy produces periodic variations in pulsar ToAs. We perform a search for evidence of ultralight dark matter in the latest PPTA data set. Finding no statistically significant signals, we place upper limits on the dark matter density: for boson mass $m \lesssim 10^{-23}$ eV, our analysis constrains the density below $6\, \text{GeV}\,\text{cm}^{-3}$ with 95\% confidence; at $m \approx 10^{-22}$ eV, our upper limits remain 3 orders of magnitude above the local dark matter density $0.4\, \text{GeV}\,\text{cm}^{-3}$ inferred from kinematics measurements of stars in the Galaxy \cite{DMdensity18}.

We derived the noise properties of PPTA data and obtain dark matter constraints using both Bayesian and Frequentist methods.
Our upper limits from the two methods are broadly consistent. We reanalyzed the NANOGrav 5-yr data set and found that the PPTA data result in a factor of 2 to 5 improvement in dark matter constraints.
We studied potential systematics due to SSE errors in our analysis and found that the search for ultralight dark matter is insensitive to such errors.
We have ignored effects from instabilities in terrestrial time standards; such clock errors produce a monopolar broad-band noise \cite{HobbsClock12}. Whereas this effect should be distinguishable from the sinusoidal ToA variations due to ultralight dark matter, one needs to include it in a future study to quantitatively assess the impact.

We forecasted the PTA sensitivity in the FAST/SKA era while accounting for realistic noise levels. We found that observing the ten best PPTA pulsars for ten years would constrain the density of FDM below $0.05\, \text{GeV}\,\text{cm}^{-3}$ for $m \lesssim 10^{-23}$ eV, about 10\% of measured total dark matter density. At $m \approx 2\times 10^{-23}$ eV, our projected limit is around $0.4\, \text{GeV}\,\text{cm}^{-3}$; for higher boson masses, the upper limits increase as $\sim m^{3}$. Above $m \approx 10^{-22}$ eV, the projected limits are more than one order of magnitude above the local dark matter density. To place interesting limits in this mass range, an ambitious timing program in which hundreds of pulsars timed with daily cadence and high precision ($\lesssim 20$ ns) for more than a decade is required. Finally, we point out that high-quality pulsars in the vicinity of the Galactic Center will be ideal tools to test the fuzzy dark matter hypothesis.

%The sensitivity can be boosted, if more frequent observations of larger set of pulsars are performed. In optimistic case scenario if 100 pulsars with similar noise properties to those listed in Table \ref{tab:PPTAfast10p} are observed for $\sim$ 2 hours per day by multiple radio astronomical facilities around the globe, the sensitivity will be increased by a factor of 20 (see Fig. \ref{fig:future_lim}). The high precision timing of stable pulsars in the vicinity to the Galactic center will lead to further enhancement of PTA sensitivity to SFDM signal. 

\section*{Acknowledgements}
The Parkes radio telescope is part of the Australia Telescope National Facility which is funded by the Commonwealth of Australia for operation as a National Facility managed by CSIRO.
N.K.P. acknowledges the support from IMPRS Bonn/Cologne and the Bonn-Cologne Graduate School (BCGS). X.Z., M.B., D.J.R., R.M.S. and L.W. are supported by ARC CE170100004. X.Z. and L.W. are additionally supported by ARC DP150102988.
L.H. is supported in part by 
NASA NXX16AB27G and DOE DE-SC0011941.
Work at NRL is supported by NASA.
P.D.L. is supported through ARC FT160100112 and ARC DP180103155.
M.B., S.O. and R.S. acknowledge support through the ARC Laureate Fellowship grant No. FL150100148.
J.W. is supported by Qing Cu Hui of Chinese Academy of Sciences (CAS).
We acknowledge the Institute for Theoretical and Experimental Physics and, in particular, Sergey Blinnikov for providing computing facilities. The authors would like to thank Maxim Pshirkov, Mikhail Ivanov, Nicolas Caballero and David Champion for fruitful discussions. We would like to thank the anonymous referees for useful comments.

\bibliography{ref}

\begin{thebibliography}{96}%
\makeatletter
\providecommand \@ifxundefined [1]{%
 \@ifx{#1\undefined}
}%
\providecommand \@ifnum [1]{%
 \ifnum #1\expandafter \@firstoftwo
 \else \expandafter \@secondoftwo
 \fi
}%
\providecommand \@ifx [1]{%
 \ifx #1\expandafter \@firstoftwo
 \else \expandafter \@secondoftwo
 \fi
}%
\providecommand \natexlab [1]{#1}%
\providecommand \enquote  [1]{``#1''}%
\providecommand \bibnamefont  [1]{#1}%
\providecommand \bibfnamefont [1]{#1}%
\providecommand \citenamefont [1]{#1}%
\providecommand \href@noop [0]{\@secondoftwo}%
\providecommand \href [0]{\begingroup \@sanitize@url \@href}%
\providecommand \@href[1]{\@@startlink{#1}\@@href}%
\providecommand \@@href[1]{\endgroup#1\@@endlink}%
\providecommand \@sanitize@url [0]{\catcode `\\12\catcode `\$12\catcode
  `\&12\catcode `\#12\catcode `\^12\catcode `\_12\catcode `\%12\relax}%
\providecommand \@@startlink[1]{}%
\providecommand \@@endlink[0]{}%
\providecommand \url  [0]{\begingroup\@sanitize@url \@url }%
\providecommand \@url [1]{\endgroup\@href {#1}{\urlprefix }}%
\providecommand \urlprefix  [0]{URL }%
\providecommand \Eprint [0]{\href }%
\providecommand \doibase [0]{http://dx.doi.org/}%
\providecommand \selectlanguage [0]{\@gobble}%
\providecommand \bibinfo  [0]{\@secondoftwo}%
\providecommand \bibfield  [0]{\@secondoftwo}%
\providecommand \translation [1]{[#1]}%
\providecommand \BibitemOpen [0]{}%
\providecommand \bibitemStop [0]{}%
\providecommand \bibitemNoStop [0]{.\EOS\space}%
\providecommand \EOS [0]{\spacefactor3000\relax}%
\providecommand \BibitemShut  [1]{\csname bibitem#1\endcsname}%
\let\auto@bib@innerbib\@empty
%</preamble>
\bibitem [{\citenamefont {{Zwicky}}(1933)}]{1933AcHPh...6..110Z}%
  \BibitemOpen
  \bibfield  {author} {\bibinfo {author} {\bibfnamefont {F.}~\bibnamefont
  {{Zwicky}}},\ }\href@noop {} {\bibfield  {journal} {\bibinfo  {journal}
  {Helvetica Physica Acta}\ }\textbf {\bibinfo {volume} {6}},\ \bibinfo {pages}
  {110} (\bibinfo {year} {1933})}\BibitemShut {NoStop}%
\bibitem [{\citenamefont {{Zwicky}}(1937)}]{1937ApJ....86..217Z}%
  \BibitemOpen
  \bibfield  {author} {\bibinfo {author} {\bibfnamefont {F.}~\bibnamefont
  {{Zwicky}}},\ }\href {\doibase 10.1086/143864} {\bibfield  {journal}
  {\bibinfo  {journal} {\apj}\ }\textbf {\bibinfo {volume} {86}},\ \bibinfo
  {pages} {217} (\bibinfo {year} {1937})}\BibitemShut {NoStop}%
\bibitem [{\citenamefont {{Smith}}(1936)}]{1936ApJ....83...23S}%
  \BibitemOpen
  \bibfield  {author} {\bibinfo {author} {\bibfnamefont {S.}~\bibnamefont
  {{Smith}}},\ }\href {\doibase 10.1086/143697} {\bibfield  {journal} {\bibinfo
   {journal} {\apj}\ }\textbf {\bibinfo {volume} {83}},\ \bibinfo {pages} {23}
  (\bibinfo {year} {1936})}\BibitemShut {NoStop}%
\bibitem [{\citenamefont {{Koopmans}}\ and\ \citenamefont
  {{Treu}}(2003)}]{2003ApJ...583..606K}%
  \BibitemOpen
  \bibfield  {author} {\bibinfo {author} {\bibfnamefont {L.~V.~E.}\
  \bibnamefont {{Koopmans}}}\ and\ \bibinfo {author} {\bibfnamefont
  {T.}~\bibnamefont {{Treu}}},\ }\href {\doibase 10.1086/345423} {\bibfield
  {journal} {\bibinfo  {journal} {\apj}\ }\textbf {\bibinfo {volume} {583}},\
  \bibinfo {pages} {606} (\bibinfo {year} {2003})},\ \Eprint
  {http://arxiv.org/abs/astro-ph/0205281} {astro-ph/0205281} \BibitemShut
  {NoStop}%
\bibitem [{\citenamefont {{Clowe}}\ \emph {et~al.}(2004)\citenamefont
  {{Clowe}}, \citenamefont {{Gonzalez}},\ and\ \citenamefont
  {{Markevitch}}}]{2004ApJ...604..596C}%
  \BibitemOpen
  \bibfield  {author} {\bibinfo {author} {\bibfnamefont {D.}~\bibnamefont
  {{Clowe}}}, \bibinfo {author} {\bibfnamefont {A.}~\bibnamefont {{Gonzalez}}},
  \ and\ \bibinfo {author} {\bibfnamefont {M.}~\bibnamefont {{Markevitch}}},\
  }\href {\doibase 10.1086/381970} {\bibfield  {journal} {\bibinfo  {journal}
  {\apj}\ }\textbf {\bibinfo {volume} {604}},\ \bibinfo {pages} {596} (\bibinfo
  {year} {2004})},\ \Eprint {http://arxiv.org/abs/astro-ph/0312273}
  {astro-ph/0312273} \BibitemShut {NoStop}%
\bibitem [{\citenamefont {{Tegmark}}\ \emph {et~al.}(2004)\citenamefont
  {{Tegmark}} \emph {et~al.}}]{2004ApJ...606..702T}%
  \BibitemOpen
  \bibfield  {author} {\bibinfo {author} {\bibfnamefont {M.}~\bibnamefont
  {{Tegmark}}} \emph {et~al.},\ }\href {\doibase 10.1086/382125} {\bibfield
  {journal} {\bibinfo  {journal} {\apj}\ }\textbf {\bibinfo {volume} {606}},\
  \bibinfo {pages} {702} (\bibinfo {year} {2004})},\ \Eprint
  {http://arxiv.org/abs/astro-ph/0310725} {astro-ph/0310725} \BibitemShut
  {NoStop}%
\bibitem [{\citenamefont {{Ade}}\ \emph {et~al.}(2016)\citenamefont {{Ade}}
  \emph {et~al.}}]{2016A&A...594A..13P}%
  \BibitemOpen
  \bibfield  {author} {\bibinfo {author} {\bibfnamefont {P.~A.~R.}\
  \bibnamefont {{Ade}}} \emph {et~al.},\ }\href {\doibase
  10.1051/0004-6361/201525830} {\bibfield  {journal} {\bibinfo  {journal}
  {A\&A}\ }\textbf {\bibinfo {volume} {594}},\ \bibinfo {eid} {A13} (\bibinfo
  {year} {2016})},\ \Eprint {http://arxiv.org/abs/1502.01589}
  {arXiv:1502.01589} \BibitemShut {NoStop}%
\bibitem [{\citenamefont {{Bertone}}\ \emph {et~al.}(2005)\citenamefont
  {{Bertone}}, \citenamefont {{Hooper}},\ and\ \citenamefont
  {{Silk}}}]{2005PhR...405..279B}%
  \BibitemOpen
  \bibfield  {author} {\bibinfo {author} {\bibfnamefont {G.}~\bibnamefont
  {{Bertone}}}, \bibinfo {author} {\bibfnamefont {D.}~\bibnamefont {{Hooper}}},
  \ and\ \bibinfo {author} {\bibfnamefont {J.}~\bibnamefont {{Silk}}},\ }\href
  {\doibase 10.1016/j.physrep.2004.08.031} {\bibfield  {journal} {\bibinfo
  {journal} {Phys. Rep.}\ }\textbf {\bibinfo {volume} {405}},\ \bibinfo {pages}
  {279} (\bibinfo {year} {2005})},\ \Eprint
  {http://arxiv.org/abs/hep-ph/0404175} {hep-ph/0404175} \BibitemShut {NoStop}%
\bibitem [{\citenamefont {{Primack}}(2012)}]{2012AnP...524..535P}%
  \BibitemOpen
  \bibfield  {author} {\bibinfo {author} {\bibfnamefont {J.~R.}\ \bibnamefont
  {{Primack}}},\ }\href {\doibase 10.1002/andp.201200077} {\bibfield  {journal}
  {\bibinfo  {journal} {Annalen der Physik}\ }\textbf {\bibinfo {volume}
  {524}},\ \bibinfo {pages} {535} (\bibinfo {year} {2012})}\BibitemShut
  {NoStop}%
\bibitem [{\citenamefont {Battaglieri}\ \emph {et~al.}(2017)\citenamefont
  {Battaglieri} \emph {et~al.}}]{Battaglieri:2017aum}%
  \BibitemOpen
  \bibfield  {author} {\bibinfo {author} {\bibfnamefont {M.}~\bibnamefont
  {Battaglieri}} \emph {et~al.},\ }\href@noop {} {\  (\bibinfo {year}
  {2017})},\ \Eprint {http://arxiv.org/abs/1707.04591} {arXiv:1707.04591
  [hep-ph]} \BibitemShut {NoStop}%
%%CITATION = ARXIV:1707.04591;%%
\bibitem [{\citenamefont {{Arvanitaki}}\ \emph {et~al.}(2010)\citenamefont
  {{Arvanitaki}}, \citenamefont {{Dimopoulos}}, \citenamefont {{Dubovsky}},
  \citenamefont {{Kaloper}},\ and\ \citenamefont
  {{March-Russell}}}]{2010PhRvD..81l3530A}%
  \BibitemOpen
  \bibfield  {author} {\bibinfo {author} {\bibfnamefont {A.}~\bibnamefont
  {{Arvanitaki}}}, \bibinfo {author} {\bibfnamefont {S.}~\bibnamefont
  {{Dimopoulos}}}, \bibinfo {author} {\bibfnamefont {S.}~\bibnamefont
  {{Dubovsky}}}, \bibinfo {author} {\bibfnamefont {N.}~\bibnamefont
  {{Kaloper}}}, \ and\ \bibinfo {author} {\bibfnamefont {J.}~\bibnamefont
  {{March-Russell}}},\ }\href {\doibase 10.1103/PhysRevD.81.123530} {\bibfield
  {journal} {\bibinfo  {journal} {\prd}\ }\textbf {\bibinfo {volume} {81}},\
  \bibinfo {eid} {123530} (\bibinfo {year} {2010})},\ \Eprint
  {http://arxiv.org/abs/0905.4720} {arXiv:0905.4720 [hep-th]} \BibitemShut
  {NoStop}%
\bibitem [{\citenamefont {{Hui}}\ \emph {et~al.}(2017)\citenamefont {{Hui}},
  \citenamefont {{Ostriker}}, \citenamefont {{Tremaine}},\ and\ \citenamefont
  {{Witten}}}]{2017PhRvD..95d3541H}%
  \BibitemOpen
  \bibfield  {author} {\bibinfo {author} {\bibfnamefont {L.}~\bibnamefont
  {{Hui}}}, \bibinfo {author} {\bibfnamefont {J.~P.}\ \bibnamefont
  {{Ostriker}}}, \bibinfo {author} {\bibfnamefont {S.}~\bibnamefont
  {{Tremaine}}}, \ and\ \bibinfo {author} {\bibfnamefont {E.}~\bibnamefont
  {{Witten}}},\ }\href {\doibase 10.1103/PhysRevD.95.043541} {\bibfield
  {journal} {\bibinfo  {journal} {\prd}\ }\textbf {\bibinfo {volume} {95}},\
  \bibinfo {eid} {043541} (\bibinfo {year} {2017})},\ \Eprint
  {http://arxiv.org/abs/1610.08297} {arXiv:1610.08297} \BibitemShut {NoStop}%
\bibitem [{\citenamefont {{Preskill}}\ \emph {et~al.}(1983)\citenamefont
  {{Preskill}}, \citenamefont {{Wise}},\ and\ \citenamefont
  {{Wilczek}}}]{Preskill:1982cy}%
  \BibitemOpen
  \bibfield  {author} {\bibinfo {author} {\bibfnamefont {J.}~\bibnamefont
  {{Preskill}}}, \bibinfo {author} {\bibfnamefont {M.~B.}\ \bibnamefont
  {{Wise}}}, \ and\ \bibinfo {author} {\bibfnamefont {F.}~\bibnamefont
  {{Wilczek}}},\ }\href {\doibase 10.1016/0370-2693(83)90637-8} {\bibfield
  {journal} {\bibinfo  {journal} {Physics Letters B}\ }\textbf {\bibinfo
  {volume} {120}},\ \bibinfo {pages} {127} (\bibinfo {year}
  {1983})}\BibitemShut {NoStop}%
\bibitem [{\citenamefont {{Abbott}}\ and\ \citenamefont
  {{Sikivie}}(1983)}]{Abbott:1982af}%
  \BibitemOpen
  \bibfield  {author} {\bibinfo {author} {\bibfnamefont {L.~F.}\ \bibnamefont
  {{Abbott}}}\ and\ \bibinfo {author} {\bibfnamefont {P.}~\bibnamefont
  {{Sikivie}}},\ }\href {\doibase 10.1016/0370-2693(83)90638-X} {\bibfield
  {journal} {\bibinfo  {journal} {Physics Letters B}\ }\textbf {\bibinfo
  {volume} {120}},\ \bibinfo {pages} {133} (\bibinfo {year}
  {1983})}\BibitemShut {NoStop}%
\bibitem [{\citenamefont {{Dine}}\ and\ \citenamefont
  {{Fischler}}(1983)}]{Dine:1982ah}%
  \BibitemOpen
  \bibfield  {author} {\bibinfo {author} {\bibfnamefont {M.}~\bibnamefont
  {{Dine}}}\ and\ \bibinfo {author} {\bibfnamefont {W.}~\bibnamefont
  {{Fischler}}},\ }\href {\doibase 10.1016/0370-2693(83)90639-1} {\bibfield
  {journal} {\bibinfo  {journal} {Physics Letters B}\ }\textbf {\bibinfo
  {volume} {120}},\ \bibinfo {pages} {137} (\bibinfo {year}
  {1983})}\BibitemShut {NoStop}%
\bibitem [{\citenamefont {Svrcek}\ and\ \citenamefont
  {Witten}(2006)}]{Svrcek:2006yi}%
  \BibitemOpen
  \bibfield  {author} {\bibinfo {author} {\bibfnamefont {P.}~\bibnamefont
  {Svrcek}}\ and\ \bibinfo {author} {\bibfnamefont {E.}~\bibnamefont
  {Witten}},\ }\href {\doibase 10.1088/1126-6708/2006/06/051} {\bibfield
  {journal} {\bibinfo  {journal} {JHEP}\ }\textbf {\bibinfo {volume} {06}},\
  \bibinfo {pages} {051} (\bibinfo {year} {2006})},\ \Eprint
  {http://arxiv.org/abs/hep-th/0605206} {arXiv:hep-th/0605206 [hep-th]}
  \BibitemShut {NoStop}%
%%CITATION = HEP-TH/0605206;%%
\bibitem [{\citenamefont {{Hu}}\ \emph {et~al.}(2000)\citenamefont {{Hu}},
  \citenamefont {{Barkana}},\ and\ \citenamefont
  {{Gruzinov}}}]{2000PhRvL..85.1158H}%
  \BibitemOpen
  \bibfield  {author} {\bibinfo {author} {\bibfnamefont {W.}~\bibnamefont
  {{Hu}}}, \bibinfo {author} {\bibfnamefont {R.}~\bibnamefont {{Barkana}}}, \
  and\ \bibinfo {author} {\bibfnamefont {A.}~\bibnamefont {{Gruzinov}}},\
  }\href {\doibase 10.1103/PhysRevLett.85.1158} {\bibfield  {journal} {\bibinfo
   {journal} {\prl}\ }\textbf {\bibinfo {volume} {85}},\ \bibinfo {pages}
  {1158} (\bibinfo {year} {2000})},\ \Eprint
  {http://arxiv.org/abs/astro-ph/0003365} {astro-ph/0003365} \BibitemShut
  {NoStop}%
\bibitem [{\citenamefont {{Bullock}}\ and\ \citenamefont
  {{Boylan-Kolchin}}(2017)}]{Bullock:2017xww}%
  \BibitemOpen
  \bibfield  {author} {\bibinfo {author} {\bibfnamefont {J.~S.}\ \bibnamefont
  {{Bullock}}}\ and\ \bibinfo {author} {\bibfnamefont {M.}~\bibnamefont
  {{Boylan-Kolchin}}},\ }\href {\doibase 10.1146/annurev-astro-091916-055313}
  {\bibfield  {journal} {\bibinfo  {journal} {Annu. Rev. Astron. Astrophys.}\
  }\textbf {\bibinfo {volume} {55}},\ \bibinfo {pages} {343} (\bibinfo {year}
  {2017})},\ \Eprint {http://arxiv.org/abs/1707.04256} {arXiv:1707.04256}
  \BibitemShut {NoStop}%
\bibitem [{\citenamefont {{Hlo{\v z}ek}}\ \emph {et~al.}(2018)\citenamefont
  {{Hlo{\v z}ek}}, \citenamefont {{Marsh}},\ and\ \citenamefont
  {{Grin}}}]{2018MNRAS.tmp..273H}%
  \BibitemOpen
  \bibfield  {author} {\bibinfo {author} {\bibfnamefont {R.}~\bibnamefont
  {{Hlo{\v z}ek}}}, \bibinfo {author} {\bibfnamefont {D.~J.~E.}\ \bibnamefont
  {{Marsh}}}, \ and\ \bibinfo {author} {\bibfnamefont {D.}~\bibnamefont
  {{Grin}}},\ }\href {\doibase 10.1093/mnras/sty271} {\bibfield  {journal}
  {\bibinfo  {journal} {Mon. Not. R. Astron. Soc.}\ }\textbf {\bibinfo {volume}
  {476}},\ \bibinfo {pages} {3063} (\bibinfo {year} {2018})},\ \Eprint
  {http://arxiv.org/abs/1708.05681} {arXiv:1708.05681} \BibitemShut {NoStop}%
\bibitem [{\citenamefont {{Ir{\v s}i{\v c}}}\ \emph {et~al.}(2017)\citenamefont
  {{Ir{\v s}i{\v c}}}, \citenamefont {{Viel}}, \citenamefont {{Haehnelt}},
  \citenamefont {{Bolton}},\ and\ \citenamefont
  {{Becker}}}]{2017PhRvL.119c1302I}%
  \BibitemOpen
  \bibfield  {author} {\bibinfo {author} {\bibfnamefont {V.}~\bibnamefont
  {{Ir{\v s}i{\v c}}}}, \bibinfo {author} {\bibfnamefont {M.}~\bibnamefont
  {{Viel}}}, \bibinfo {author} {\bibfnamefont {M.~G.}\ \bibnamefont
  {{Haehnelt}}}, \bibinfo {author} {\bibfnamefont {J.~S.}\ \bibnamefont
  {{Bolton}}}, \ and\ \bibinfo {author} {\bibfnamefont {G.~D.}\ \bibnamefont
  {{Becker}}},\ }\href {\doibase 10.1103/PhysRevLett.119.031302} {\bibfield
  {journal} {\bibinfo  {journal} {\prl}\ }\textbf {\bibinfo {volume} {119}},\
  \bibinfo {eid} {031302} (\bibinfo {year} {2017})},\ \Eprint
  {http://arxiv.org/abs/1703.04683} {arXiv:1703.04683} \BibitemShut {NoStop}%
\bibitem [{\citenamefont {{Kobayashi}}\ \emph {et~al.}(2017)\citenamefont
  {{Kobayashi}}, \citenamefont {{Murgia}}, \citenamefont {{De Simone}},
  \citenamefont {{Ir{\v s}i{\v c}}},\ and\ \citenamefont
  {{Viel}}}]{2017PhRvD..96l3514K}%
  \BibitemOpen
  \bibfield  {author} {\bibinfo {author} {\bibfnamefont {T.}~\bibnamefont
  {{Kobayashi}}}, \bibinfo {author} {\bibfnamefont {R.}~\bibnamefont
  {{Murgia}}}, \bibinfo {author} {\bibfnamefont {A.}~\bibnamefont {{De
  Simone}}}, \bibinfo {author} {\bibfnamefont {V.}~\bibnamefont {{Ir{\v s}i{\v
  c}}}}, \ and\ \bibinfo {author} {\bibfnamefont {M.}~\bibnamefont {{Viel}}},\
  }\href {\doibase 10.1103/PhysRevD.96.123514} {\bibfield  {journal} {\bibinfo
  {journal} {\prd}\ }\textbf {\bibinfo {volume} {96}},\ \bibinfo {eid} {123514}
  (\bibinfo {year} {2017})},\ \Eprint {http://arxiv.org/abs/1708.00015}
  {arXiv:1708.00015} \BibitemShut {NoStop}%
\bibitem [{\citenamefont {{Bowman}}\ \emph {et~al.}(2018)\citenamefont
  {{Bowman}}, \citenamefont {{Rogers}}, \citenamefont {{Monsalve}},
  \citenamefont {{Mozdzen}},\ and\ \citenamefont {{Mahesh}}}]{Bowman:2018yin}%
  \BibitemOpen
  \bibfield  {author} {\bibinfo {author} {\bibfnamefont {J.~D.}\ \bibnamefont
  {{Bowman}}}, \bibinfo {author} {\bibfnamefont {A.~E.~E.}\ \bibnamefont
  {{Rogers}}}, \bibinfo {author} {\bibfnamefont {R.~A.}\ \bibnamefont
  {{Monsalve}}}, \bibinfo {author} {\bibfnamefont {T.~J.}\ \bibnamefont
  {{Mozdzen}}}, \ and\ \bibinfo {author} {\bibfnamefont {N.}~\bibnamefont
  {{Mahesh}}},\ }\href {\doibase 10.1038/nature25792} {\bibfield  {journal}
  {\bibinfo  {journal} {Nature}\ }\textbf {\bibinfo {volume} {555}},\ \bibinfo
  {pages} {67} (\bibinfo {year} {2018})}\BibitemShut {NoStop}%
\bibitem [{\citenamefont {{Schneider}}(2018)}]{Schneider:2018xba}%
  \BibitemOpen
  \bibfield  {author} {\bibinfo {author} {\bibfnamefont {A.}~\bibnamefont
  {{Schneider}}},\ }\href {\doibase 10.1103/PhysRevD.98.063021} {\bibfield
  {journal} {\bibinfo  {journal} {\prd}\ }\textbf {\bibinfo {volume} {98}},\
  \bibinfo {eid} {063021} (\bibinfo {year} {2018})},\ \Eprint
  {http://arxiv.org/abs/1805.00021} {arXiv:1805.00021} \BibitemShut {NoStop}%
\bibitem [{\citenamefont {{Lidz}}\ and\ \citenamefont
  {{Hui}}(2018)}]{Lidz:2018fqo}%
  \BibitemOpen
  \bibfield  {author} {\bibinfo {author} {\bibfnamefont {A.}~\bibnamefont
  {{Lidz}}}\ and\ \bibinfo {author} {\bibfnamefont {L.}~\bibnamefont {{Hui}}},\
  }\href {\doibase 10.1103/PhysRevD.98.023011} {\bibfield  {journal} {\bibinfo
  {journal} {\prd}\ }\textbf {\bibinfo {volume} {98}},\ \bibinfo {eid} {023011}
  (\bibinfo {year} {2018})},\ \Eprint {http://arxiv.org/abs/1805.01253}
  {arXiv:1805.01253} \BibitemShut {NoStop}%
\bibitem [{\citenamefont {{Sullivan}}\ \emph {et~al.}(2018)\citenamefont
  {{Sullivan}}, \citenamefont {{Hirano}},\ and\ \citenamefont
  {{Bromm}}}]{2018arXiv180901679S}%
  \BibitemOpen
  \bibfield  {author} {\bibinfo {author} {\bibfnamefont {J.~M.}\ \bibnamefont
  {{Sullivan}}}, \bibinfo {author} {\bibfnamefont {S.}~\bibnamefont
  {{Hirano}}}, \ and\ \bibinfo {author} {\bibfnamefont {V.}~\bibnamefont
  {{Bromm}}},\ }\href {\doibase 10.1093/mnrasl/sly164} {\bibfield  {journal}
  {\bibinfo  {journal} {Mon. Not. R. Astron. Soc.}\ }\textbf {\bibinfo {volume}
  {481}},\ \bibinfo {pages} {L69} (\bibinfo {year} {2018})},\ \Eprint
  {http://arxiv.org/abs/1809.01679} {arXiv:1809.01679} \BibitemShut {NoStop}%
\bibitem [{\citenamefont {{Calabrese}}\ and\ \citenamefont
  {{Spergel}}(2016)}]{Calabrese:2016hmp}%
  \BibitemOpen
  \bibfield  {author} {\bibinfo {author} {\bibfnamefont {E.}~\bibnamefont
  {{Calabrese}}}\ and\ \bibinfo {author} {\bibfnamefont {D.~N.}\ \bibnamefont
  {{Spergel}}},\ }\href {\doibase 10.1093/mnras/stw1256} {\bibfield  {journal}
  {\bibinfo  {journal} {Mon. Not. R. Astron. Soc.}\ }\textbf {\bibinfo {volume}
  {460}},\ \bibinfo {pages} {4397} (\bibinfo {year} {2016})},\ \Eprint
  {http://arxiv.org/abs/1603.07321} {arXiv:1603.07321} \BibitemShut {NoStop}%
\bibitem [{\citenamefont {{Deng}}\ \emph {et~al.}(2018)\citenamefont {{Deng}},
  \citenamefont {{Hertzberg}}, \citenamefont {{Namjoo}},\ and\ \citenamefont
  {{Masoumi}}}]{Deng:2018jjz}%
  \BibitemOpen
  \bibfield  {author} {\bibinfo {author} {\bibfnamefont {H.}~\bibnamefont
  {{Deng}}}, \bibinfo {author} {\bibfnamefont {M.~P.}\ \bibnamefont
  {{Hertzberg}}}, \bibinfo {author} {\bibfnamefont {M.~H.}\ \bibnamefont
  {{Namjoo}}}, \ and\ \bibinfo {author} {\bibfnamefont {A.}~\bibnamefont
  {{Masoumi}}},\ }\href {\doibase 10.1103/PhysRevD.98.023513} {\bibfield
  {journal} {\bibinfo  {journal} {\prd}\ }\textbf {\bibinfo {volume} {98}},\
  \bibinfo {eid} {023513} (\bibinfo {year} {2018})},\ \Eprint
  {http://arxiv.org/abs/1804.05921} {arXiv:1804.05921} \BibitemShut {NoStop}%
\bibitem [{\citenamefont {{Bar}}\ \emph {et~al.}(2018)\citenamefont {{Bar}},
  \citenamefont {{Blas}}, \citenamefont {{Blum}},\ and\ \citenamefont
  {{Sibiryakov}}}]{2018arXiv180500122B}%
  \BibitemOpen
  \bibfield  {author} {\bibinfo {author} {\bibfnamefont {N.}~\bibnamefont
  {{Bar}}}, \bibinfo {author} {\bibfnamefont {D.}~\bibnamefont {{Blas}}},
  \bibinfo {author} {\bibfnamefont {K.}~\bibnamefont {{Blum}}}, \ and\ \bibinfo
  {author} {\bibfnamefont {S.}~\bibnamefont {{Sibiryakov}}},\ }\href@noop {}
  {\bibfield  {journal} {\bibinfo  {journal} {ArXiv e-prints}\ } (\bibinfo
  {year} {2018})},\ \Eprint {http://arxiv.org/abs/1805.00122}
  {arXiv:1805.00122} \BibitemShut {NoStop}%
\bibitem [{\citenamefont {{Schive}}\ \emph {et~al.}(2014)\citenamefont
  {{Schive}}, \citenamefont {{Chiueh}},\ and\ \citenamefont
  {{Broadhurst}}}]{2014NatPh..10..496S}%
  \BibitemOpen
  \bibfield  {author} {\bibinfo {author} {\bibfnamefont {H.-Y.}\ \bibnamefont
  {{Schive}}}, \bibinfo {author} {\bibfnamefont {T.}~\bibnamefont {{Chiueh}}},
  \ and\ \bibinfo {author} {\bibfnamefont {T.}~\bibnamefont {{Broadhurst}}},\
  }\href {\doibase 10.1038/nphys2996} {\bibfield  {journal} {\bibinfo
  {journal} {Nature Physics}\ }\textbf {\bibinfo {volume} {10}},\ \bibinfo
  {pages} {496} (\bibinfo {year} {2014})},\ \Eprint
  {http://arxiv.org/abs/1406.6586} {arXiv:1406.6586} \BibitemShut {NoStop}%
\bibitem [{\citenamefont {{Mocz}}\ and\ \citenamefont
  {{Succi}}(2015)}]{Mocz:2015sda}%
  \BibitemOpen
  \bibfield  {author} {\bibinfo {author} {\bibfnamefont {P.}~\bibnamefont
  {{Mocz}}}\ and\ \bibinfo {author} {\bibfnamefont {S.}~\bibnamefont
  {{Succi}}},\ }\href {\doibase 10.1103/PhysRevE.91.053304} {\bibfield
  {journal} {\bibinfo  {journal} {\pre}\ }\textbf {\bibinfo {volume} {91}},\
  \bibinfo {eid} {053304} (\bibinfo {year} {2015})},\ \Eprint
  {http://arxiv.org/abs/1503.03869} {arXiv:1503.03869 [physics.comp-ph]}
  \BibitemShut {NoStop}%
\bibitem [{\citenamefont {{Veltmaat}}\ and\ \citenamefont
  {{Niemeyer}}(2016)}]{Veltmaat:2016rxo}%
  \BibitemOpen
  \bibfield  {author} {\bibinfo {author} {\bibfnamefont {J.}~\bibnamefont
  {{Veltmaat}}}\ and\ \bibinfo {author} {\bibfnamefont {J.~C.}\ \bibnamefont
  {{Niemeyer}}},\ }\href {\doibase 10.1103/PhysRevD.94.123523} {\bibfield
  {journal} {\bibinfo  {journal} {\prd}\ }\textbf {\bibinfo {volume} {94}},\
  \bibinfo {eid} {123523} (\bibinfo {year} {2016})},\ \Eprint
  {http://arxiv.org/abs/1608.00802} {arXiv:1608.00802} \BibitemShut {NoStop}%
\bibitem [{\citenamefont {{Nori}}\ and\ \citenamefont
  {{Baldi}}(2018)}]{Nori:2018hud}%
  \BibitemOpen
  \bibfield  {author} {\bibinfo {author} {\bibfnamefont {M.}~\bibnamefont
  {{Nori}}}\ and\ \bibinfo {author} {\bibfnamefont {M.}~\bibnamefont
  {{Baldi}}},\ }\href {\doibase 10.1093/mnras/sty1224} {\bibfield  {journal}
  {\bibinfo  {journal} {Mon. Not. R. Astron. Soc.}\ }\textbf {\bibinfo {volume}
  {478}},\ \bibinfo {pages} {3935} (\bibinfo {year} {2018})},\ \Eprint
  {http://arxiv.org/abs/1801.08144} {arXiv:1801.08144} \BibitemShut {NoStop}%
\bibitem [{\citenamefont {{Veltmaat}}\ \emph {et~al.}(2018)\citenamefont
  {{Veltmaat}}, \citenamefont {{Niemeyer}},\ and\ \citenamefont
  {{Schwabe}}}]{Veltmaat:2018dfz}%
  \BibitemOpen
  \bibfield  {author} {\bibinfo {author} {\bibfnamefont {J.}~\bibnamefont
  {{Veltmaat}}}, \bibinfo {author} {\bibfnamefont {J.~C.}\ \bibnamefont
  {{Niemeyer}}}, \ and\ \bibinfo {author} {\bibfnamefont {B.}~\bibnamefont
  {{Schwabe}}},\ }\href {\doibase 10.1103/PhysRevD.98.043509} {\bibfield
  {journal} {\bibinfo  {journal} {\prd}\ }\textbf {\bibinfo {volume} {98}},\
  \bibinfo {eid} {043509} (\bibinfo {year} {2018})},\ \Eprint
  {http://arxiv.org/abs/1804.09647} {arXiv:1804.09647} \BibitemShut {NoStop}%
\bibitem [{\citenamefont {{Khmelnitsky}}\ and\ \citenamefont
  {{Rubakov}}(2014)}]{2014JCAP...02..019K}%
  \BibitemOpen
  \bibfield  {author} {\bibinfo {author} {\bibfnamefont {A.}~\bibnamefont
  {{Khmelnitsky}}}\ and\ \bibinfo {author} {\bibfnamefont {V.}~\bibnamefont
  {{Rubakov}}},\ }\href {\doibase 10.1088/1475-7516/2014/02/019} {\bibfield
  {journal} {\bibinfo  {journal} {JCAP}\ }\textbf {\bibinfo {volume} {2}},\
  \bibinfo {eid} {019} (\bibinfo {year} {2014})},\ \Eprint
  {http://arxiv.org/abs/1309.5888} {arXiv:1309.5888} \BibitemShut {NoStop}%
\bibitem [{\citenamefont {{Porayko}}\ and\ \citenamefont
  {{Postnov}}(2014)}]{Natasha14}%
  \BibitemOpen
  \bibfield  {author} {\bibinfo {author} {\bibfnamefont {N.~K.}\ \bibnamefont
  {{Porayko}}}\ and\ \bibinfo {author} {\bibfnamefont {K.~A.}\ \bibnamefont
  {{Postnov}}},\ }\href {\doibase 10.1103/PhysRevD.90.062008} {\bibfield
  {journal} {\bibinfo  {journal} {\prd}\ }\textbf {\bibinfo {volume} {90}},\
  \bibinfo {eid} {062008} (\bibinfo {year} {2014})},\ \Eprint
  {http://arxiv.org/abs/1408.4670} {arXiv:1408.4670} \BibitemShut {NoStop}%
\bibitem [{\citenamefont {{De Martino}}\ \emph {et~al.}(2017)\citenamefont {{De
  Martino}}, \citenamefont {{Broadhurst}}, \citenamefont {{Tye}}, \citenamefont
  {{Chiueh}}, \citenamefont {{Schive}},\ and\ \citenamefont
  {{Lazkoz}}}]{DeMartino:2017qsa}%
  \BibitemOpen
  \bibfield  {author} {\bibinfo {author} {\bibfnamefont {I.}~\bibnamefont {{De
  Martino}}}, \bibinfo {author} {\bibfnamefont {T.}~\bibnamefont
  {{Broadhurst}}}, \bibinfo {author} {\bibfnamefont {S.-H.~H.}\ \bibnamefont
  {{Tye}}}, \bibinfo {author} {\bibfnamefont {T.}~\bibnamefont {{Chiueh}}},
  \bibinfo {author} {\bibfnamefont {H.-Y.}\ \bibnamefont {{Schive}}}, \ and\
  \bibinfo {author} {\bibfnamefont {R.}~\bibnamefont {{Lazkoz}}},\ }\href
  {\doibase 10.1103/PhysRevLett.119.221103} {\bibfield  {journal} {\bibinfo
  {journal} {\prl}\ }\textbf {\bibinfo {volume} {119}},\ \bibinfo {eid}
  {221103} (\bibinfo {year} {2017})},\ \Eprint
  {http://arxiv.org/abs/1705.04367} {arXiv:1705.04367} \BibitemShut {NoStop}%
\bibitem [{\citenamefont {{Blas}}\ \emph {et~al.}(2017)\citenamefont {{Blas}},
  \citenamefont {{Nacir}},\ and\ \citenamefont {{Sibiryakov}}}]{Blas2017}%
  \BibitemOpen
  \bibfield  {author} {\bibinfo {author} {\bibfnamefont {D.}~\bibnamefont
  {{Blas}}}, \bibinfo {author} {\bibfnamefont {D.~L.}\ \bibnamefont {{Nacir}}},
  \ and\ \bibinfo {author} {\bibfnamefont {S.}~\bibnamefont {{Sibiryakov}}},\
  }\href {\doibase 10.1103/PhysRevLett.118.261102} {\bibfield  {journal}
  {\bibinfo  {journal} {\prl}\ }\textbf {\bibinfo {volume} {118}},\ \bibinfo
  {eid} {261102} (\bibinfo {year} {2017})},\ \Eprint
  {http://arxiv.org/abs/1612.06789} {arXiv:1612.06789 [hep-ph]} \BibitemShut
  {NoStop}%
\bibitem [{\citenamefont {{Sazhin}}(1978)}]{1978SvA....22...36S}%
  \BibitemOpen
  \bibfield  {author} {\bibinfo {author} {\bibfnamefont {M.~V.}\ \bibnamefont
  {{Sazhin}}},\ }\href@noop {} {\bibfield  {journal} {\bibinfo  {journal}
  {Soviet Astronomy}\ }\textbf {\bibinfo {volume} {22}},\ \bibinfo {pages} {36}
  (\bibinfo {year} {1978})}\BibitemShut {NoStop}%
\bibitem [{\citenamefont {{Detweiler}}(1979)}]{Detweiler79}%
  \BibitemOpen
  \bibfield  {author} {\bibinfo {author} {\bibfnamefont {S.}~\bibnamefont
  {{Detweiler}}},\ }\href {\doibase 10.1086/157593} {\bibfield  {journal}
  {\bibinfo  {journal} {\apj}\ }\textbf {\bibinfo {volume} {234}},\ \bibinfo
  {pages} {1100} (\bibinfo {year} {1979})}\BibitemShut {NoStop}%
\bibitem [{\citenamefont {{Hellings}}\ and\ \citenamefont
  {{Downs}}(1983)}]{Hellings_Downs}%
  \BibitemOpen
  \bibfield  {author} {\bibinfo {author} {\bibfnamefont {R.~W.}\ \bibnamefont
  {{Hellings}}}\ and\ \bibinfo {author} {\bibfnamefont {G.~S.}\ \bibnamefont
  {{Downs}}},\ }\href {\doibase 10.1086/183954} {\bibfield  {journal} {\bibinfo
   {journal} {\apj}\ }\textbf {\bibinfo {volume} {265}},\ \bibinfo {pages}
  {L39} (\bibinfo {year} {1983})}\BibitemShut {NoStop}%
\bibitem [{\citenamefont {{Foster}}\ and\ \citenamefont
  {{Backer}}(1990)}]{1990ApJ...361..300F}%
  \BibitemOpen
  \bibfield  {author} {\bibinfo {author} {\bibfnamefont {R.~S.}\ \bibnamefont
  {{Foster}}}\ and\ \bibinfo {author} {\bibfnamefont {D.~C.}\ \bibnamefont
  {{Backer}}},\ }\href {\doibase 10.1086/169195} {\bibfield  {journal}
  {\bibinfo  {journal} {\apj}\ }\textbf {\bibinfo {volume} {361}},\ \bibinfo
  {pages} {300} (\bibinfo {year} {1990})}\BibitemShut {NoStop}%
\bibitem [{\citenamefont {{Manchester}}\ \emph {et~al.}(2013)\citenamefont
  {{Manchester}} \emph {et~al.}}]{2013PASA...30...17M}%
  \BibitemOpen
  \bibfield  {author} {\bibinfo {author} {\bibfnamefont {R.~N.}\ \bibnamefont
  {{Manchester}}} \emph {et~al.},\ }\href {\doibase 10.1017/pasa.2012.017}
  {\bibfield  {journal} {\bibinfo  {journal} {Publ. Astron. Soc. Aust.}\
  }\textbf {\bibinfo {volume} {30}},\ \bibinfo {eid} {e017} (\bibinfo {year}
  {2013})},\ \Eprint {http://arxiv.org/abs/1210.6130} {arXiv:1210.6130
  [astro-ph.IM]} \BibitemShut {NoStop}%
\bibitem [{\citenamefont {{McLaughlin}}(2013)}]{2013CQGra..30v4008M}%
  \BibitemOpen
  \bibfield  {author} {\bibinfo {author} {\bibfnamefont {M.~A.}\ \bibnamefont
  {{McLaughlin}}},\ }\href {\doibase 10.1088/0264-9381/30/22/224008} {\bibfield
   {journal} {\bibinfo  {journal} {Classical and Quantum Gravity}\ }\textbf
  {\bibinfo {volume} {30}},\ \bibinfo {eid} {224008} (\bibinfo {year}
  {2013})},\ \Eprint {http://arxiv.org/abs/1310.0758} {arXiv:1310.0758
  [astro-ph.IM]} \BibitemShut {NoStop}%
\bibitem [{\citenamefont {{Kramer}}\ and\ \citenamefont
  {{Champion}}(2013)}]{2013CQGra..30v4009K}%
  \BibitemOpen
  \bibfield  {author} {\bibinfo {author} {\bibfnamefont {M.}~\bibnamefont
  {{Kramer}}}\ and\ \bibinfo {author} {\bibfnamefont {D.~J.}\ \bibnamefont
  {{Champion}}},\ }\href {\doibase 10.1088/0264-9381/30/22/224009} {\bibfield
  {journal} {\bibinfo  {journal} {Classical and Quantum Gravity}\ }\textbf
  {\bibinfo {volume} {30}},\ \bibinfo {eid} {224009} (\bibinfo {year}
  {2013})}\BibitemShut {NoStop}%
\bibitem [{\citenamefont {{Hobbs}}\ \emph
  {et~al.}(2010{\natexlab{a}})\citenamefont {{Hobbs}} \emph
  {et~al.}}]{2010CQGra..27h4013H}%
  \BibitemOpen
  \bibfield  {author} {\bibinfo {author} {\bibfnamefont {G.}~\bibnamefont
  {{Hobbs}}} \emph {et~al.},\ }\href {\doibase 10.1088/0264-9381/27/8/084013}
  {\bibfield  {journal} {\bibinfo  {journal} {Classical and Quantum Gravity}\
  }\textbf {\bibinfo {volume} {27}},\ \bibinfo {eid} {084013} (\bibinfo {year}
  {2010}{\natexlab{a}})},\ \Eprint {http://arxiv.org/abs/0911.5206}
  {arXiv:0911.5206 [astro-ph.SR]} \BibitemShut {NoStop}%
\bibitem [{\citenamefont {{Verbiest}}\ \emph {et~al.}(2016)\citenamefont
  {{Verbiest}} \emph {et~al.}}]{IPTA16}%
  \BibitemOpen
  \bibfield  {author} {\bibinfo {author} {\bibfnamefont {J.~P.~W.}\
  \bibnamefont {{Verbiest}}} \emph {et~al.},\ }\href {\doibase
  10.1093/mnras/stw347} {\bibfield  {journal} {\bibinfo  {journal} {Mon. Not.
  R. Astron. Soc.}\ }\textbf {\bibinfo {volume} {458}},\ \bibinfo {pages}
  {1267} (\bibinfo {year} {2016})},\ \Eprint {http://arxiv.org/abs/1602.03640}
  {arXiv:1602.03640 [astro-ph.IM]} \BibitemShut {NoStop}%
\bibitem [{\citenamefont {{Shannon}}\ \emph {et~al.}(2013)\citenamefont
  {{Shannon}} \emph {et~al.}}]{2013Sci...342..334S}%
  \BibitemOpen
  \bibfield  {author} {\bibinfo {author} {\bibfnamefont {R.~M.}\ \bibnamefont
  {{Shannon}}} \emph {et~al.},\ }\href {\doibase 10.1126/science.1238012}
  {\bibfield  {journal} {\bibinfo  {journal} {Science}\ }\textbf {\bibinfo
  {volume} {342}},\ \bibinfo {pages} {334} (\bibinfo {year} {2013})},\ \Eprint
  {http://arxiv.org/abs/1310.4569} {arXiv:1310.4569} \BibitemShut {NoStop}%
\bibitem [{\citenamefont {{Zhu}}\ \emph {et~al.}(2014)\citenamefont {{Zhu}}
  \emph {et~al.}}]{2014MNRAS.444.3709Z}%
  \BibitemOpen
  \bibfield  {author} {\bibinfo {author} {\bibfnamefont {X.-J.}\ \bibnamefont
  {{Zhu}}} \emph {et~al.},\ }\href {\doibase 10.1093/mnras/stu1717} {\bibfield
  {journal} {\bibinfo  {journal} {Mon. Not. R. Astron. Soc.}\ }\textbf
  {\bibinfo {volume} {444}},\ \bibinfo {pages} {3709} (\bibinfo {year}
  {2014})},\ \Eprint {http://arxiv.org/abs/1408.5129} {arXiv:1408.5129}
  \BibitemShut {NoStop}%
\bibitem [{\citenamefont {{Wang}}\ \emph
  {et~al.}(2015{\natexlab{a}})\citenamefont {{Wang}} \emph
  {et~al.}}]{2015MNRAS.446.1657W}%
  \BibitemOpen
  \bibfield  {author} {\bibinfo {author} {\bibfnamefont {J.~B.}\ \bibnamefont
  {{Wang}}} \emph {et~al.},\ }\href {\doibase 10.1093/mnras/stu2137} {\bibfield
   {journal} {\bibinfo  {journal} {Mon. Not. R. Astron. Soc.}\ }\textbf
  {\bibinfo {volume} {446}},\ \bibinfo {pages} {1657} (\bibinfo {year}
  {2015}{\natexlab{a}})},\ \Eprint {http://arxiv.org/abs/1410.3323}
  {arXiv:1410.3323} \BibitemShut {NoStop}%
\bibitem [{\citenamefont {{Shannon}}\ \emph {et~al.}(2015)\citenamefont
  {{Shannon}} \emph {et~al.}}]{2015Sci...349.1522S}%
  \BibitemOpen
  \bibfield  {author} {\bibinfo {author} {\bibfnamefont {R.~M.}\ \bibnamefont
  {{Shannon}}} \emph {et~al.},\ }\href {\doibase 10.1126/science.aab1910}
  {\bibfield  {journal} {\bibinfo  {journal} {Science}\ }\textbf {\bibinfo
  {volume} {349}},\ \bibinfo {pages} {1522} (\bibinfo {year} {2015})},\ \Eprint
  {http://arxiv.org/abs/1509.07320} {arXiv:1509.07320} \BibitemShut {NoStop}%
\bibitem [{\citenamefont {{Demorest}}\ \emph {et~al.}(2013)\citenamefont
  {{Demorest}} \emph {et~al.}}]{NANOGrav5yr}%
  \BibitemOpen
  \bibfield  {author} {\bibinfo {author} {\bibfnamefont {P.~B.}\ \bibnamefont
  {{Demorest}}} \emph {et~al.},\ }\href {\doibase 10.1088/0004-637X/762/2/94}
  {\bibfield  {journal} {\bibinfo  {journal} {\apj}\ }\textbf {\bibinfo
  {volume} {762}},\ \bibinfo {eid} {94} (\bibinfo {year} {2013})},\ \Eprint
  {http://arxiv.org/abs/1201.6641} {arXiv:1201.6641} \BibitemShut {NoStop}%
\bibitem [{\citenamefont {Ellis}\ and\ \citenamefont {van
  Haasteren}(2017)}]{PAL2}%
  \BibitemOpen
  \bibfield  {author} {\bibinfo {author} {\bibfnamefont {J.}~\bibnamefont
  {Ellis}}\ and\ \bibinfo {author} {\bibfnamefont {R.}~\bibnamefont {van
  Haasteren}},\ }\href {\doibase 10.5281/zenodo.251456} {\enquote {\bibinfo
  {title} {jellis18/pal2: Pal2},}\ } (\bibinfo {year} {2017})\BibitemShut
  {NoStop}%
\bibitem [{\citenamefont {Taylor}\ and\ \citenamefont
  {Baker}(2017)}]{steve_taylor_2017_250258}%
  \BibitemOpen
  \bibfield  {author} {\bibinfo {author} {\bibfnamefont {S.}~\bibnamefont
  {Taylor}}\ and\ \bibinfo {author} {\bibfnamefont {P.~T.}\ \bibnamefont
  {Baker}},\ }\href {\doibase 10.5281/zenodo.250258} {\enquote {\bibinfo
  {title} {stevertaylor/nx01 1.2},}\ } (\bibinfo {year} {2017})\BibitemShut
  {NoStop}%
\bibitem [{\citenamefont {{Krnjaic}}\ \emph {et~al.}(2018)\citenamefont
  {{Krnjaic}}, \citenamefont {{Machado}},\ and\ \citenamefont
  {{Necib}}}]{2018PhRvD..97g5017K}%
  \BibitemOpen
  \bibfield  {author} {\bibinfo {author} {\bibfnamefont {G.}~\bibnamefont
  {{Krnjaic}}}, \bibinfo {author} {\bibfnamefont {P.~A.~N.}\ \bibnamefont
  {{Machado}}}, \ and\ \bibinfo {author} {\bibfnamefont {L.}~\bibnamefont
  {{Necib}}},\ }\href {\doibase 10.1103/PhysRevD.97.075017} {\bibfield
  {journal} {\bibinfo  {journal} {\prd}\ }\textbf {\bibinfo {volume} {97}},\
  \bibinfo {eid} {075017} (\bibinfo {year} {2018})}\BibitemShut {NoStop}%
\bibitem [{\citenamefont {{Brdar}}\ \emph {et~al.}(2018)\citenamefont
  {{Brdar}}, \citenamefont {{Kopp}}, \citenamefont {{Liu}}, \citenamefont
  {{Prass}},\ and\ \citenamefont {{Wang}}}]{2018PhRvD..97d3001B}%
  \BibitemOpen
  \bibfield  {author} {\bibinfo {author} {\bibfnamefont {V.}~\bibnamefont
  {{Brdar}}}, \bibinfo {author} {\bibfnamefont {J.}~\bibnamefont {{Kopp}}},
  \bibinfo {author} {\bibfnamefont {J.}~\bibnamefont {{Liu}}}, \bibinfo
  {author} {\bibfnamefont {P.}~\bibnamefont {{Prass}}}, \ and\ \bibinfo
  {author} {\bibfnamefont {X.-P.}\ \bibnamefont {{Wang}}},\ }\href {\doibase
  10.1103/PhysRevD.97.043001} {\bibfield  {journal} {\bibinfo  {journal}
  {\prd}\ }\textbf {\bibinfo {volume} {97}},\ \bibinfo {eid} {043001} (\bibinfo
  {year} {2018})},\ \Eprint {http://arxiv.org/abs/1705.09455} {arXiv:1705.09455
  [hep-ph]} \BibitemShut {NoStop}%
\bibitem [{\citenamefont {{Zhu}}\ \emph {et~al.}(2016)\citenamefont {{Zhu}},
  \citenamefont {{Wen}}, \citenamefont {{Xiong}}, \citenamefont {{Xu}},
  \citenamefont {{Wang}}, \citenamefont {{Mohanty}}, \citenamefont {{Hobbs}},\
  and\ \citenamefont {{Manchester}}}]{ZhuPTA16}%
  \BibitemOpen
  \bibfield  {author} {\bibinfo {author} {\bibfnamefont {X.-J.}\ \bibnamefont
  {{Zhu}}}, \bibinfo {author} {\bibfnamefont {L.}~\bibnamefont {{Wen}}},
  \bibinfo {author} {\bibfnamefont {J.}~\bibnamefont {{Xiong}}}, \bibinfo
  {author} {\bibfnamefont {Y.}~\bibnamefont {{Xu}}}, \bibinfo {author}
  {\bibfnamefont {Y.}~\bibnamefont {{Wang}}}, \bibinfo {author} {\bibfnamefont
  {S.~D.}\ \bibnamefont {{Mohanty}}}, \bibinfo {author} {\bibfnamefont
  {G.}~\bibnamefont {{Hobbs}}}, \ and\ \bibinfo {author} {\bibfnamefont
  {R.~N.}\ \bibnamefont {{Manchester}}},\ }\href {\doibase
  10.1093/mnras/stw1446} {\bibfield  {journal} {\bibinfo  {journal} {Mon. Not.
  R. Astron. Soc.}\ }\textbf {\bibinfo {volume} {461}},\ \bibinfo {pages}
  {1317} (\bibinfo {year} {2016})},\ \Eprint {http://arxiv.org/abs/1606.04539}
  {arXiv:1606.04539 [astro-ph.IM]} \BibitemShut {NoStop}%
\bibitem [{\citenamefont {{Bovy}}\ and\ \citenamefont
  {{Tremaine}}(2012)}]{2012ApJ...756...89B}%
  \BibitemOpen
  \bibfield  {author} {\bibinfo {author} {\bibfnamefont {J.}~\bibnamefont
  {{Bovy}}}\ and\ \bibinfo {author} {\bibfnamefont {S.}~\bibnamefont
  {{Tremaine}}},\ }\href {\doibase 10.1088/0004-637X/756/1/89} {\bibfield
  {journal} {\bibinfo  {journal} {\apj}\ }\textbf {\bibinfo {volume} {756}},\
  \bibinfo {eid} {89} (\bibinfo {year} {2012})},\ \Eprint
  {http://arxiv.org/abs/1205.4033} {arXiv:1205.4033 [astro-ph.GA]} \BibitemShut
  {NoStop}%
\bibitem [{\citenamefont {{Read}}(2014)}]{DMdensityReview14}%
  \BibitemOpen
  \bibfield  {author} {\bibinfo {author} {\bibfnamefont {J.~I.}\ \bibnamefont
  {{Read}}},\ }\href {\doibase 10.1088/0954-3899/41/6/063101} {\bibfield
  {journal} {\bibinfo  {journal} {Journal of Physics G Nuclear Physics}\
  }\textbf {\bibinfo {volume} {41}},\ \bibinfo {eid} {063101} (\bibinfo {year}
  {2014})},\ \Eprint {http://arxiv.org/abs/1404.1938} {arXiv:1404.1938}
  \BibitemShut {NoStop}%
\bibitem [{\citenamefont {{Sivertsson}}\ \emph {et~al.}(2018)\citenamefont
  {{Sivertsson}}, \citenamefont {{Silverwood}}, \citenamefont {{Read}},
  \citenamefont {{Bertone}},\ and\ \citenamefont {{Steger}}}]{DMdensity18}%
  \BibitemOpen
  \bibfield  {author} {\bibinfo {author} {\bibfnamefont {S.}~\bibnamefont
  {{Sivertsson}}}, \bibinfo {author} {\bibfnamefont {H.}~\bibnamefont
  {{Silverwood}}}, \bibinfo {author} {\bibfnamefont {J.~I.}\ \bibnamefont
  {{Read}}}, \bibinfo {author} {\bibfnamefont {G.}~\bibnamefont {{Bertone}}}, \
  and\ \bibinfo {author} {\bibfnamefont {P.}~\bibnamefont {{Steger}}},\ }\href
  {\doibase 10.1093/mnras/sty977} {\bibfield  {journal} {\bibinfo  {journal}
  {Mon. Not. R. Astron. Soc.}\ }\textbf {\bibinfo {volume} {478}},\ \bibinfo
  {pages} {1677} (\bibinfo {year} {2018})},\ \Eprint
  {http://arxiv.org/abs/1708.07836} {arXiv:1708.07836} \BibitemShut {NoStop}%
\bibitem [{\citenamefont {{Lentati}}\ \emph {et~al.}(2016)\citenamefont
  {{Lentati}} \emph {et~al.}}]{2016MNRAS.458.2161L}%
  \BibitemOpen
  \bibfield  {author} {\bibinfo {author} {\bibfnamefont {L.}~\bibnamefont
  {{Lentati}}} \emph {et~al.},\ }\href {\doibase 10.1093/mnras/stw395}
  {\bibfield  {journal} {\bibinfo  {journal} {Mon. Not. R. Astron. Soc.}\
  }\textbf {\bibinfo {volume} {458}},\ \bibinfo {pages} {2161} (\bibinfo {year}
  {2016})},\ \Eprint {http://arxiv.org/abs/1602.05570} {arXiv:1602.05570
  [astro-ph.IM]} \BibitemShut {NoStop}%
\bibitem [{\citenamefont {{Manchester}}\ \emph {et~al.}(2005)\citenamefont
  {{Manchester}}, \citenamefont {{Hobbs}}, \citenamefont {{Teoh}},\ and\
  \citenamefont {{Hobbs}}}]{ATNF05Pulsar}%
  \BibitemOpen
  \bibfield  {author} {\bibinfo {author} {\bibfnamefont {R.~N.}\ \bibnamefont
  {{Manchester}}}, \bibinfo {author} {\bibfnamefont {G.~B.}\ \bibnamefont
  {{Hobbs}}}, \bibinfo {author} {\bibfnamefont {A.}~\bibnamefont {{Teoh}}}, \
  and\ \bibinfo {author} {\bibfnamefont {M.}~\bibnamefont {{Hobbs}}},\ }\href
  {\doibase 10.1086/428488} {\bibfield  {journal} {\bibinfo  {journal} {AJ}\
  }\textbf {\bibinfo {volume} {129}},\ \bibinfo {pages} {1993} (\bibinfo {year}
  {2005})},\ \Eprint {http://arxiv.org/abs/arXiv:astro-ph/0412641}
  {arXiv:astro-ph/0412641} \BibitemShut {NoStop}%
\bibitem [{\citenamefont {Folkner}\ \emph {et~al.}(2007)\citenamefont
  {Folkner}, \citenamefont {Standish}, \citenamefont {Williams},\ and\
  \citenamefont {Boggs}}]{DE418}%
  \BibitemOpen
  \bibfield  {author} {\bibinfo {author} {\bibfnamefont {W.}~\bibnamefont
  {Folkner}}, \bibinfo {author} {\bibfnamefont {E.}~\bibnamefont {Standish}},
  \bibinfo {author} {\bibfnamefont {J.}~\bibnamefont {Williams}}, \ and\
  \bibinfo {author} {\bibfnamefont {D.}~\bibnamefont {Boggs}},\ }\href@noop {}
  {\bibfield  {journal} {\bibinfo  {journal} {Jet Propulsion Laboratory,
  Memorandum IOM 343R-08-003}\ } (\bibinfo {year} {2007})}\BibitemShut
  {NoStop}%
\bibitem [{\citenamefont {{Hobbs}}\ \emph {et~al.}(2006)\citenamefont
  {{Hobbs}}, \citenamefont {{Edwards}},\ and\ \citenamefont
  {{Manchester}}}]{TEMPO2}%
  \BibitemOpen
  \bibfield  {author} {\bibinfo {author} {\bibfnamefont {G.~B.}\ \bibnamefont
  {{Hobbs}}}, \bibinfo {author} {\bibfnamefont {R.~T.}\ \bibnamefont
  {{Edwards}}}, \ and\ \bibinfo {author} {\bibfnamefont {R.~N.}\ \bibnamefont
  {{Manchester}}},\ }\href {\doibase 10.1111/j.1365-2966.2006.10302.x}
  {\bibfield  {journal} {\bibinfo  {journal} {Mon. Not. R. Astron. Soc.}\
  }\textbf {\bibinfo {volume} {369}},\ \bibinfo {pages} {655} (\bibinfo {year}
  {2006})},\ \Eprint {http://arxiv.org/abs/astro-ph/0603381} {astro-ph/0603381}
  \BibitemShut {NoStop}%
\bibitem [{\citenamefont {{Edwards}}\ \emph {et~al.}(2006)\citenamefont
  {{Edwards}}, \citenamefont {{Hobbs}},\ and\ \citenamefont
  {{Manchester}}}]{TEMPO2model}%
  \BibitemOpen
  \bibfield  {author} {\bibinfo {author} {\bibfnamefont {R.~T.}\ \bibnamefont
  {{Edwards}}}, \bibinfo {author} {\bibfnamefont {G.~B.}\ \bibnamefont
  {{Hobbs}}}, \ and\ \bibinfo {author} {\bibfnamefont {R.~N.}\ \bibnamefont
  {{Manchester}}},\ }\href {\doibase 10.1111/j.1365-2966.2006.10870.x}
  {\bibfield  {journal} {\bibinfo  {journal} {Mon. Not. R. Astron. Soc.}\
  }\textbf {\bibinfo {volume} {372}},\ \bibinfo {pages} {1549} (\bibinfo {year}
  {2006})},\ \Eprint {http://arxiv.org/abs/astro-ph/0607664} {astro-ph/0607664}
  \BibitemShut {NoStop}%
\bibitem [{\citenamefont {{van Haasteren}}\ \emph {et~al.}(2009)\citenamefont
  {{van Haasteren}}, \citenamefont {{Levin}}, \citenamefont {{McDonald}},\ and\
  \citenamefont {{Lu}}}]{2009MNRAS.395.1005V}%
  \BibitemOpen
  \bibfield  {author} {\bibinfo {author} {\bibfnamefont {R.}~\bibnamefont {{van
  Haasteren}}}, \bibinfo {author} {\bibfnamefont {Y.}~\bibnamefont {{Levin}}},
  \bibinfo {author} {\bibfnamefont {P.}~\bibnamefont {{McDonald}}}, \ and\
  \bibinfo {author} {\bibfnamefont {T.}~\bibnamefont {{Lu}}},\ }\href {\doibase
  10.1111/j.1365-2966.2009.14590.x} {\bibfield  {journal} {\bibinfo  {journal}
  {Mon. Not. R. Astron. Soc.}\ }\textbf {\bibinfo {volume} {395}},\ \bibinfo
  {pages} {1005} (\bibinfo {year} {2009})},\ \Eprint
  {http://arxiv.org/abs/0809.0791} {arXiv:0809.0791} \BibitemShut {NoStop}%
\bibitem [{\citenamefont {{van Haasteren}}\ and\ \citenamefont
  {{Levin}}(2013)}]{vanHaasteren13}%
  \BibitemOpen
  \bibfield  {author} {\bibinfo {author} {\bibfnamefont {R.}~\bibnamefont {{van
  Haasteren}}}\ and\ \bibinfo {author} {\bibfnamefont {Y.}~\bibnamefont
  {{Levin}}},\ }\href {\doibase 10.1093/mnras/sts097} {\bibfield  {journal}
  {\bibinfo  {journal} {Mon. Not. R. Astron. Soc.}\ }\textbf {\bibinfo {volume}
  {428}},\ \bibinfo {pages} {1147} (\bibinfo {year} {2013})},\ \Eprint
  {http://arxiv.org/abs/1202.5932} {arXiv:1202.5932 [astro-ph.IM]} \BibitemShut
  {NoStop}%
\bibitem [{\citenamefont {Press}\ \emph {et~al.}(1996)\citenamefont {Press},
  \citenamefont {Teukolsky}, \citenamefont {Vetterling},\ and\ \citenamefont
  {Flannery}}]{Numerical_Recipes}%
  \BibitemOpen
  \bibfield  {author} {\bibinfo {author} {\bibfnamefont {W.~H.}\ \bibnamefont
  {Press}}, \bibinfo {author} {\bibfnamefont {S.~A.}\ \bibnamefont
  {Teukolsky}}, \bibinfo {author} {\bibfnamefont {W.~T.}\ \bibnamefont
  {Vetterling}}, \ and\ \bibinfo {author} {\bibfnamefont {B.~P.}\ \bibnamefont
  {Flannery}},\ }\href@noop {} {\emph {\bibinfo {title} {Numerical Recipes in
  C}}},\ Vol.~\bibinfo {volume} {2}\ (\bibinfo  {publisher} {Cambridge
  university press Cambridge},\ \bibinfo {year} {1996})\BibitemShut {NoStop}%
\bibitem [{\citenamefont {{Boynton}}\ \emph {et~al.}(1972)\citenamefont
  {{Boynton}}, \citenamefont {{Groth}}, \citenamefont {{Hutchinson}},
  \citenamefont {{Nanos}}, \citenamefont {{Partridge}},\ and\ \citenamefont
  {{Wilkinson}}}]{1972ApJ...175..217B}%
  \BibitemOpen
  \bibfield  {author} {\bibinfo {author} {\bibfnamefont {P.~E.}\ \bibnamefont
  {{Boynton}}}, \bibinfo {author} {\bibfnamefont {E.~J.}\ \bibnamefont
  {{Groth}}}, \bibinfo {author} {\bibfnamefont {D.~P.}\ \bibnamefont
  {{Hutchinson}}}, \bibinfo {author} {\bibfnamefont {G.~P.}\ \bibnamefont
  {{Nanos}}, \bibfnamefont {Jr.}}, \bibinfo {author} {\bibfnamefont {R.~B.}\
  \bibnamefont {{Partridge}}}, \ and\ \bibinfo {author} {\bibfnamefont {D.~T.}\
  \bibnamefont {{Wilkinson}}},\ }\href {\doibase 10.1086/151550} {\bibfield
  {journal} {\bibinfo  {journal} {\apj}\ }\textbf {\bibinfo {volume} {175}},\
  \bibinfo {pages} {217} (\bibinfo {year} {1972})}\BibitemShut {NoStop}%
\bibitem [{\citenamefont {{Blandford}}\ \emph {et~al.}(1984)\citenamefont
  {{Blandford}}, \citenamefont {{Narayan}},\ and\ \citenamefont
  {{Romani}}}]{Blandford84rednoise}%
  \BibitemOpen
  \bibfield  {author} {\bibinfo {author} {\bibfnamefont {R.}~\bibnamefont
  {{Blandford}}}, \bibinfo {author} {\bibfnamefont {R.}~\bibnamefont
  {{Narayan}}}, \ and\ \bibinfo {author} {\bibfnamefont {R.~W.}\ \bibnamefont
  {{Romani}}},\ }\href {\doibase 10.1007/BF02714466} {\bibfield  {journal}
  {\bibinfo  {journal} {Journal of Astrophysics and Astronomy}\ }\textbf
  {\bibinfo {volume} {5}},\ \bibinfo {pages} {369} (\bibinfo {year}
  {1984})}\BibitemShut {NoStop}%
\bibitem [{\citenamefont {{Hobbs}}\ \emph
  {et~al.}(2010{\natexlab{b}})\citenamefont {{Hobbs}}, \citenamefont {{Lyne}},\
  and\ \citenamefont {{Kramer}}}]{2010MNRAS.402.1027H}%
  \BibitemOpen
  \bibfield  {author} {\bibinfo {author} {\bibfnamefont {G.}~\bibnamefont
  {{Hobbs}}}, \bibinfo {author} {\bibfnamefont {A.~G.}\ \bibnamefont {{Lyne}}},
  \ and\ \bibinfo {author} {\bibfnamefont {M.}~\bibnamefont {{Kramer}}},\
  }\href {\doibase 10.1111/j.1365-2966.2009.15938.x} {\bibfield  {journal}
  {\bibinfo  {journal} {Mon. Not. R. Astron. Soc.}\ }\textbf {\bibinfo {volume}
  {402}},\ \bibinfo {pages} {1027} (\bibinfo {year} {2010}{\natexlab{b}})},\
  \Eprint {http://arxiv.org/abs/0912.4537} {arXiv:0912.4537} \BibitemShut
  {NoStop}%
\bibitem [{\citenamefont {William}(1989)}]{Hager}%
  \BibitemOpen
  \bibfield  {author} {\bibinfo {author} {\bibfnamefont {H.~W.}\ \bibnamefont
  {William}},\ }\href {\doibase https://doi.org/10.1137/1031049} {\bibfield
  {journal} {\bibinfo  {journal} {SIAM Revew}\ }\textbf {\bibinfo {volume}
  {31}},\ \bibinfo {pages} {221} (\bibinfo {year} {1989})}\BibitemShut
  {NoStop}%
\bibitem [{\citenamefont {{van Haasteren}}\ and\ \citenamefont
  {{Vallisneri}}(2015)}]{2015MNRAS.446.1170V}%
  \BibitemOpen
  \bibfield  {author} {\bibinfo {author} {\bibfnamefont {R.}~\bibnamefont {{van
  Haasteren}}}\ and\ \bibinfo {author} {\bibfnamefont {M.}~\bibnamefont
  {{Vallisneri}}},\ }\href {\doibase 10.1093/mnras/stu2157} {\bibfield
  {journal} {\bibinfo  {journal} {Mon. Not. R. Astron. Soc.}\ }\textbf
  {\bibinfo {volume} {446}},\ \bibinfo {pages} {1170} (\bibinfo {year}
  {2015})},\ \Eprint {http://arxiv.org/abs/1407.6710} {arXiv:1407.6710
  [astro-ph.IM]} \BibitemShut {NoStop}%
\bibitem [{\citenamefont {{Arzoumanian}}\ \emph {et~al.}(2014)\citenamefont
  {{Arzoumanian}} \emph {et~al.}}]{2014ApJ...794..141A}%
  \BibitemOpen
  \bibfield  {author} {\bibinfo {author} {\bibfnamefont {Z.}~\bibnamefont
  {{Arzoumanian}}} \emph {et~al.},\ }\href {\doibase
  10.1088/0004-637X/794/2/141} {\bibfield  {journal} {\bibinfo  {journal}
  {\apj}\ }\textbf {\bibinfo {volume} {794}},\ \bibinfo {eid} {141} (\bibinfo
  {year} {2014})},\ \Eprint {http://arxiv.org/abs/1404.1267} {arXiv:1404.1267}
  \BibitemShut {NoStop}%
\bibitem [{\citenamefont {{Lentati}}\ \emph {et~al.}(2013)\citenamefont
  {{Lentati}}, \citenamefont {{Alexander}}, \citenamefont {{Hobson}},
  \citenamefont {{Taylor}}, \citenamefont {{Gair}}, \citenamefont {{Balan}},\
  and\ \citenamefont {{van Haasteren}}}]{2013PhRvD..87j4021L}%
  \BibitemOpen
  \bibfield  {author} {\bibinfo {author} {\bibfnamefont {L.}~\bibnamefont
  {{Lentati}}}, \bibinfo {author} {\bibfnamefont {P.}~\bibnamefont
  {{Alexander}}}, \bibinfo {author} {\bibfnamefont {M.~P.}\ \bibnamefont
  {{Hobson}}}, \bibinfo {author} {\bibfnamefont {S.}~\bibnamefont {{Taylor}}},
  \bibinfo {author} {\bibfnamefont {J.}~\bibnamefont {{Gair}}}, \bibinfo
  {author} {\bibfnamefont {S.~T.}\ \bibnamefont {{Balan}}}, \ and\ \bibinfo
  {author} {\bibfnamefont {R.}~\bibnamefont {{van Haasteren}}},\ }\href
  {\doibase 10.1103/PhysRevD.87.104021} {\bibfield  {journal} {\bibinfo
  {journal} {\prd}\ }\textbf {\bibinfo {volume} {87}},\ \bibinfo {eid} {104021}
  (\bibinfo {year} {2013})},\ \Eprint {http://arxiv.org/abs/1210.3578}
  {arXiv:1210.3578 [astro-ph.IM]} \BibitemShut {NoStop}%
\bibitem [{\citenamefont {{Lentati}}\ \emph {et~al.}(2014)\citenamefont
  {{Lentati}}, \citenamefont {{Alexander}}, \citenamefont {{Hobson}},
  \citenamefont {{Feroz}}, \citenamefont {{van Haasteren}}, \citenamefont
  {{Lee}},\ and\ \citenamefont {{Shannon}}}]{Lentati14}%
  \BibitemOpen
  \bibfield  {author} {\bibinfo {author} {\bibfnamefont {L.}~\bibnamefont
  {{Lentati}}}, \bibinfo {author} {\bibfnamefont {P.}~\bibnamefont
  {{Alexander}}}, \bibinfo {author} {\bibfnamefont {M.~P.}\ \bibnamefont
  {{Hobson}}}, \bibinfo {author} {\bibfnamefont {F.}~\bibnamefont {{Feroz}}},
  \bibinfo {author} {\bibfnamefont {R.}~\bibnamefont {{van Haasteren}}},
  \bibinfo {author} {\bibfnamefont {K.~J.}\ \bibnamefont {{Lee}}}, \ and\
  \bibinfo {author} {\bibfnamefont {R.~M.}\ \bibnamefont {{Shannon}}},\ }\href
  {\doibase 10.1093/mnras/stt2122} {\bibfield  {journal} {\bibinfo  {journal}
  {Mon. Not. R. Astron. Soc.}\ }\textbf {\bibinfo {volume} {437}},\ \bibinfo
  {pages} {3004} (\bibinfo {year} {2014})},\ \Eprint
  {http://arxiv.org/abs/1310.2120} {arXiv:1310.2120 [astro-ph.IM]} \BibitemShut
  {NoStop}%
\bibitem [{\citenamefont {{Feroz}}\ \emph {et~al.}(2009)\citenamefont
  {{Feroz}}, \citenamefont {{Hobson}},\ and\ \citenamefont
  {{Bridges}}}]{MULTINEST09}%
  \BibitemOpen
  \bibfield  {author} {\bibinfo {author} {\bibfnamefont {F.}~\bibnamefont
  {{Feroz}}}, \bibinfo {author} {\bibfnamefont {M.~P.}\ \bibnamefont
  {{Hobson}}}, \ and\ \bibinfo {author} {\bibfnamefont {M.}~\bibnamefont
  {{Bridges}}},\ }\href {\doibase 10.1111/j.1365-2966.2009.14548.x} {\bibfield
  {journal} {\bibinfo  {journal} {Mon. Not. R. Astron. Soc.}\ }\textbf
  {\bibinfo {volume} {398}},\ \bibinfo {pages} {1601} (\bibinfo {year}
  {2009})},\ \Eprint {http://arxiv.org/abs/0809.3437} {arXiv:0809.3437}
  \BibitemShut {NoStop}%
\bibitem [{\citenamefont {{Feroz}}\ \emph {et~al.}(2013)\citenamefont
  {{Feroz}}, \citenamefont {{Hobson}}, \citenamefont {{Cameron}},\ and\
  \citenamefont {{Pettitt}}}]{2013arXiv1306.2144F}%
  \BibitemOpen
  \bibfield  {author} {\bibinfo {author} {\bibfnamefont {F.}~\bibnamefont
  {{Feroz}}}, \bibinfo {author} {\bibfnamefont {M.~P.}\ \bibnamefont
  {{Hobson}}}, \bibinfo {author} {\bibfnamefont {E.}~\bibnamefont {{Cameron}}},
  \ and\ \bibinfo {author} {\bibfnamefont {A.~N.}\ \bibnamefont {{Pettitt}}},\
  }\href@noop {} {\bibfield  {journal} {\bibinfo  {journal} {ArXiv e-prints}\ }
  (\bibinfo {year} {2013})},\ \Eprint {http://arxiv.org/abs/1306.2144}
  {arXiv:1306.2144 [astro-ph.IM]} \BibitemShut {NoStop}%
\bibitem [{\citenamefont {{Keith}}\ \emph {et~al.}(2013)\citenamefont {{Keith}}
  \emph {et~al.}}]{KeithDM13}%
  \BibitemOpen
  \bibfield  {author} {\bibinfo {author} {\bibfnamefont {M.~J.}\ \bibnamefont
  {{Keith}}} \emph {et~al.},\ }\href {\doibase 10.1093/mnras/sts486} {\bibfield
   {journal} {\bibinfo  {journal} {Mon. Not. R. Astron. Soc.}\ }\textbf
  {\bibinfo {volume} {429}},\ \bibinfo {pages} {2161} (\bibinfo {year}
  {2013})},\ \Eprint {http://arxiv.org/abs/1211.5887} {arXiv:1211.5887}
  \BibitemShut {NoStop}%
\bibitem [{\citenamefont {{Coles}}\ \emph {et~al.}(2015)\citenamefont {{Coles}}
  \emph {et~al.}}]{ColesESE}%
  \BibitemOpen
  \bibfield  {author} {\bibinfo {author} {\bibfnamefont {W.~A.}\ \bibnamefont
  {{Coles}}} \emph {et~al.},\ }\href {\doibase 10.1088/0004-637X/808/2/113}
  {\bibfield  {journal} {\bibinfo  {journal} {\apj}\ }\textbf {\bibinfo
  {volume} {808}},\ \bibinfo {eid} {113} (\bibinfo {year} {2015})},\ \Eprint
  {http://arxiv.org/abs/1506.07948} {arXiv:1506.07948 [astro-ph.SR]}
  \BibitemShut {NoStop}%
\bibitem [{\citenamefont {{You}}\ \emph {et~al.}(2007)\citenamefont {{You}}
  \emph {et~al.}}]{YouDM07}%
  \BibitemOpen
  \bibfield  {author} {\bibinfo {author} {\bibfnamefont {X.~P.}\ \bibnamefont
  {{You}}} \emph {et~al.},\ }\href {\doibase 10.1111/j.1365-2966.2007.11617.x}
  {\bibfield  {journal} {\bibinfo  {journal} {Mon. Not. R. Astron. Soc.}\
  }\textbf {\bibinfo {volume} {378}},\ \bibinfo {pages} {493} (\bibinfo {year}
  {2007})},\ \Eprint {http://arxiv.org/abs/astro-ph/0702366} {astro-ph/0702366}
  \BibitemShut {NoStop}%
\bibitem [{\citenamefont {{Coles}}\ \emph {et~al.}(2011)\citenamefont
  {{Coles}}, \citenamefont {{Hobbs}}, \citenamefont {{Champion}}, \citenamefont
  {{Manchester}},\ and\ \citenamefont {{Verbiest}}}]{Coles11red}%
  \BibitemOpen
  \bibfield  {author} {\bibinfo {author} {\bibfnamefont {W.}~\bibnamefont
  {{Coles}}}, \bibinfo {author} {\bibfnamefont {G.}~\bibnamefont {{Hobbs}}},
  \bibinfo {author} {\bibfnamefont {D.~J.}\ \bibnamefont {{Champion}}},
  \bibinfo {author} {\bibfnamefont {R.~N.}\ \bibnamefont {{Manchester}}}, \
  and\ \bibinfo {author} {\bibfnamefont {J.~P.~W.}\ \bibnamefont
  {{Verbiest}}},\ }\href {\doibase 10.1111/j.1365-2966.2011.19505.x} {\bibfield
   {journal} {\bibinfo  {journal} {Mon. Not. R. Astron. Soc.}\ }\textbf
  {\bibinfo {volume} {418}},\ \bibinfo {pages} {561} (\bibinfo {year}
  {2011})},\ \Eprint {http://arxiv.org/abs/1107.5366} {arXiv:1107.5366
  [astro-ph.IM]} \BibitemShut {NoStop}%
\bibitem [{\citenamefont {{Babak}}\ \emph {et~al.}(2016)\citenamefont {{Babak}}
  \emph {et~al.}}]{2016MNRAS.455.1665B}%
  \BibitemOpen
  \bibfield  {author} {\bibinfo {author} {\bibfnamefont {S.}~\bibnamefont
  {{Babak}}} \emph {et~al.},\ }\href {\doibase 10.1093/mnras/stv2092}
  {\bibfield  {journal} {\bibinfo  {journal} {Mon. Not. R. Astron. Soc.}\
  }\textbf {\bibinfo {volume} {455}},\ \bibinfo {pages} {1665} (\bibinfo {year}
  {2016})},\ \Eprint {http://arxiv.org/abs/1509.02165} {arXiv:1509.02165}
  \BibitemShut {NoStop}%
\bibitem [{\citenamefont {{Arzoumanian}}\ \emph {et~al.}(2018)\citenamefont
  {{Arzoumanian}} \emph {et~al.}}]{NANOGrav11yrLim}%
  \BibitemOpen
  \bibfield  {author} {\bibinfo {author} {\bibfnamefont {Z.}~\bibnamefont
  {{Arzoumanian}}} \emph {et~al.},\ }\href {\doibase 10.3847/1538-4357/aabd3b}
  {\bibfield  {journal} {\bibinfo  {journal} {\apj}\ }\textbf {\bibinfo
  {volume} {859}},\ \bibinfo {eid} {47} (\bibinfo {year} {2018})},\ \Eprint
  {http://arxiv.org/abs/1801.02617} {arXiv:1801.02617 [astro-ph.HE]}
  \BibitemShut {NoStop}%
\bibitem [{\citenamefont {{Tiburzi}}\ \emph {et~al.}(2016)\citenamefont
  {{Tiburzi}}, \citenamefont {{Hobbs}}, \citenamefont {{Kerr}}, \citenamefont
  {{Coles}}, \citenamefont {{Dai}}, \citenamefont {{Manchester}}, \citenamefont
  {{Possenti}}, \citenamefont {{Shannon}},\ and\ \citenamefont
  {{You}}}]{2016MNRAS.455.4339T}%
  \BibitemOpen
  \bibfield  {author} {\bibinfo {author} {\bibfnamefont {C.}~\bibnamefont
  {{Tiburzi}}}, \bibinfo {author} {\bibfnamefont {G.}~\bibnamefont {{Hobbs}}},
  \bibinfo {author} {\bibfnamefont {M.}~\bibnamefont {{Kerr}}}, \bibinfo
  {author} {\bibfnamefont {W.~A.}\ \bibnamefont {{Coles}}}, \bibinfo {author}
  {\bibfnamefont {S.}~\bibnamefont {{Dai}}}, \bibinfo {author} {\bibfnamefont
  {R.~N.}\ \bibnamefont {{Manchester}}}, \bibinfo {author} {\bibfnamefont
  {A.}~\bibnamefont {{Possenti}}}, \bibinfo {author} {\bibfnamefont {R.~M.}\
  \bibnamefont {{Shannon}}}, \ and\ \bibinfo {author} {\bibfnamefont {X.~P.}\
  \bibnamefont {{You}}},\ }\href {\doibase 10.1093/mnras/stv2143} {\bibfield
  {journal} {\bibinfo  {journal} {Mon. Not. R. Astron. Soc.}\ }\textbf
  {\bibinfo {volume} {455}},\ \bibinfo {pages} {4339} (\bibinfo {year}
  {2016})},\ \Eprint {http://arxiv.org/abs/1510.02363} {arXiv:1510.02363
  [astro-ph.IM]} \BibitemShut {NoStop}%
\bibitem [{\citenamefont {Eberhart}\ and\ \citenamefont
  {Kennedy}(1995)}]{PSO95}%
  \BibitemOpen
  \bibfield  {author} {\bibinfo {author} {\bibfnamefont {R.~C.}\ \bibnamefont
  {Eberhart}}\ and\ \bibinfo {author} {\bibfnamefont {J.}~\bibnamefont
  {Kennedy}},\ }in\ \href@noop {} {\emph {\bibinfo {booktitle} {Proceedings of
  the sixth international symposium on micro machine and human science}}},\
  Vol.~\bibinfo {volume} {1},\ \bibinfo {organization} {New York, NY}\
  (\bibinfo  {publisher} {IEEE},\ \bibinfo {year} {1995})\ pp.\ \bibinfo
  {pages} {39--43}\BibitemShut {NoStop}%
\bibitem [{\citenamefont {{Wang}}\ \emph
  {et~al.}(2015{\natexlab{b}})\citenamefont {{Wang}}, \citenamefont
  {{Mohanty}},\ and\ \citenamefont {{Jenet}}}]{WangYan15pta}%
  \BibitemOpen
  \bibfield  {author} {\bibinfo {author} {\bibfnamefont {Y.}~\bibnamefont
  {{Wang}}}, \bibinfo {author} {\bibfnamefont {S.~D.}\ \bibnamefont
  {{Mohanty}}}, \ and\ \bibinfo {author} {\bibfnamefont {F.~A.}\ \bibnamefont
  {{Jenet}}},\ }\href {\doibase 10.1088/0004-637X/815/2/125} {\bibfield
  {journal} {\bibinfo  {journal} {\apj}\ }\textbf {\bibinfo {volume} {815}},\
  \bibinfo {eid} {125} (\bibinfo {year} {2015}{\natexlab{b}})},\ \Eprint
  {http://arxiv.org/abs/1506.01526} {arXiv:1506.01526 [astro-ph.IM]}
  \BibitemShut {NoStop}%
\bibitem [{\citenamefont {{Nan}}\ \emph {et~al.}(2011)\citenamefont {{Nan}},
  \citenamefont {{Li}}, \citenamefont {{Jin}}, \citenamefont {{Wang}},
  \citenamefont {{Zhu}}, \citenamefont {{Zhu}}, \citenamefont {{Zhang}},
  \citenamefont {{Yue}},\ and\ \citenamefont {{Qian}}}]{FAST11}%
  \BibitemOpen
  \bibfield  {author} {\bibinfo {author} {\bibfnamefont {R.}~\bibnamefont
  {{Nan}}}, \bibinfo {author} {\bibfnamefont {D.}~\bibnamefont {{Li}}},
  \bibinfo {author} {\bibfnamefont {C.}~\bibnamefont {{Jin}}}, \bibinfo
  {author} {\bibfnamefont {Q.}~\bibnamefont {{Wang}}}, \bibinfo {author}
  {\bibfnamefont {L.}~\bibnamefont {{Zhu}}}, \bibinfo {author} {\bibfnamefont
  {W.}~\bibnamefont {{Zhu}}}, \bibinfo {author} {\bibfnamefont
  {H.}~\bibnamefont {{Zhang}}}, \bibinfo {author} {\bibfnamefont
  {Y.}~\bibnamefont {{Yue}}}, \ and\ \bibinfo {author} {\bibfnamefont
  {L.}~\bibnamefont {{Qian}}},\ }\href {\doibase 10.1142/S0218271811019335}
  {\bibfield  {journal} {\bibinfo  {journal} {International Journal of Modern
  Physics D}\ }\textbf {\bibinfo {volume} {20}},\ \bibinfo {pages} {989}
  (\bibinfo {year} {2011})},\ \Eprint {http://arxiv.org/abs/1105.3794}
  {arXiv:1105.3794 [astro-ph.IM]} \BibitemShut {NoStop}%
\bibitem [{\citenamefont {{Bailes}}\ \emph {et~al.}(2018)\citenamefont
  {{Bailes}} \emph {et~al.}}]{MeerTime}%
  \BibitemOpen
  \bibfield  {author} {\bibinfo {author} {\bibfnamefont {M.}~\bibnamefont
  {{Bailes}}} \emph {et~al.},\ }in\ \href@noop {} {\emph {\bibinfo {booktitle}
  {Proceedings of MeerKAT Science2016}}}\ (\bibinfo {year} {2018})\ p.~\bibinfo
  {pages} {11},\ \Eprint {http://arxiv.org/abs/1803.07424} {arXiv:1803.07424
  [astro-ph.IM]} \BibitemShut {NoStop}%
\bibitem [{\citenamefont {{Lazio}}(2013)}]{LazioSKAPTA}%
  \BibitemOpen
  \bibfield  {author} {\bibinfo {author} {\bibfnamefont {T.~J.~W.}\
  \bibnamefont {{Lazio}}},\ }\href {\doibase 10.1088/0264-9381/30/22/224011}
  {\bibfield  {journal} {\bibinfo  {journal} {Classical and Quantum Gravity}\
  }\textbf {\bibinfo {volume} {30}},\ \bibinfo {eid} {224011} (\bibinfo {year}
  {2013})}\BibitemShut {NoStop}%
\bibitem [{\citenamefont {{Hobbs}}\ \emph {et~al.}(2014)\citenamefont
  {{Hobbs}}, \citenamefont {{Dai}}, \citenamefont {{Manchester}}, \citenamefont
  {{Shannon}}, \citenamefont {{Kerr}}, \citenamefont {{Lee}},\ and\
  \citenamefont {{Xu}}}]{HobbsFAST}%
  \BibitemOpen
  \bibfield  {author} {\bibinfo {author} {\bibfnamefont {G.}~\bibnamefont
  {{Hobbs}}}, \bibinfo {author} {\bibfnamefont {S.}~\bibnamefont {{Dai}}},
  \bibinfo {author} {\bibfnamefont {R.~N.}\ \bibnamefont {{Manchester}}},
  \bibinfo {author} {\bibfnamefont {R.~M.}\ \bibnamefont {{Shannon}}}, \bibinfo
  {author} {\bibfnamefont {M.}~\bibnamefont {{Kerr}}}, \bibinfo {author}
  {\bibfnamefont {K.~J.}\ \bibnamefont {{Lee}}}, \ and\ \bibinfo {author}
  {\bibfnamefont {R.}~\bibnamefont {{Xu}}},\ }\href@noop {} {\bibfield
  {journal} {\bibinfo  {journal} {ArXiv e-prints}\ } (\bibinfo {year}
  {2014})},\ \Eprint {http://arxiv.org/abs/1407.0435} {arXiv:1407.0435
  [astro-ph.IM]} \BibitemShut {NoStop}%
\bibitem [{\citenamefont {{Os{\l}owski}}\ \emph {et~al.}(2011)\citenamefont
  {{Os{\l}owski}}, \citenamefont {{van Straten}}, \citenamefont {{Hobbs}},
  \citenamefont {{Bailes}},\ and\ \citenamefont {{Demorest}}}]{StefanJitter}%
  \BibitemOpen
  \bibfield  {author} {\bibinfo {author} {\bibfnamefont {S.}~\bibnamefont
  {{Os{\l}owski}}}, \bibinfo {author} {\bibfnamefont {W.}~\bibnamefont {{van
  Straten}}}, \bibinfo {author} {\bibfnamefont {G.~B.}\ \bibnamefont
  {{Hobbs}}}, \bibinfo {author} {\bibfnamefont {M.}~\bibnamefont {{Bailes}}}, \
  and\ \bibinfo {author} {\bibfnamefont {P.}~\bibnamefont {{Demorest}}},\
  }\href {\doibase 10.1111/j.1365-2966.2011.19578.x} {\bibfield  {journal}
  {\bibinfo  {journal} {Mon. Not. R. Astron. Soc.}\ }\textbf {\bibinfo {volume}
  {418}},\ \bibinfo {pages} {1258} (\bibinfo {year} {2011})},\ \Eprint
  {http://arxiv.org/abs/1108.0812} {arXiv:1108.0812 [astro-ph.GA]} \BibitemShut
  {NoStop}%
\bibitem [{\citenamefont {{Shannon}}\ and\ \citenamefont
  {{Cordes}}(2012)}]{ShannonJitter12}%
  \BibitemOpen
  \bibfield  {author} {\bibinfo {author} {\bibfnamefont {R.~M.}\ \bibnamefont
  {{Shannon}}}\ and\ \bibinfo {author} {\bibfnamefont {J.~M.}\ \bibnamefont
  {{Cordes}}},\ }\href {\doibase 10.1088/0004-637X/761/1/64} {\bibfield
  {journal} {\bibinfo  {journal} {\apj}\ }\textbf {\bibinfo {volume} {761}},\
  \bibinfo {eid} {64} (\bibinfo {year} {2012})},\ \Eprint
  {http://arxiv.org/abs/1210.7021} {arXiv:1210.7021 [astro-ph.SR]} \BibitemShut
  {NoStop}%
\bibitem [{\citenamefont {{Navarro}}\ \emph {et~al.}(1996)\citenamefont
  {{Navarro}}, \citenamefont {{Frenk}},\ and\ \citenamefont
  {{White}}}]{NFWdarkmatter}%
  \BibitemOpen
  \bibfield  {author} {\bibinfo {author} {\bibfnamefont {J.~F.}\ \bibnamefont
  {{Navarro}}}, \bibinfo {author} {\bibfnamefont {C.~S.}\ \bibnamefont
  {{Frenk}}}, \ and\ \bibinfo {author} {\bibfnamefont {S.~D.~M.}\ \bibnamefont
  {{White}}},\ }\href {\doibase 10.1086/177173} {\bibfield  {journal} {\bibinfo
   {journal} {\apj}\ }\textbf {\bibinfo {volume} {462}},\ \bibinfo {pages}
  {563} (\bibinfo {year} {1996})},\ \Eprint
  {http://arxiv.org/abs/astro-ph/9508025} {astro-ph/9508025} \BibitemShut
  {NoStop}%
\bibitem [{\citenamefont {{Sofue}}(2012)}]{2012PASJ...64...75S}%
  \BibitemOpen
  \bibfield  {author} {\bibinfo {author} {\bibfnamefont {Y.}~\bibnamefont
  {{Sofue}}},\ }\href {\doibase 10.1093/pasj/64.4.75} {\bibfield  {journal}
  {\bibinfo  {journal} {Pub. of the Astr. Soc. of Jap.}\ }\textbf {\bibinfo
  {volume} {64}},\ \bibinfo {eid} {75} (\bibinfo {year} {2012})},\ \Eprint
  {http://arxiv.org/abs/1110.4431} {arXiv:1110.4431} \BibitemShut {NoStop}%
\bibitem [{\citenamefont {{Kramer}}\ \emph {et~al.}(2004)\citenamefont
  {{Kramer}}, \citenamefont {{Backer}}, \citenamefont {{Cordes}}, \citenamefont
  {{Lazio}}, \citenamefont {{Stappers}},\ and\ \citenamefont
  {{Johnston}}}]{2004NewAR..48..993K}%
  \BibitemOpen
  \bibfield  {author} {\bibinfo {author} {\bibfnamefont {M.}~\bibnamefont
  {{Kramer}}}, \bibinfo {author} {\bibfnamefont {D.~C.}\ \bibnamefont
  {{Backer}}}, \bibinfo {author} {\bibfnamefont {J.~M.}\ \bibnamefont
  {{Cordes}}}, \bibinfo {author} {\bibfnamefont {T.~J.~W.}\ \bibnamefont
  {{Lazio}}}, \bibinfo {author} {\bibfnamefont {B.~W.}\ \bibnamefont
  {{Stappers}}}, \ and\ \bibinfo {author} {\bibfnamefont {S.}~\bibnamefont
  {{Johnston}}},\ }\href {\doibase 10.1016/j.newar.2004.09.020} {\bibfield
  {journal} {\bibinfo  {journal} {New Astr. Rev.}\ }\textbf {\bibinfo {volume}
  {48}},\ \bibinfo {pages} {993} (\bibinfo {year} {2004})},\ \Eprint
  {http://arxiv.org/abs/astro-ph/0409379} {astro-ph/0409379} \BibitemShut
  {NoStop}%
\bibitem [{\citenamefont {{Hobbs}}\ \emph {et~al.}(2012)\citenamefont {{Hobbs}}
  \emph {et~al.}}]{HobbsClock12}%
  \BibitemOpen
  \bibfield  {author} {\bibinfo {author} {\bibfnamefont {G.}~\bibnamefont
  {{Hobbs}}} \emph {et~al.},\ }\href {\doibase
  10.1111/j.1365-2966.2012.21946.x} {\bibfield  {journal} {\bibinfo  {journal}
  {Mon. Not. R. Astron. Soc.}\ }\textbf {\bibinfo {volume} {427}},\ \bibinfo
  {pages} {2780} (\bibinfo {year} {2012})},\ \Eprint
  {http://arxiv.org/abs/1208.3560} {arXiv:1208.3560 [astro-ph.IM]} \BibitemShut
  {NoStop}%
\end{thebibliography}%
\appendix

\begin{figure*}
\centering
\begin{subfigure}
\centering\includegraphics[width=0.45\textwidth]{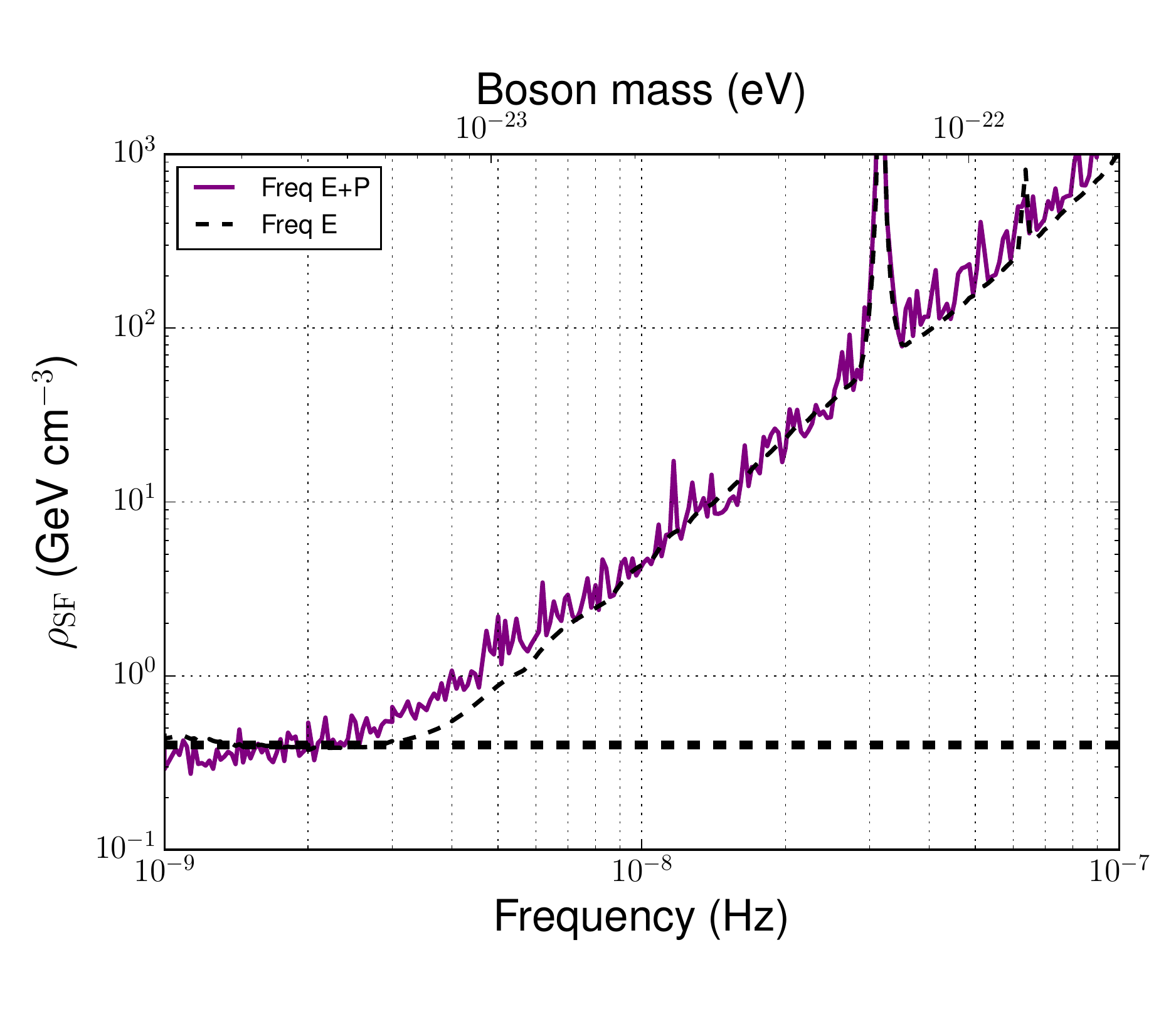}
\end{subfigure}%
\begin{subfigure}
\centering\includegraphics[width=0.45\textwidth]{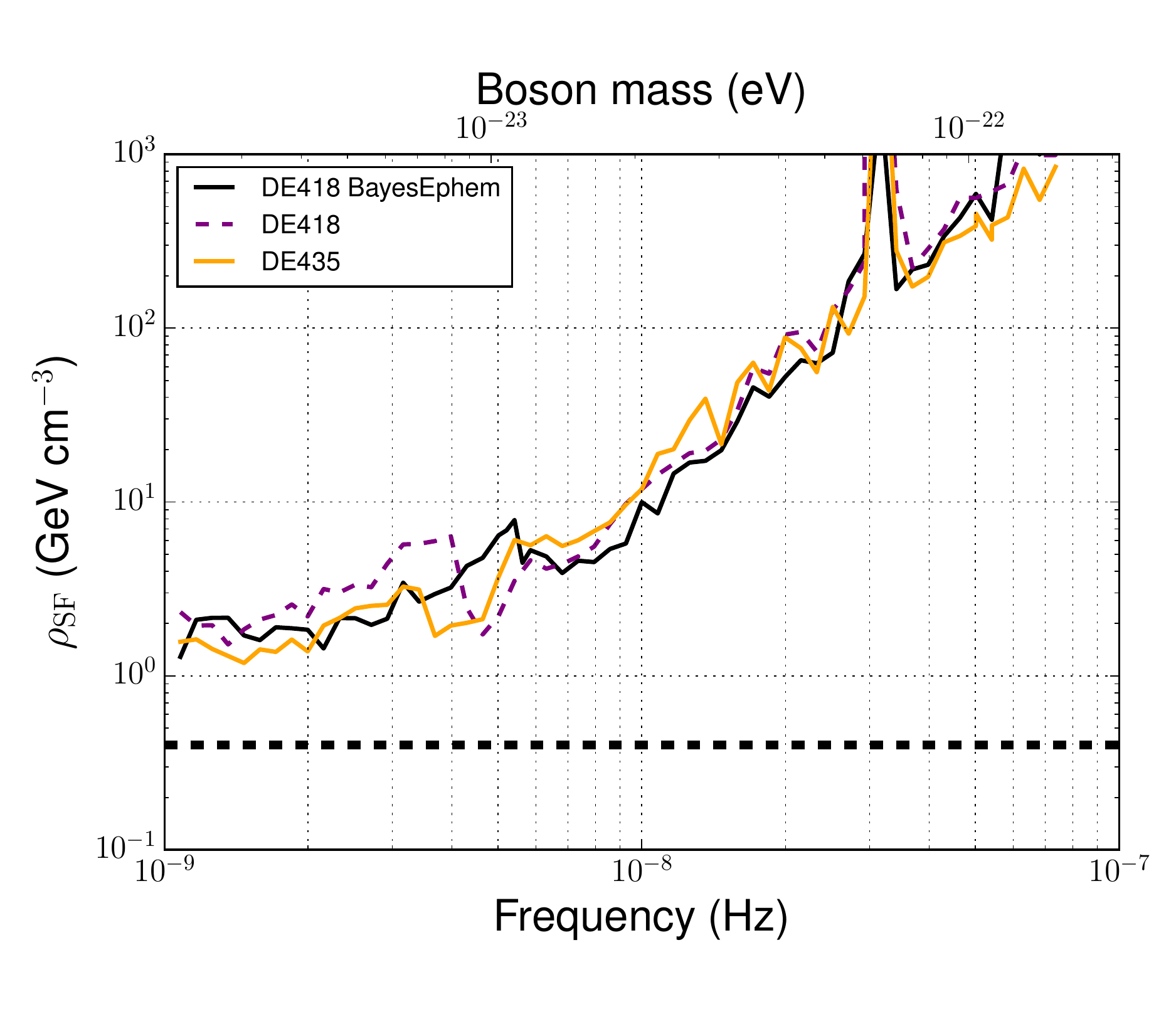}
\end{subfigure}
\caption{Upper limits on the density of fuzzy dark matter $\rho_{\text{SF}}$ in the Galaxy, as a function of frequency (boson mass). \textit{Left}: results from Frequentist analysis when only the Earth term is included (Freq E) or both terms are used (Freq E+P). \textit{Right}: Bayesian upper limits when SSE parameters are included in the search (BayesEphem), or using fixed DE418 and DE435 planet ephemerids. The horizontal black dashed line marks the measured local dark matter density $0.4\,\text{GeV}\,\text{cm}^{-3}$ \cite{DMdensity18}.}\label{fig:upper_lim2}
\end{figure*}

\section{Earth-term limits and effects of SSE}
\label{app:b}

When searching for continuous GWs in PTA data, it is common to use only the Earth terms. Similarly, for the case of scalar field dark matter, we can include in the analysis only Earth terms in Eq.~(\ref{eq:signalt}).
Although pulsar and Earth terms lie in the same frequency bin, we expect that for a sufficiently large set of pulsars, pulsar terms will be averaged out, as they all have different phases. In the left panel of Fig. \ref{fig:upper_lim2}, we compare the Frequentist upper limits on the density of scalar field dark matter $\rho_{\text{SF}}$ when only Earth terms are considered (black dashed) and when the full signal is used (purple solid). We find that both limits are comparable to each other. The noisy features in the (purple) solid curve are due to the amplitude modulation of pulsar terms; see Eq. (\ref{eq:signalt1}).

We also demonstrate the effects of SSE errors. In the right panel of Fig. \ref{fig:upper_lim2}, we show the upper limits obtained when DE418 and DE435 planetary ephemeris are used. The results with fixed ephemeris are overplotted with upper limits obtained with \texttt{BayesEphem} model, which accounts for uncertainties in the SSE. We see that the results are comparable, indicating the search for FDM signal, or continuous waves in general, is insensitive to SSE errors.

%The results imply that PPTA DR2 is sensitive to the choice of SSE model. BayesEphem can  successfully remove the systematic bias, associated with SSE uncertainties, and improve the upper limits in the low-frequency regime by a factor of 2.

\section{Noise properties for six PPTA pulsars}
\label{app:a}
\begin{figure*}
    \includegraphics[width=0.81\textwidth]{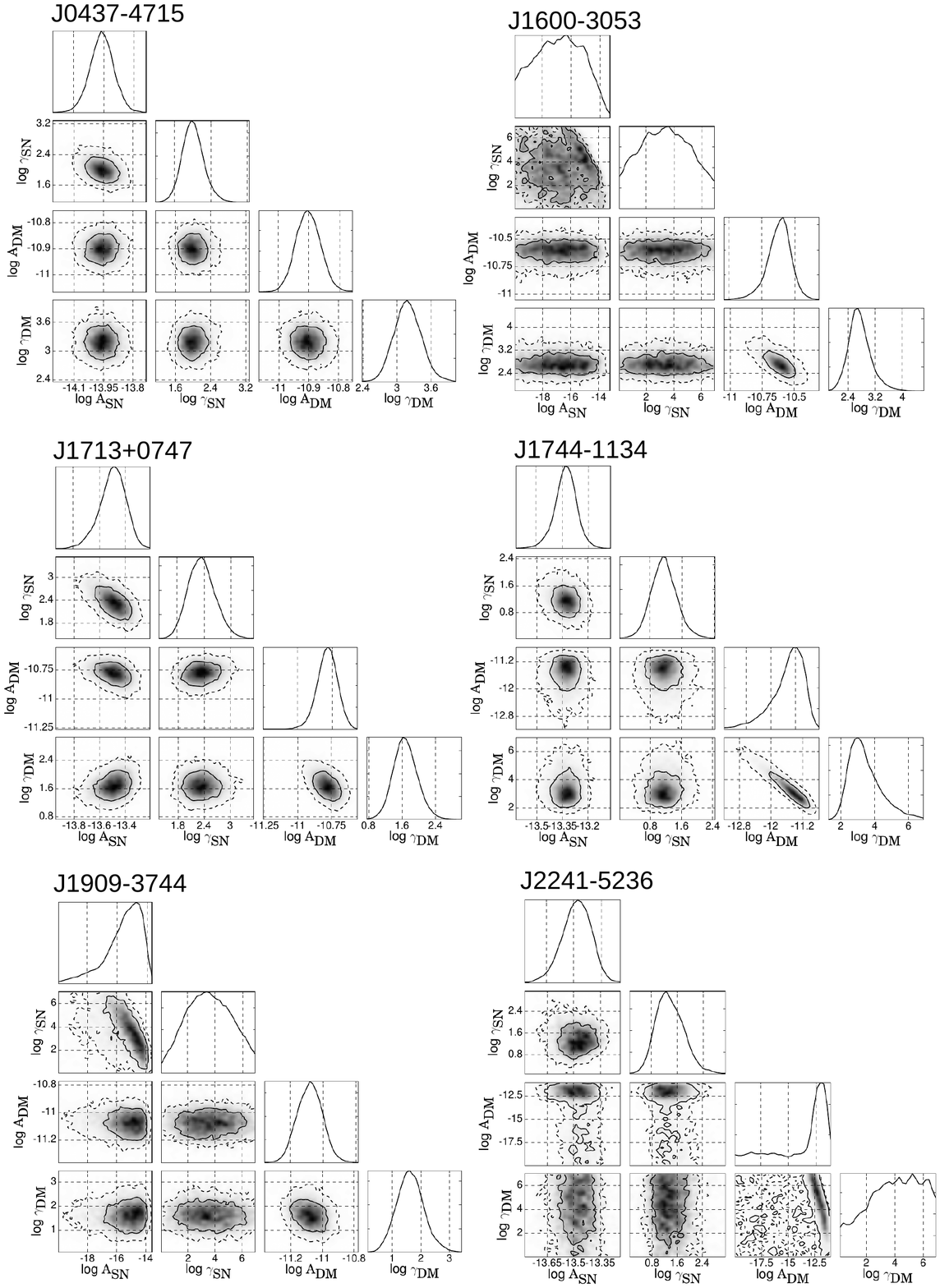}
    \caption{The one- and two-dimensional marginalized posterior distributions for the log-amplitude and slope of the DM and spin noises for the six best pulsars in the current PPTA data set.}
    \label{fig:dm_noise}
\end{figure*}

Figure \ref{fig:dm_noise} shows results of the Bayesian noise parameter estimation, described in Sec. \ref{sec:bayes1}, for the six most sensitive pulsars in the current PPTA data set.

\end{document}